\documentclass[preprint,prc,showpacs,preprintnumbers,
               superscriptaddress,amsmath,amssymb,floatfix]{revtex4}

\usepackage{graphicx}
\usepackage{dcolumn}
\usepackage{bm}
\usepackage{longtable}
\usepackage{multirow} 
\usepackage[figuresright]{rotating}
\usepackage{subfigure}
\newcommand{\beq}{\begin{equation}}
\newcommand{\eeq}{\end{equation}}
\newcommand{\beqn}{\begin{eqnarray}}
\newcommand{\eeqn}{\end{eqnarray}}
\newcommand{\btab}{\begin{tabular}}
\newcommand{\etab}{\end{tabular}}
\newcommand{\bcen}{\begin{center}}
\newcommand{\ecen}{\end{center}}

\newcommand{\ks}{$\blacksquare$}

\begin{document}

\title{Magic numbers for superheavy nuclei
in relativistic continuum Hartree-Bogoliubov theory}
\author{W. Zhang}
 \affiliation{School of Physics, Peking University, Beijing 100871}
\author{J. Meng}\thanks{e-mail: mengj@pku.edu.cn}
 \affiliation{School of Physics, Peking University, Beijing 100871}
 \affiliation{Institute of Theoretical Physics, Chinese Academy of
Science, Beijing 100080}
 \affiliation{Center of Theoretical Nuclear Physics, National Laboratory of
Heavy Ion Accelerator, Lanzhou 730000}
\author{S. Q. Zhang}
 \affiliation{School of Physics, Peking University, Beijing 100871}
\author{L. S. Geng}
 \affiliation{School of Physics, Peking University, Beijing 100871}
 \affiliation{Research Center for Nuclear Physics (RCNP),
 Osaka University, Ibaraki, Osaka, 567-0047, Japan}
\author{H. Toki}
 \affiliation{Research Center for Nuclear Physics (RCNP),
 Osaka University, Ibaraki, Osaka, 567-0047, Japan}

\date{\today}

\begin{abstract}
The magic proton and neutron numbers are searched in the
superheavy region with proton number $Z$=100 - 140 and neutron
number $N$= ($Z$+30) - (2$Z$+32) by the relativistic continuum
Hartree-Bogoliubov (RCHB) theory with interactions NL1, NL3, NLSH,
TM1, TW99, DD-ME1, PK1, and PK1R. Based on the two-nucleon
separation energies $S_{2p}$ and $S_{2n}$, the two-nucleon gaps
$\delta_{2p}$ and $\delta_{2n}$, the shell correction energies
$E_{shell}^{p}$ and $E_{shell}^{n}$, the pairing energies
$E_{pair}^{p}$ and $E_{pair}^{n}$, and the pairing gaps
$\Delta_{p}$ and $\Delta_{n}$, $Z$=120, 132, and 138 and $N$=172,
184, 198, 228, 238, and 258 are suggested to be the magic numbers
within the present approach. The $\alpha$-decay half-lives are
also discussed. In addition, the potential energy surfaces of
possible doubly magic nuclei are obtained by the
deformation-constrained relativistic mean field (RMF) theory, and
the shell effects stabilizing the nuclei are investigated.
Furthermore, the formation cross sections of $^{292}_{172}$120 and
$^{304}_{184}$120 at the optimal excitation energy are estimated
by a phenomenological cold fusion reactions model with the
structure information extracted from the constrained RMF
calculation.
\end{abstract}


\maketitle

\renewcommand{\baselinestretch}{1.5} \small \normalsize
\section{Introduction}

The structure of heavy and superheavy nuclei has been an
interesting field of nuclear physics research during the last
decades. Since the stability of superheavy nuclei is mainly
determined by the shell effects, it is important to find out the
regions in the ($Z$, $N$) plane where the shell effects are strong
and the long lifetime of the superheavy nuclei can be expected.
Exploring the limit of nuclear charge and mass is a long-term goal
of nuclear physics, and searching for the superheavy nuclei has
been enlivened by the hope of creating nuclei with masses and
charges much larger than those we are familiar with. The limit on
stability in superheavy nuclear region is essential for
understanding not only the nuclear structure, but also the
structure of the stars and the evolution of the universe. In the
classical droplet theory, the superheavy nuclei can not exist due
to their large Coulomb potentials. It is the shell effects that
play a major role for the very existence of nuclei with magic
numbers and provide higher stability and also higher abundance as
compared to their neighbors. A semblance of the same would work
also for superheavy nuclei, if there were magic numbers in this
region. Consequently, these nuclei will be guarded against a
faster decay by fission as compared to their non-magic
counterparts, and have more opportunities to be bound.

The pioneer theoretical work on the superheavy elements can be
found in the 1960's \cite{Myers1966, Meldner1967, Nilsson1969,
Mosel1969}. A series of calculations based on the
macroscopic-microscopic method (Nilsson-Strutinsky approach) with
the folded-Yukawa deformed single-particle potential
\cite{Moller1995} and with the Woods-Saxon deformed
single-particle potential \cite{Cwiok1983, Sobiczewski1989,
Patyk1991} are successful in reproducing the well-known
$\alpha$-decay half-lives of heavy elements. The existence of
island of superheavy nuclei would be separated in neutron and
proton number from the known heavy elements by a region of much
higher instability. Experimentally, the superheavy elements up to
$Z$=116 are synthesized or claimed to be synthesized
\cite{Hofmann1995-1, Hofmann1995-2, Hofmann1996, Ghiorso1995,
Lazarev1994, Lazarev1995, Lazarev1996, Oganessian1999,
Oganessian2000-1, Oganessian2000-2, Oganessian2004} by different
heavy ion reaction types including the cold fusion reactions. The
usual view is that although these isotopes are very heavy indeed,
they are not examples of the originally sought island of
superheavy elements.

There is no consensus among different theories with regard to the
center of the island of superheavy nuclei. Based upon
phenomenological models such as finite-range droplet model (FRDM),
the shell closure were predicted at $Z$=114 and $N$=184
\cite{Moller1994}. Additionally, the FRDM also predicts larger
shell gaps at $Z$=104, 106, 108, 110 and at $N$=162, 164
\cite{Moller1994}. Based on Nilsson-Strutinsky scheme, a similar
pattern of deformed nuclei have been predicted about $Z$=108 and
$N$=162 as in FRDM \cite{Patyk1991, Sobiczewski1994}. However, the
main obstacle is the question whether the macroscopic approaches
which apply to the region of $\beta$-stability line can be
extrapolated to the superheavy nuclei. Recently, microscopic
calculations \cite{Wu1996, Lalazissis1996, Cwiok1996, Rutz1997,
Bender1999, Meng2000, Long2002, Geng2003} are also attempted in
describing the superheavy nuclei. In the framework of relativistic
Hartree-Bogoliubov (RHB) theory, calculation with a finite-range
pairing force of Gogny interaction D1 and effective interaction
NLSH show that $Z$=114 and $N$=160, $N$=166, $N$=184 exhibit
stability compared to their neighbors and indications for a doubly
magic character at $Z$=106 and $N$=160 are also observed
\cite{Lalazissis1996}. The Skyrme Hartree-Fock (SHF) method with
interactions SkP and SLy7 predicts magic numbers at $Z$=126 and
$N$=184, and also predicts the increased stability due to the
deformed shell effects at $N$=162 \cite{Cwiok1996}. Considering
non-relativistic SHF effective interactions SkM*, SkP, SLy6, SkI1,
SkI3, SkI4 and relativistic mean field (RMF) effective
interactions PL-40, NLSH, NL-Z, TM1, Ref. \cite{Rutz1997} gave the
doubly magic spherical nuclei $_{184}$114, $_{172}$120 and
$_{184}$126 based on two-nucleon gaps $\delta_{2p}$ and
$\delta_{2n}$. The common point of such studies is that the magic
numbers depend on effective interactions. The study of superheavy
nuclei remains a challenge for interaction, which have to be
rigorously tested by a wide variety of nuclear properties
throughout the periodic table.

The extrapolation towards superheavy nuclei challenges the
predictive power of nuclear structure models. The relativistic
mean field (RMF) theory has achieved a great success in describing
many nuclear phenomena \cite{Ring1996} and it is believed as one
of the best candidates for application to superheavy nuclei. The
relativistic continuum Hartree-Bogoliubov (RCHB) theory is the
extension of the RMF and the Bogoliubov transformation in the
coordinate representation \cite{Meng1998a}. As the RCHB formalism
allows for the proper description of the coupling between the
bound states and the continuum by the pairing force, therefore it
is suitable not only for stable nuclei but also for the nuclei
near the drip line \cite{Meng1996, Meng1998b, Meng1998c,
Meng2002a, Meng2002b}. The RCHB theory has shown a remarkable
success in the description of nuclei with unusual $N/Z$ ratio,
e.g., the halo in $^{11}$Li \cite{Meng1996} and the description of
the giant halo at the neutron drip line in Zr and Ca isotopes
\cite{Meng1998b, Meng2002b}. A systematical study on $^{259}$Db
and its $\alpha$-daughter nuclei were also achieved
\cite{Long2002}. In principle, only a calculation in a large
multidimensional deformation space can definitively decide the
appropriate ground-state shape. However, as the traditional
superheavy nuclei are expected to be spherical and are located on
the nuclear chart around a spherical doubly magic nucleus next to
$^{208}_{126}$Pb \cite{Smolanczuk1997}, the RCHB theory with the
assumption of spherical shape can be applied to this preliminary
scan for magic numbers. It is the aim of this paper to predict
doubly magic superheavy nuclei within the RCHB theory using a
variant of the RMF interactions. As most of the presently found
superheavy nuclei have large deformations, we will study the
potential energy surface with the deformed-constrained RMF theory
in order to confirm that the spherical configuration exists in
these doubly magic nuclei.

The material contained in this study is organized as follows. The
framework of RCHB and constrained RMF theory are outlined in
Section II. Section III contains the predictions made from the
two-nucleon separation energies $S_{2p}$ and $S_{2n}$, the
two-nucleon gaps $\delta_{2p}$ and $\delta_{2n}$, the shell
correction energies $E_{shell}^{p}$ and $E_{shell}^{n}$, the
pairing energies $E_{pair}^{p}$ and $E_{pair}^{n}$, and the
pairing gap $\Delta_{p}$ and $\Delta_{n}$, obtained in the RCHB
theory with interactions NL1 \cite{Joon1986}, NL3
\cite{Lalazissis1997}, NLSH \cite{Sharma1992}, TM1
\cite{Sugahara1994}, TW99 \cite{Typel1999}, DD-ME1
\cite{Niksic2002}, PK1 and PK1R \cite{Long2003}. The
$\alpha$-decay half-lives are also discussed there. The binding
energy and the corresponding macroscopic energy as a function of
deformation obtained in the deformation-constrained RMF
calculation for $Z$=120 are presented in Sec. IV. Additionally,
the formation cross sections for the possible doubly magic nuclei
$^{292}$120 and $^{304}$120 at the optimal excitation energy are
estimated by a phenomenological model proposed by Smola\'nczuk
\cite{Smolanczuk1999} with the input from the constrained RMF
theory. Finally, Section V contains the main conclusions of this
work.

\section{Framework}

\subsection{Relativistic continuum Hartree-Bogoliubov theory}
The basic ansatz of the RMF theory is a Lagrangian density where
nucleons are described as Dirac particles which interact via the
exchange of various mesons. The Lagrangian density considered has
the form:
 \beqn \nonumber \cal L&=&\bar\psi \left[
i\gamma^\mu\partial_\mu-M-g_\sigma\sigma-g_\omega\gamma^\mu\omega_\mu
-g_\rho\gamma^\mu \vec\tau \cdot \vec\rho_\mu - e\gamma^\mu\dfrac{1-\tau_3}{2}A_\mu \right] \psi\\
\nonumber
&&+\dfrac{1}{2}\partial^\mu\sigma\partial_\mu\sigma-\dfrac{1}{2} m_\sigma^2\sigma^2
-\dfrac{1}{3}g_2\sigma^3-\dfrac{1}{4}g_3\sigma^4\\
\nonumber
&&-\dfrac{1}{4}\Omega^{\mu\nu}\Omega_{\mu\nu}+\dfrac{1}{2}m_\omega^2\omega^\mu\omega_\mu
+\dfrac{1}{4}c_3 (\omega^\mu\omega_\mu)^2\\
\nonumber
&&-\dfrac{1}{4}\vec R^{\mu\nu}\cdot\vec R_{\mu\nu}+\dfrac{1}{2}m_\rho^2\vec\rho^\mu\cdot\vec\rho_\mu
+\dfrac{1}{4}d_3(\vec\rho^\mu\cdot\vec\rho_\mu)^2\\
&&-\dfrac{1}{4}F^{\mu\nu}F_{\mu\nu}
 \label{Eq:lagrangian}
 \eeqn
where $\psi$ is Dirac spinor and $\bar\psi=\psi^\dag\gamma^0$.
$M$, $m_{\sigma}$, $m_{\omega}$, and $m_{\rho}$ are the nucleon,
$\sigma$, $\omega$ and $\rho$ meson masses respectively, while
$g_{\sigma}$, $g_{2}$, $g_{3}$, $g_{\omega}$, $c_{3}$, $g_{\rho}$,
$d_{3}$, and $e^2/4\pi$ = 1/137 are the corresponding coupling
constants for the mesons and the photon. The field tensors of the
vector mesons ($\omega$ and $\rho$ mesons) and of the
electromagnetic fields take the following form:
\begin{eqnarray}
\Omega^{\mu\nu}
 &=& \partial^\mu\omega^\nu - \partial^\nu\omega^\mu \nonumber \\
\vec R^{\mu\nu}
 &=& \partial^\mu \vec\rho^\nu - \partial^\nu \vec\rho^\mu
    - 2g_\rho\vec\rho^\mu\times\vec\rho^\nu \\
F^{\mu\nu}
 &=& \partial^{\mu}A^{\nu} - \partial^{\nu}A^{\mu} \nonumber
\end{eqnarray}
With the classical variational principle, one can obtain the
coupled equations of motion, which contain the Dirac equation for
the nucleons and the Klein-Gordon type equations for the mesons
and the photon. The coupled equations are self-consistently solved
by iterations.

For the RMF theory, however, as the classical meson fields are
used, the equations of motion for nucleons derived from Eq.
(\ref{Eq:lagrangian}) do not contain pairing interaction. In order
to have two-body interaction, one has to quantize the meson fields
which leads to a Hamiltonian with two-body interaction. Following
the standard procedure of Bogoliubov transformation, a Dirac
Hartree-Bogoliubov equation could be derived and then a unified
description of the mean field and pairing correlation in nuclei
could be achieved \cite{Kucharek1991}. The Dirac
Hartree-Bogoliubov equations are as following:
 \beq
   \int d^3r'
   \left( \begin{array}{cc}
          h-\lambda &   \Delta \\
          \Delta    &  - h+\lambda
          \end{array} \right)
   \binom { \psi_U }{\psi_V }   ~
   = ~ E ~ \binom{ \psi_U}{ \psi_V } ,
\label{Eq:ghfb}
 \eeq
 where, $ h = {\vec \alpha} \cdot {\vec p}
     + g_\omega \omega + g_\rho \tau_3 \rho
     + \beta (M+ g_\sigma \sigma)
$ is the Dirac Hamiltonian and the Fock term has been neglected as
is usually done in RMF. A Lagrange multiplier $\lambda$ is
introduced to adjust the proper particle number of neutrons and
protons. The pairing potential $\Delta$ in Eq. (\ref{Eq:ghfb}) is
:
\begin{eqnarray}
   \Delta_{ab} = \frac{1}{2} \sum_{cd}V_{abcd}^{pp} \kappa_{cd}.
\label{Eq:gap}
\end{eqnarray}
It is obtained from one-meson exchange interaction $V_{abcd}$ in
the $pp$-channel and the pairing tensor $\kappa=\psi_U^*\psi_V^T$.
The interaction $V_{abcd}$ used for the pairing potential in Eq.
(\ref{Eq:gap}) is either the density-dependent two-body force of
zero range or Gogny-type finite range force as shown in Ref.
\cite{Meng1998a}. Considering the consuming computational time, we
prefer to use the density-dependent zero range $\delta$ force with
the interaction strength $V_0$ and the nuclear matter density
$\rho_0$ as
 \beq
   V(\vec{r}_1,\vec{r}_2) = V_0
     \delta(\vec{r}_1-\vec{r}_2)
     \frac{1}{4}\left[1-
     \vec{\mbox{$\sigma$}}_1\cdot\vec{\mbox{$\sigma$}}_2\right]
     \left(1 - \frac{\rho(r)}{\rho_0}\right).
\label{Eq:vpp}
 \eeq

The relativistic continuum Hartree-Bogoliubov theory solves the
RHB equations in the coordinate representation in order to
describe both the continuum and the bound states self-consistently
\cite{Meng1998a}. It is then applicable to not only stable nuclei
but also exotic nuclei. When the $\delta$-force in the pairing
channel is used, the RCHB equations are a set of four coupled
differential equations, which are solved in a self-consistent way
by the shooting method and the Runge-Kutta algorithm. More details
can be found in Ref. \cite{Meng1998a}.

In this paper, the RCHB equations are solved in a spherical box
with radius $R$=25 fm, the step size of 0.1 fm, and proper
boundary conditions. It has been shown in Refs. \cite{Meng1998a,
Meng1998b} that the results do not depend on the box size when
$R\geq15$ fm. The strength $V_0$ of the pairing force for proton
and neutron is fixed at -650 MeV$\cdot$fm$^{-3}$. For $\rho_0$ we
use the nuclear matter density 0.152 fm$^{-3}$.

\subsection{Deformation-constrained RMF theory}

To examine the ground state geometrical configuration of certain
heavy or superheavy nucleus, as well as its fission barrier, one
has to obtain the potential energy surface. The potential energy
curve can be calculated microscopically by the
deformation-constrained RMF theory. The binding energy at certain
deformation value is obtained by constraining the quadrupole
moment $\langle Q_2 \rangle$ to a given value $\mu_2$ in the
expectation value of the Hamiltonian \cite{Ring1980},
 \beq
  \langle H'\rangle~
   =~\langle H\rangle
    +\displaystyle\frac{1}{2} C_{\mu} \left(\langle Q_2\rangle -\mu_2\right)^2,
 \eeq
where $C_{\mu}$ is the constraint multiplier. The expected
value of the quadrupole moment operator is
$<Q_2>=<Q_2^n>+<Q_2^p>$, where the quadrupole moment operator for
neutrons (protons) is calculated by $<Q_2^{n(p)}>=<2 r^2
P_2(\cos\theta)>_{n (p)}$. The equations of motion are solved by
expanding the Dirac spinors in harmonic oscillator basis in the
cylindrical coordinates. The size of the basis is crucial and
responsible for the reliability of the result. By varying $\mu_2$,
the binding energy at different deformation can be obtained. The
pairing is taken into account by the constant gap approximation
(BCS) in which the pairing gap is taken as $\dfrac{12}{\sqrt{A}}$
for even number nucleons. An enormous advantage of microscopic
constrained RMF calculations compared with the phenomenological
calculations is that in the microscopic constrained calculation we
need only follow a one-parameter line, e.g. $\beta_2$, while in
the phenomenological methods we must calculate multidimensional
energy surfaces \cite{Ring1980}.

\section{Magic numbers in superheavy nuclei}

\begin{figure}[htbp]
\centering
\includegraphics[width=10cm]{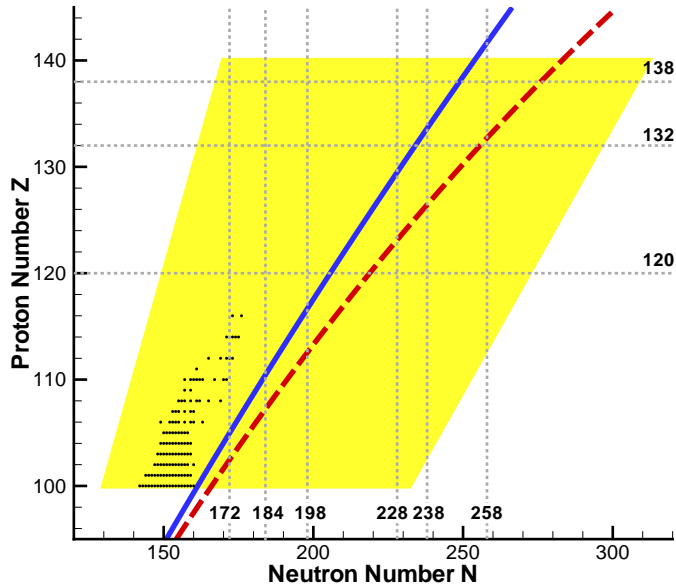}
\caption{Nuclear chart in the superheavy region. The long-dashed
and solid lines represent the $\beta$-stability lines by formula
$N-Z=6\times 10^{-3}\cdot A^{5/3}$ \cite{Bohr1969} and
$Z=\dfrac{A}{1.98+0.0155\cdot A^{2/3}}$ \cite{Marmier1971}
respectively. The dots lying at the lower-left corner represent
the superheavy nuclei observed or declared to be observed
experimentally. } \label{nuch}
\end{figure}

The doubly magic nuclei are searched for the even-even nuclei
(more than 1200 nuclei) in the nuclear chart with proton number
$Z$=100 - 140 and neutron number $N$= ($Z$+30) - (2$Z$+32) (i.e.,
$N$=130 - 312) by the RCHB calculation with the effective
interactions NL1 \cite{Joon1986}, NL3 \cite{Lalazissis1997}, NLSH
\cite{Sharma1992}, TM1 \cite{Sugahara1994}, TW99 \cite{Typel1999},
DD-ME1 \cite{Niksic2002}, PK1 and PK1R \cite{Long2003}. These
nuclei are presented as shaded area in Fig. \ref{nuch}. The
long-dashed line and solid line in Fig. \ref{nuch} represent the
$\beta$-stability line of the parametrization \beq N-Z=6\times
10^{-3}\cdot A^{5/3} \label{beta1} \eeq from Ref. \cite{Bohr1969}
and \beq Z=\dfrac{A}{1.98+0.0155\cdot A^{2/3}} \label{beta2} \eeq
from Ref. \cite{Marmier1971}, respectively. The black dots lying
at the lower-left corner represent the superheavy nuclei which
have been observed or declared to be observed experimentally
\cite{NPR2003}. The vertical and horizontal lines in Fig.
\ref{nuch} represent the possible magic proton and neutron numbers
which will be discussed later.

For RMF calculations, there exist many effective interactions,
which provide nearly equal quality of description for stable
nuclei. However, when they are applied to the exotic nuclei, the
differences appear. In view of the uncertainties, the effective
interactions considered here are four frequently used interactions
NL1, NL3, NLSH, TM1, density-dependent interactions TW99 and
DD-ME1, as well as the newly developed PK1 and PK1R which were
adjusted with the elaborate consideration of the cent-of-mass
correction. The effective interactions PK1 and PK1R are listed in
Table \ref{pk}.

In the following four subsections, the signatures of shell closure
on the two-nucleon separation energies $S_{2p}$ and $S_{2n}$, the
two-nucleon gaps $\delta_{2p}$ and $\delta_{2n}$, the shell
correction energies $E_{shell}^{p}$ and $E_{shell}^{n}$, the
pairing energies $E_{pair}^{p}$ and $E_{pair}^{n}$ and the pairing
gaps $\Delta_{p}$ and $\Delta_{n}$ obtained in RCHB calculation
will be analyzed consecutively. The $\alpha$-decay half-lives
$T_{\alpha}$ will be discussed afterwards. At last, the prediction
of the magic proton and neutron numbers in superheavy nuclei will
be summarized.

\subsection{Two-nucleon separation energies}

The two-nucleon separation energy, \beq
S_{2n}(N,Z)=E_B(N,Z)-E_B(N-2,Z),~~~~S_{2p}(N,Z)=E_B(N,Z)-E_B(N,Z-2)
\eeq is better to quantify shell effects than the single-nucleon
separation energy due to the absence of odd-even effects, here
$E_B(N,Z)$ is the binding energy of the nucleus $^{N+Z}$Z. For an
isotonic chain, the $S_{2p}$ become smaller with the proton number
$Z$. A jump of $S_{2p}$ indicates the occurrence of a proton shell
closure.

\begin{figure}[htbp]
\centering
\includegraphics[scale=0.35]{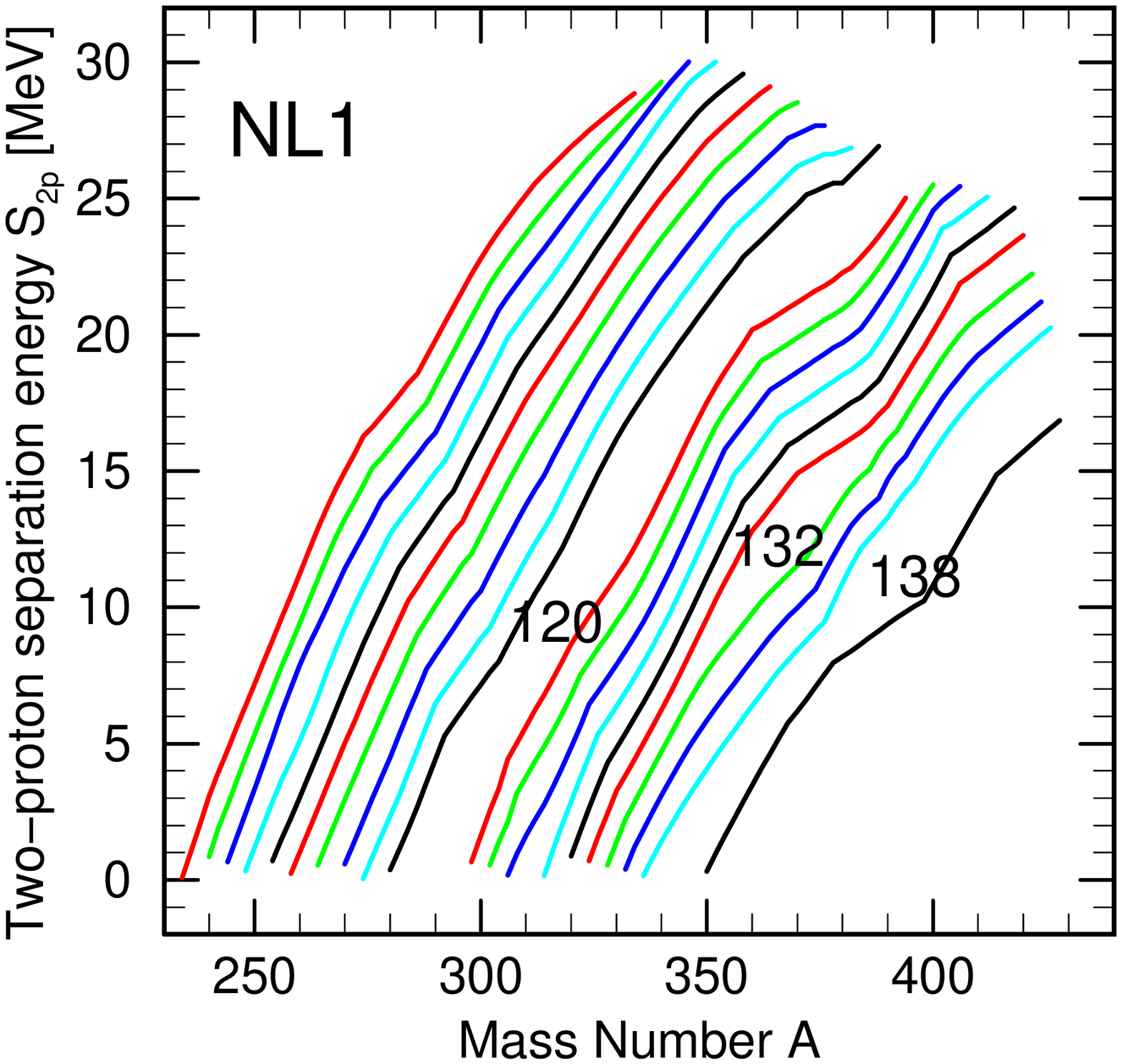}\hspace{0.5cm}
\includegraphics[scale=0.35]{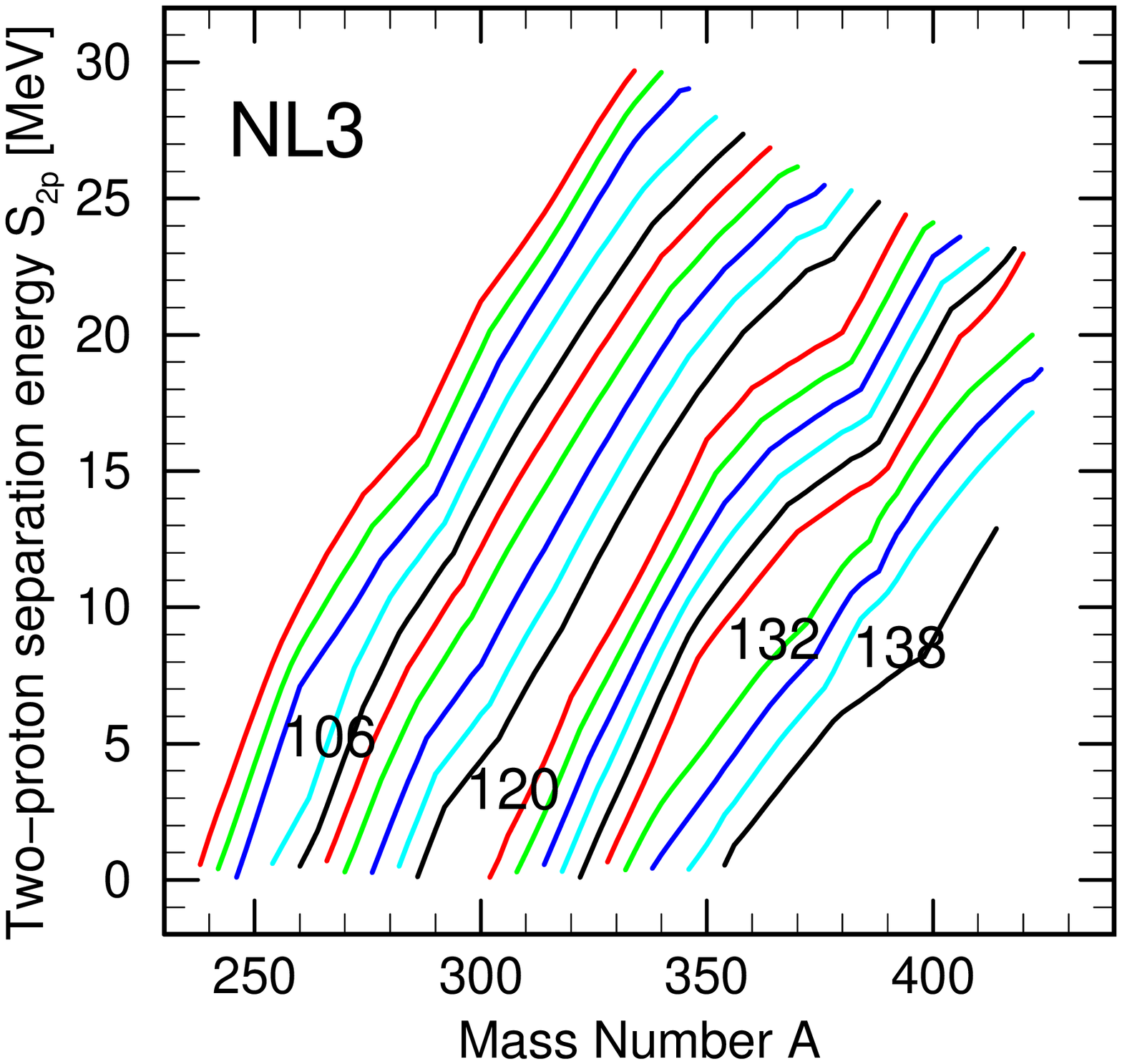}
\includegraphics[scale=0.35]{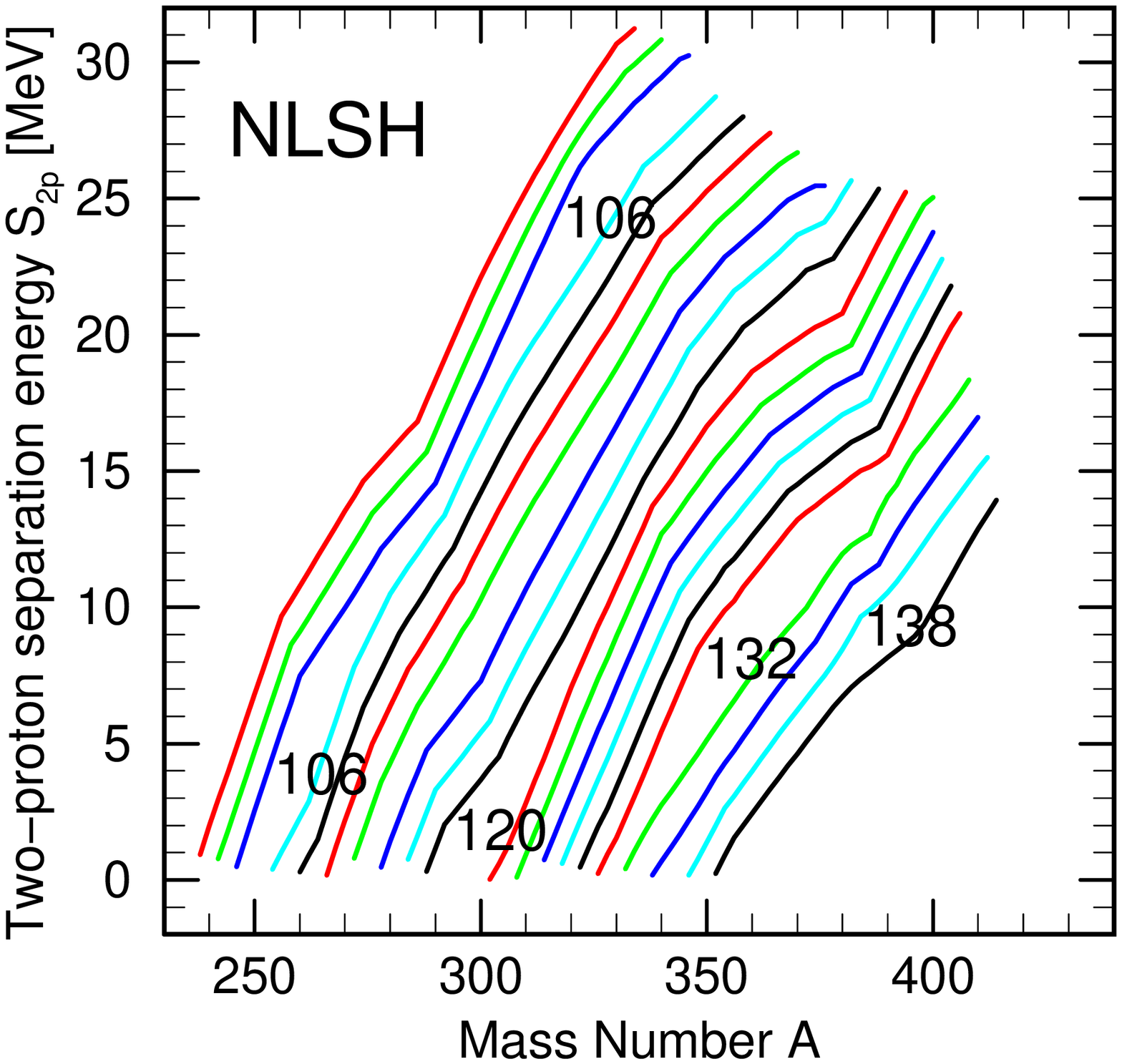}\hspace{0.5cm}
\includegraphics[scale=0.35]{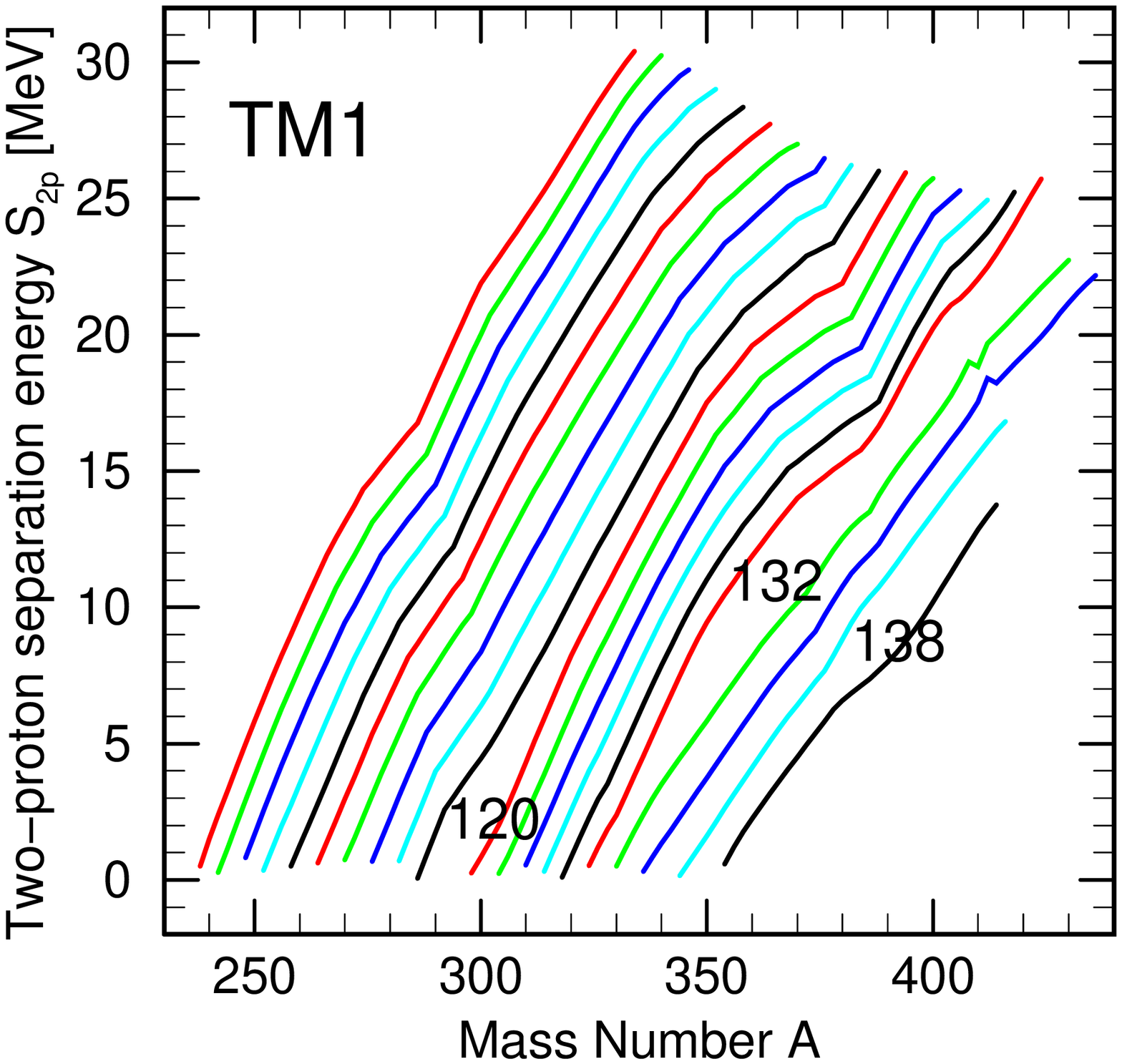}
\includegraphics[scale=0.35]{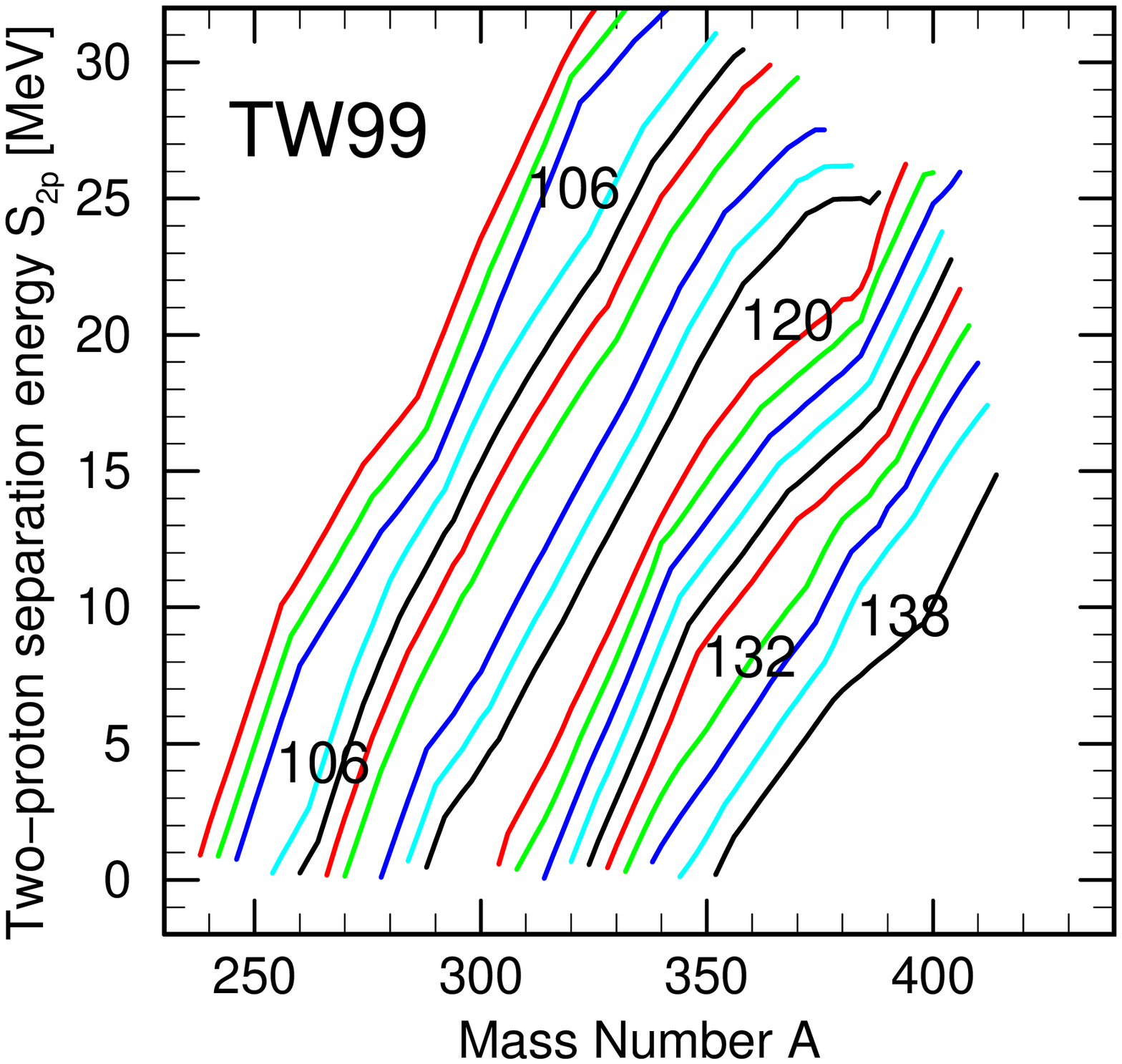}\hspace{0.5cm}
\includegraphics[scale=0.35]{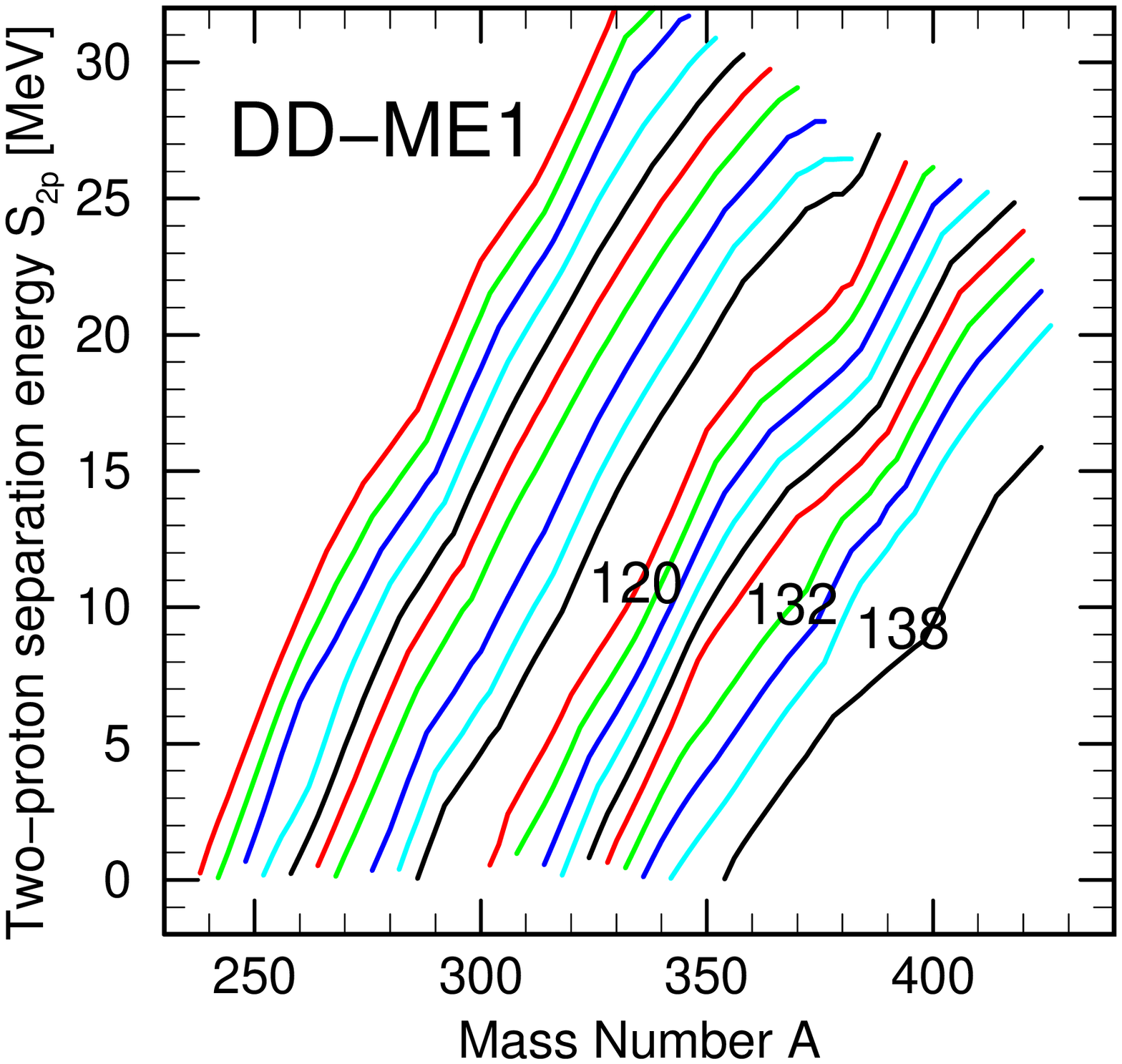}
\includegraphics[scale=0.35]{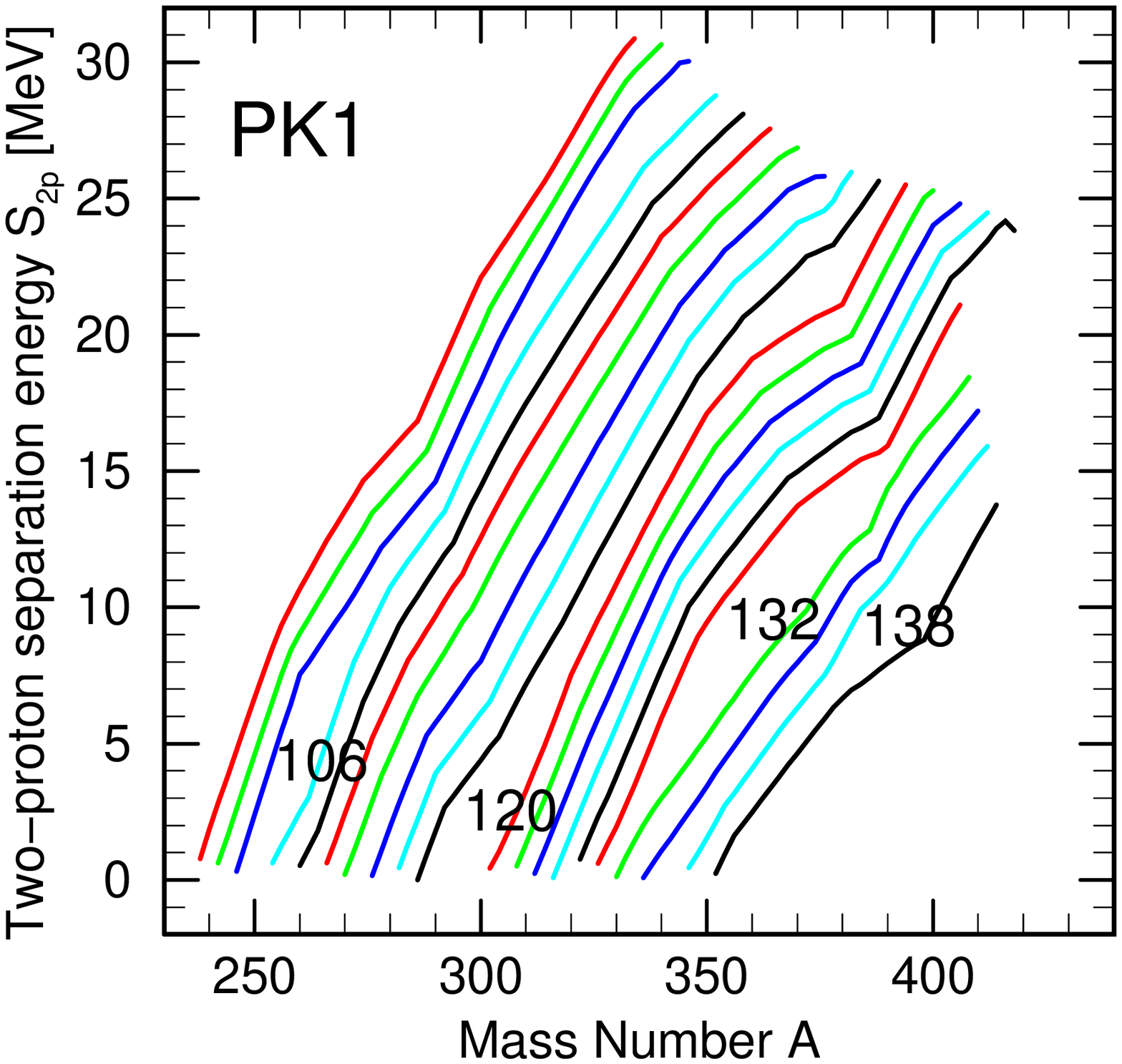}\hspace{0.5cm}
\includegraphics[scale=0.35]{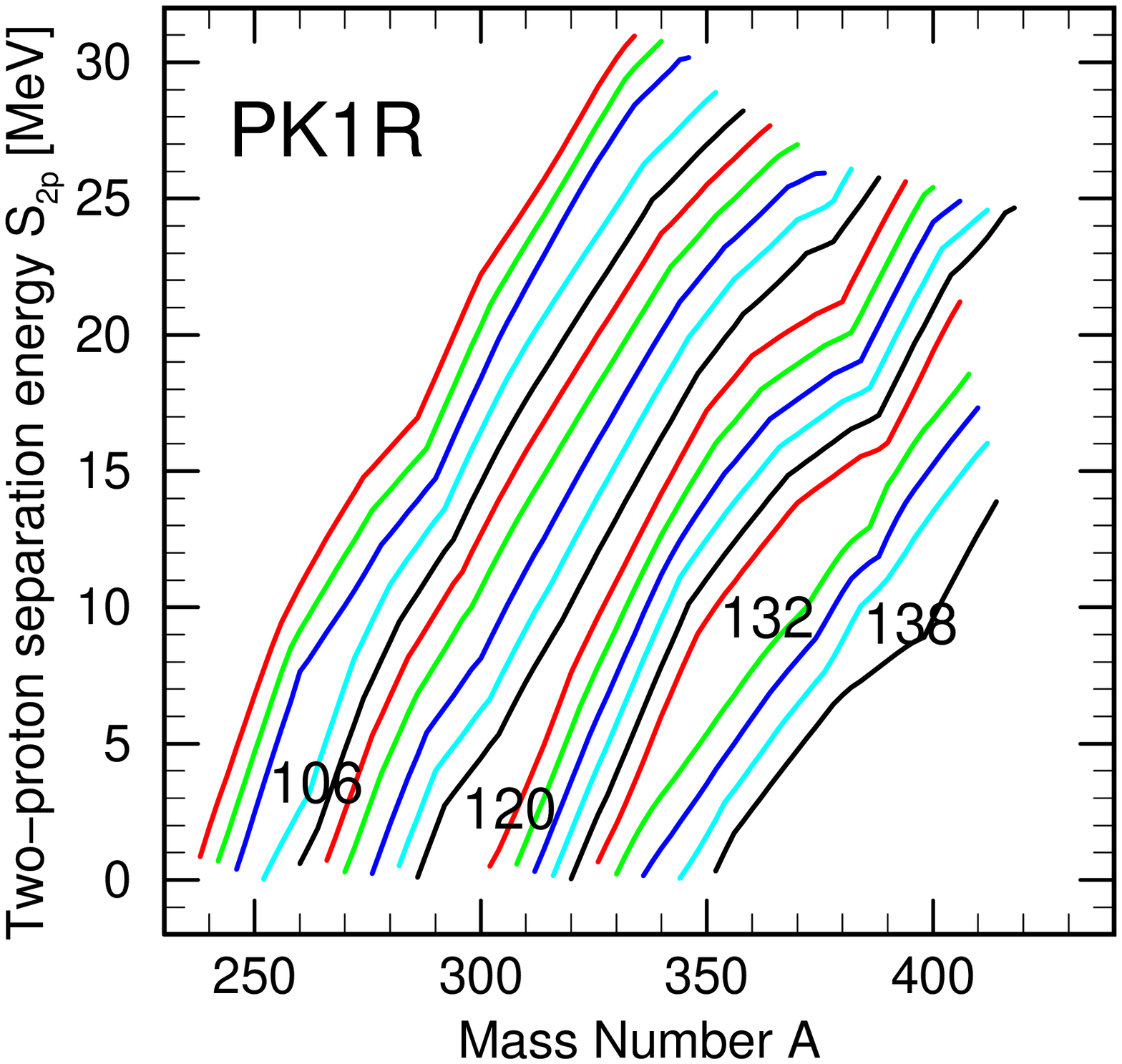}
\caption{The two-proton separation energies
$S_{2p}(N,Z)=E_B(N,Z)-E_B(N,Z-2)$ as a function of mass number $A$
obtained by RCHB calculation with effective interactions NL1, NL3,
NL-SH, TM1, TW-99, DD-ME1, PK1, and PK1R, respectively.}
\label{s2p}
\end{figure}

For the sake of a clear presentation of the results, the
two-proton separation energies $S_{2p}$ from the RCHB calculation
for even-even nuclei with $Z$=102 - 140 as a function of mass
number $A$ with the effective interactions NL1, NL3, NLSH, TM1,
TW99, DD-ME1, PK1, and PK1R are shown in Fig. \ref{s2p}. In the
figure, $S_{2p}(N,Z)$ for the same $Z$ are connected to form a
curve so that the jumps of $S_{2p}$ compose a gap between the
curves. The 20 $S_{2p}$ curves in each subfigure correspond to
isotopic chains for $Z$ ranging from 102 to 140 with a step of
two. The curve at the upper-left corner is for $Z$=102 and the
neighboring one is for $Z$=104, and so on.

For an isotonic chain, the $S_{2p}$ gradually decrease from about
25 MeV to zero for the curve at the upper-left corner to the
lower-right corner. Each curve is roughly in parallel with the
other. A gap is enchased if a magic proton number exists. Take
$S_{2p}$ for the effective interaction NL1 as an example: the gap
between $Z$=120 and $Z$=122 (abbr. at $Z$=120) is large and
stable, the gap at $Z$=132 is visible just in the region $N$=220 -
250, and the gap at $Z$=138 is relatively large but unstable. The
gap at $Z$=138 becomes weak at $N$$\approx$ 240.

For all effective interactions, the magic proton numbers $Z$=120,
132, and 138 are common while $Z$=106 is observed only for NL3,
NLSH, TW99, PK1, and PK1R. Generally, the gaps do not exist for
all isotopes, e.g., the gap at $Z$=132 for NL1 exists only for
$N$=220 - 250, and the gap at $Z$=138 for PK1 exists only for
$N$=230 - 270.

\begin{figure}[htbp]
\centering
\includegraphics[scale=0.3]{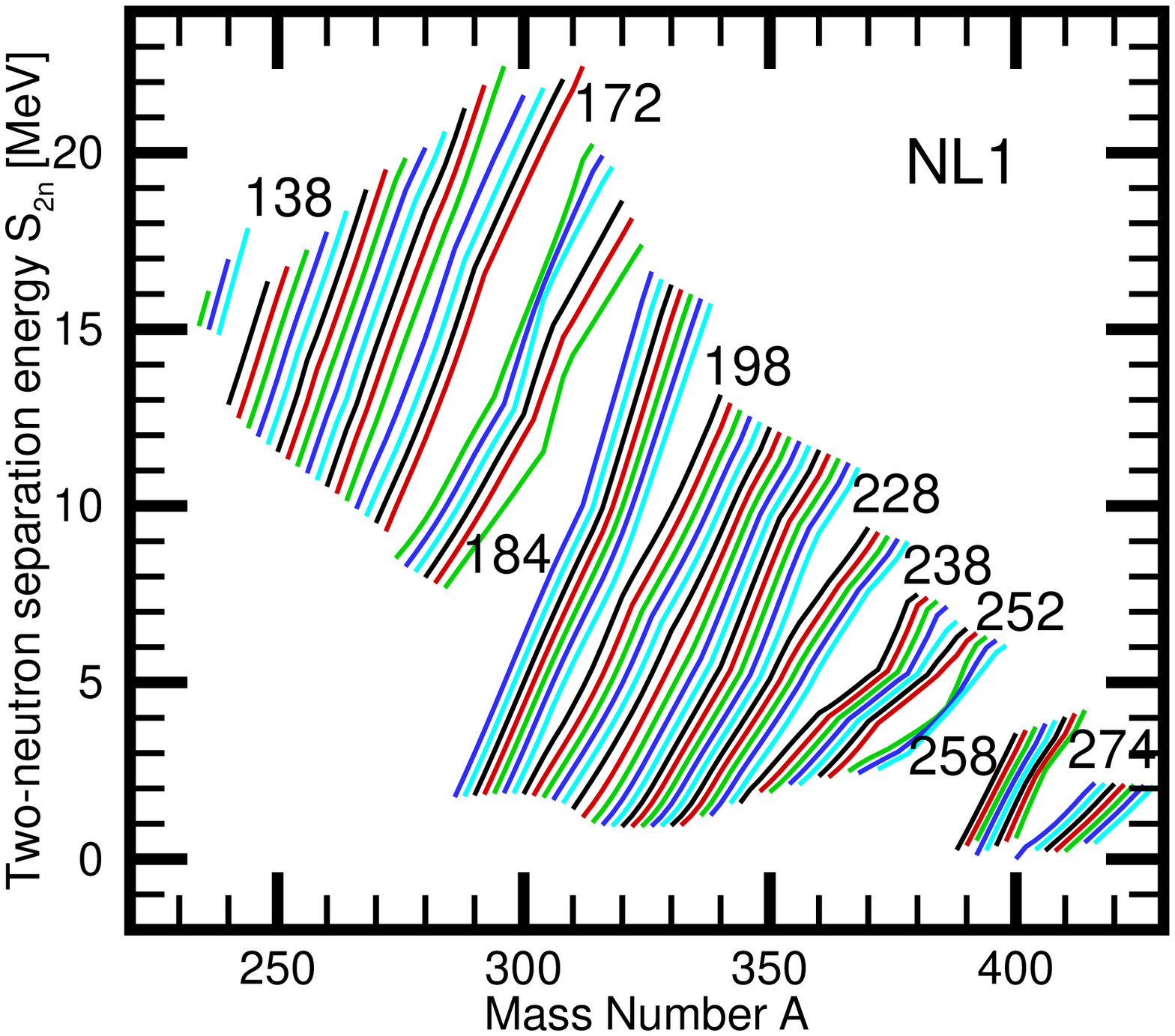}
\includegraphics[scale=0.3]{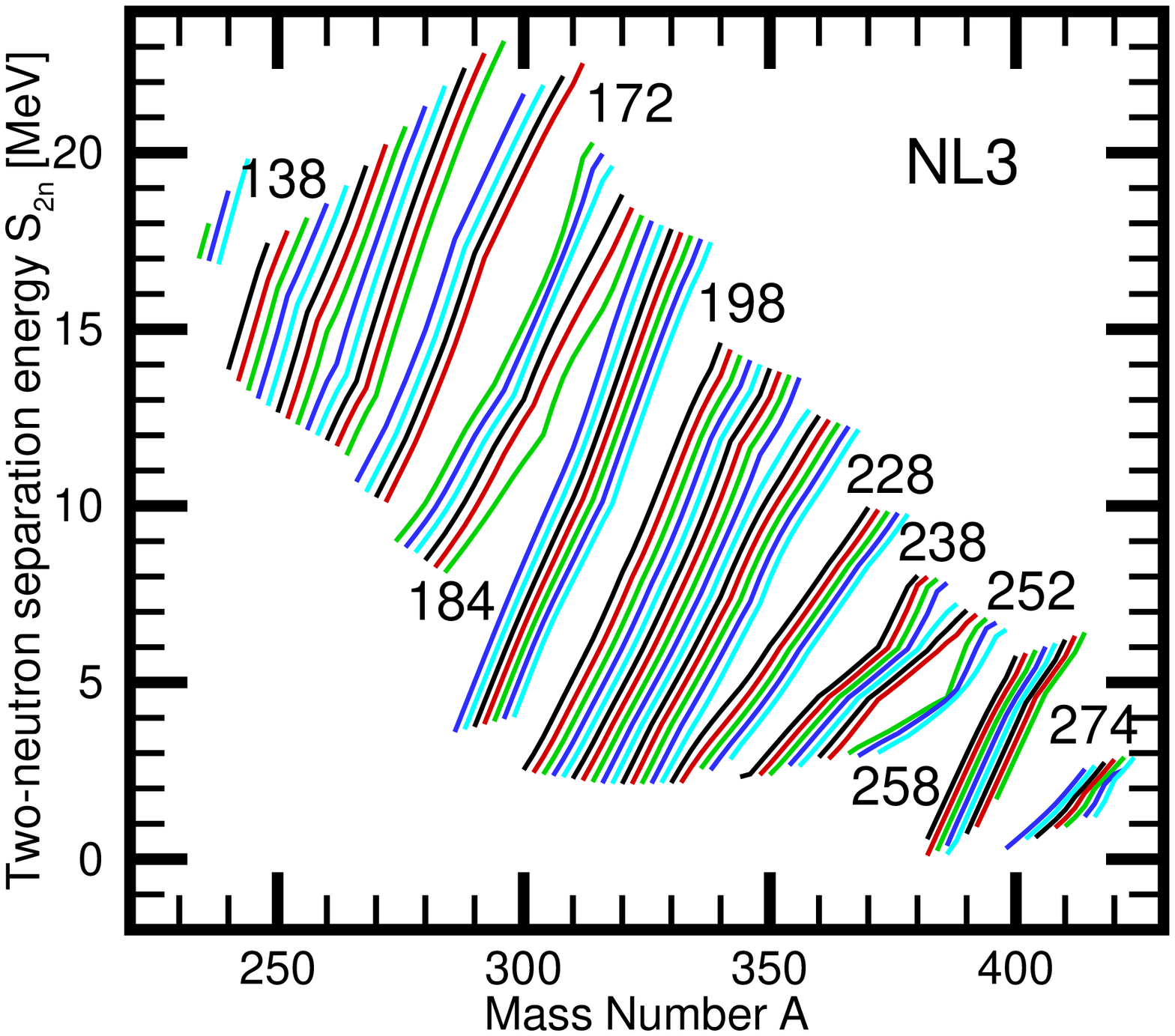}
\includegraphics[scale=0.3]{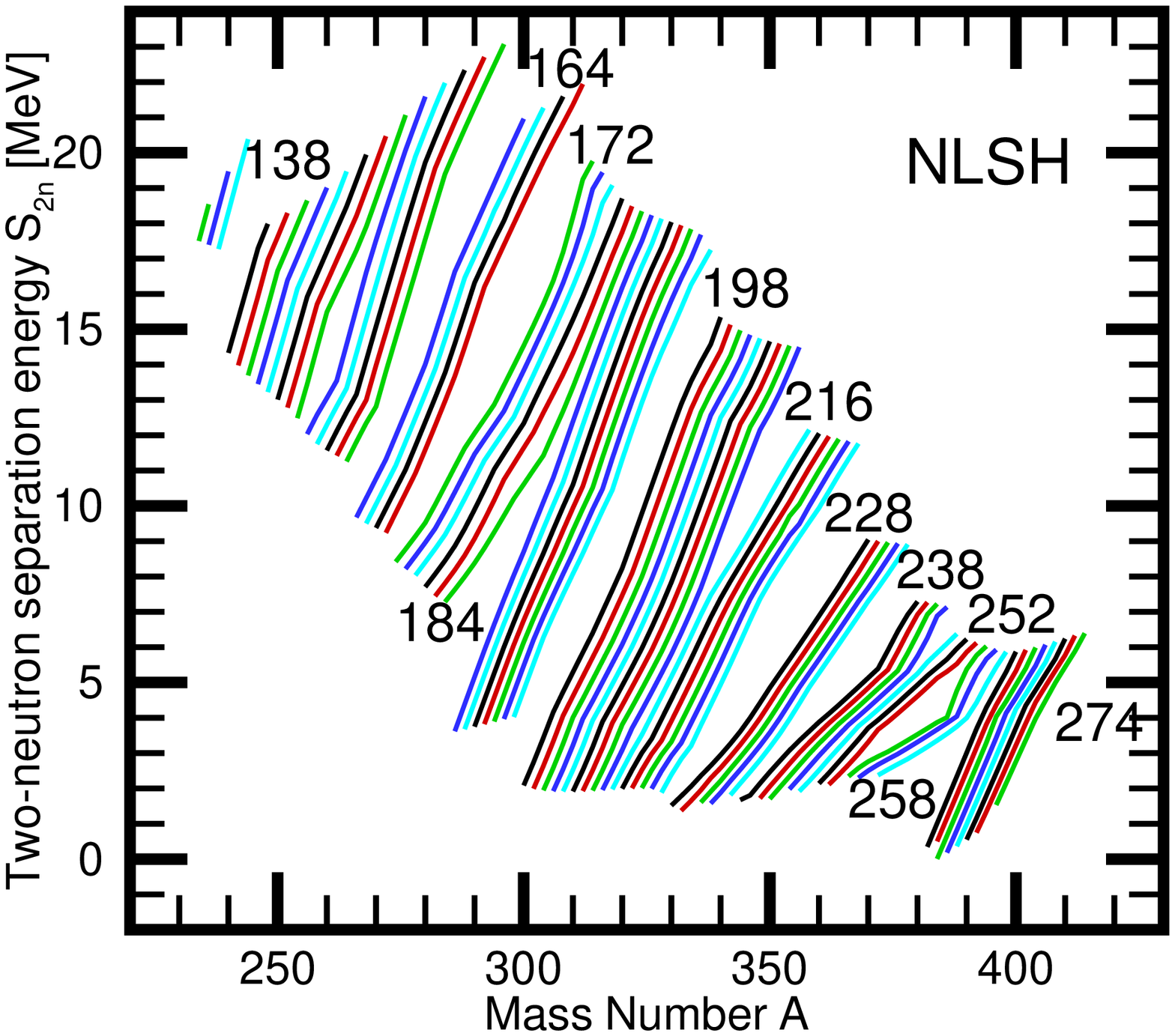}
\includegraphics[scale=0.3]{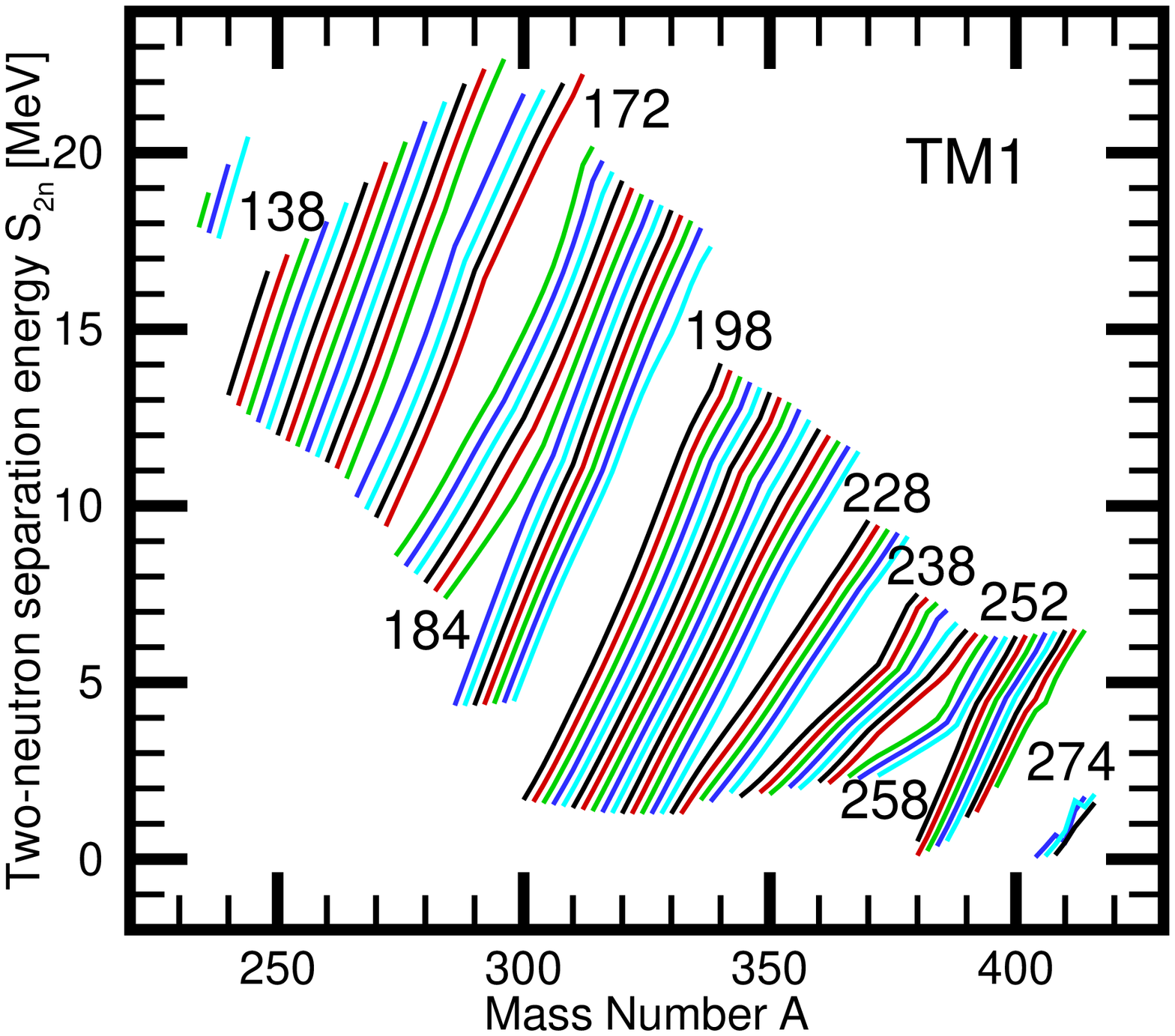}
\includegraphics[scale=0.3]{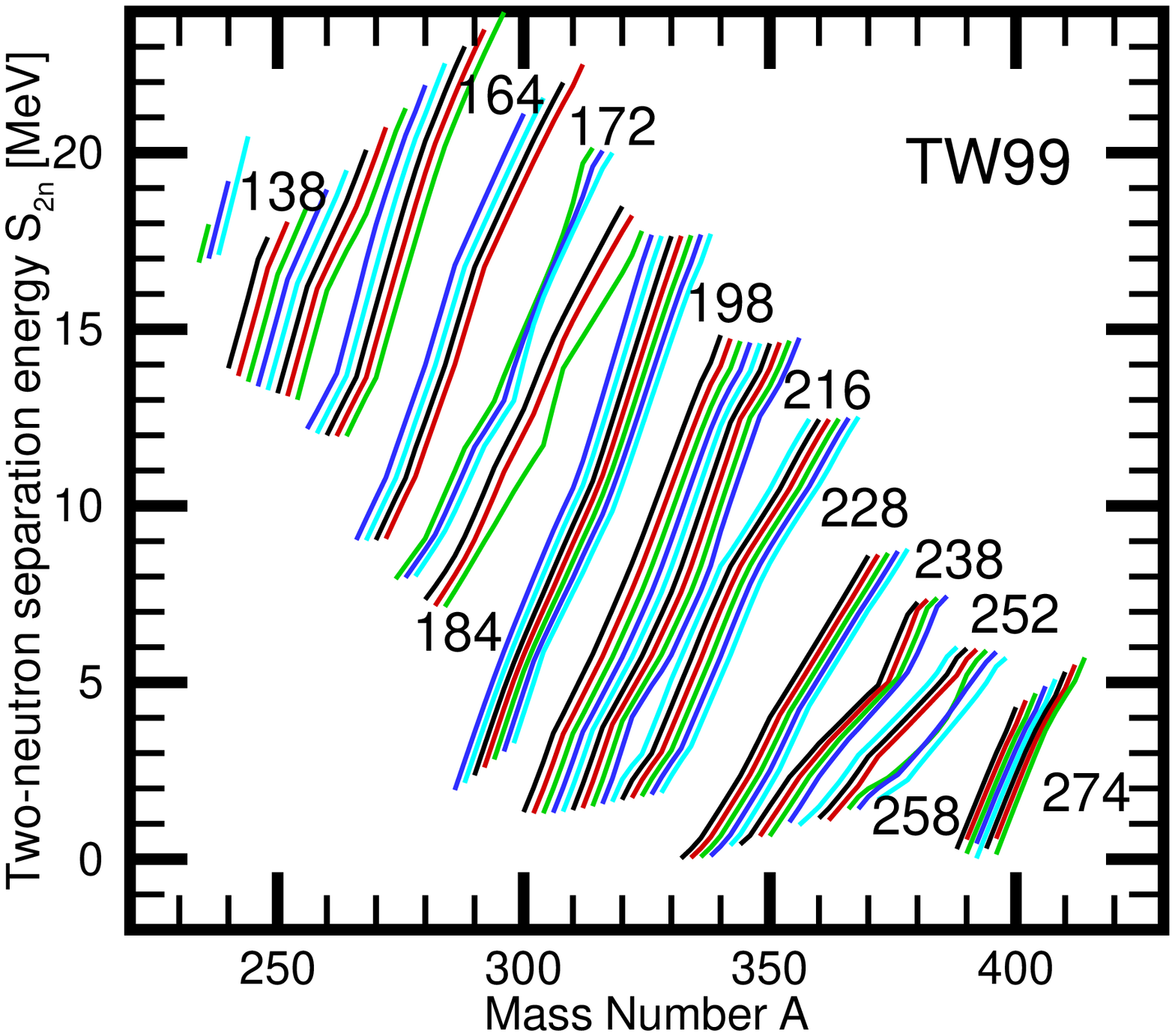}
\includegraphics[scale=0.3]{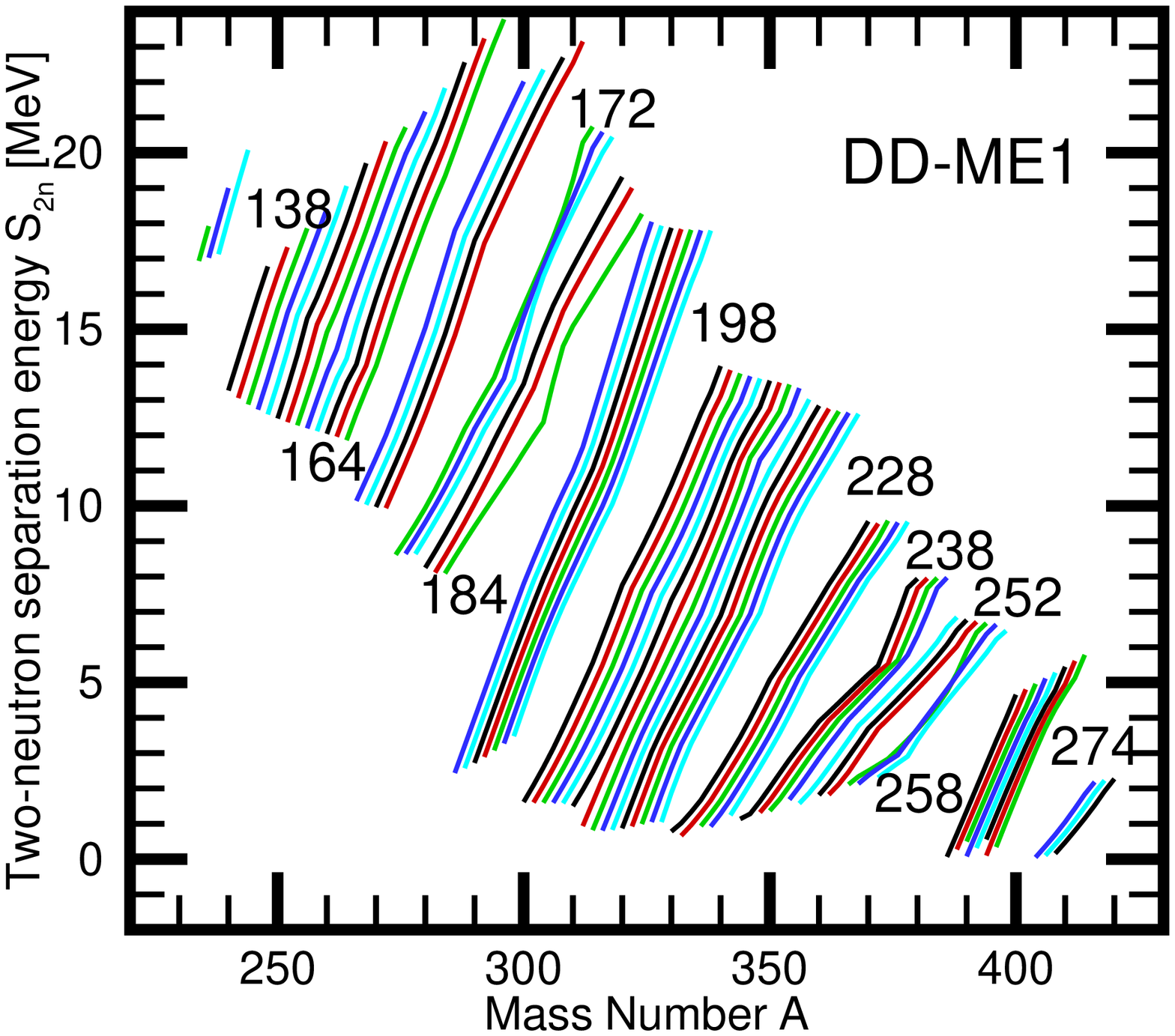}
\includegraphics[scale=0.3]{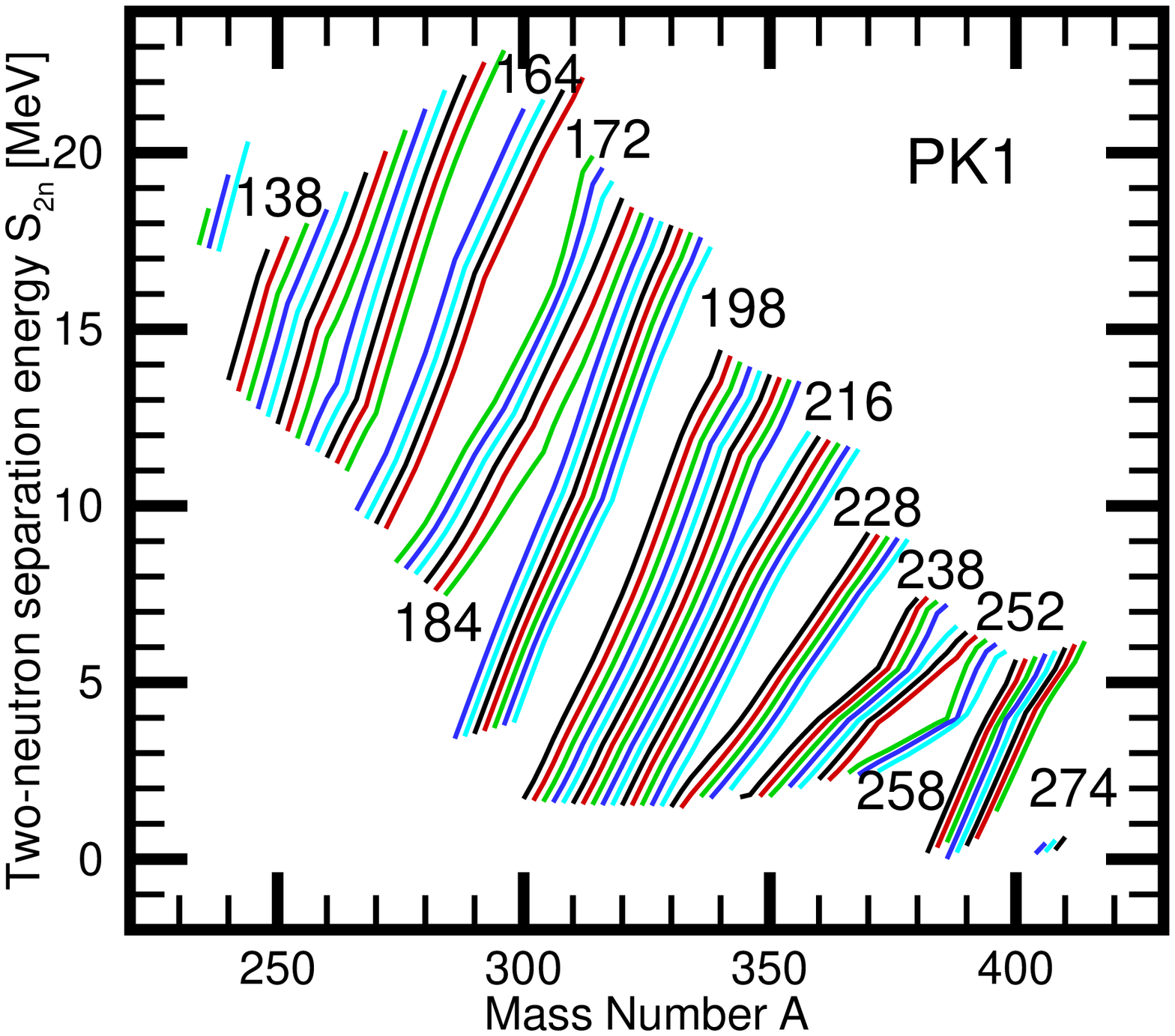}
\includegraphics[scale=0.3]{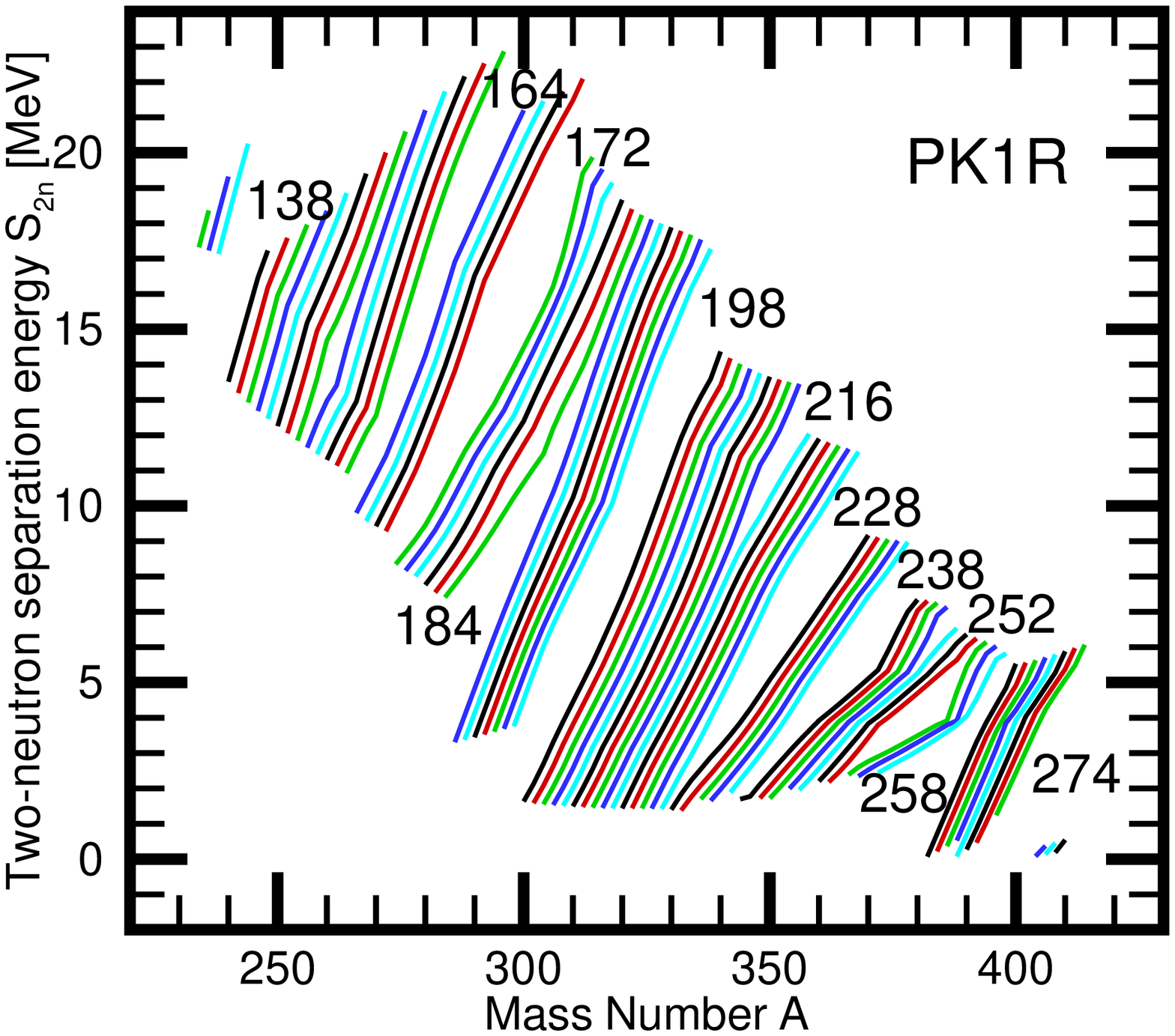}
\caption{The two-neutron separation energies
$S_{2n}(N,Z)=E_B(N,Z)-E_B(N-2,Z)$ as a function of mass number $A$
obtained by RCHB calculation with effective interactions NL1, NL3,
NL-SH, TM1, TW-99, DD-ME1, PK1, and PK1R, respectively.}
\label{s2n}
\end{figure}

Fig. \ref{s2n} is the same as Fig. \ref{s2p}, but for $S_{2n}$
with $N$=132 - 312. Each curve corresponds to an isotonic chain.
The upper-left curve is for $N$=132 and the difference of the
neutron number between the neighboring curves is 2.

The gaps of $S_{2n}$ are more complicated than those of $S_{2p}$.
For fixed $Z$, $S_{2n}$ decrease gradually with $N$. For effective
interaction NL1, there are nine gaps at $N$=138, 172, 184, 198,
228, 238, 252, 258, and 274. The size and the shape of the gaps
differ notablely: The gaps at $N$=172, 184, and 258 are relatively
larger than the other gaps.

For all effective interactions, $S_{2n}$ show distinguishable gaps
at $N$=138, 172, 184, 198, 228, 238, 252, 258, and 274. Such gaps
mark the magic neutron number. Apart from the above mentioned
common gaps, there are also other gaps which appear only for some
effective interactions, i.e., N=164 for NLSH, TW99, DD-ME1, PK1
and PK1R, and N=216 for NLSH, TW99, PK1, and PK1R. For the magic
numbers marked either by the common gaps or the
interaction-dependent gaps, the shell quenching phenomena, i.e.,
the gaps at $N$=184, 198 (NL1), 216(NLSH, TW99, PK1, and PK1R),
228 (NL1, NL3, NLSH, TM1, PK1, and PK1R), 238, 252, and 258 (NLSH,
TM1, PK1, and PK1R) appear only for certain $Z$, are observed.

Summarizing the above results, we may say that based on
two-nucleon separation energies $S_{2p}$ and $S_{2n}$, $Z$=120,
132, and 138 ,and $N$=138, 172, 184, 198, 228, 238, 252, 258, and
274, are the magic numbers predicted in the RCHB calculation. They
are independent of the effective interactions used in the
calculation. There are also some other magic numbers, e.g.,  $Z$=
106 and $N$= 164 and 216, which appear only for some effective
interactions.

\subsection{Two-nucleon gaps}
The changes of the two-nucleon separation energies can also be
quantified by the second difference of the binding energies, i.e.,
the two-nucleon gaps:
 \beqn
\delta_{2n}(N,Z)&=&2E_B(N,Z)-E_B(N+2,Z)-E_B(N-2,Z)=S_{2n}(N,Z)-S_{2n}(N+2,Z)\\
\delta_{2p}(N,Z)&=&2E_B(N,Z)-E_B(N,Z+2)-E_B(N,Z-2)=S_{2p}(N,Z)-S_{2p}(N,Z+2).
 \eeqn
A peak of the two-nucleon gap implicates the drastic changes of
the two-nucleon separation energies, which can be used as the
evidence of the magic number existence.

\begin{sidewaysfigure}[htbp]
\centering
\includegraphics[scale=0.5]{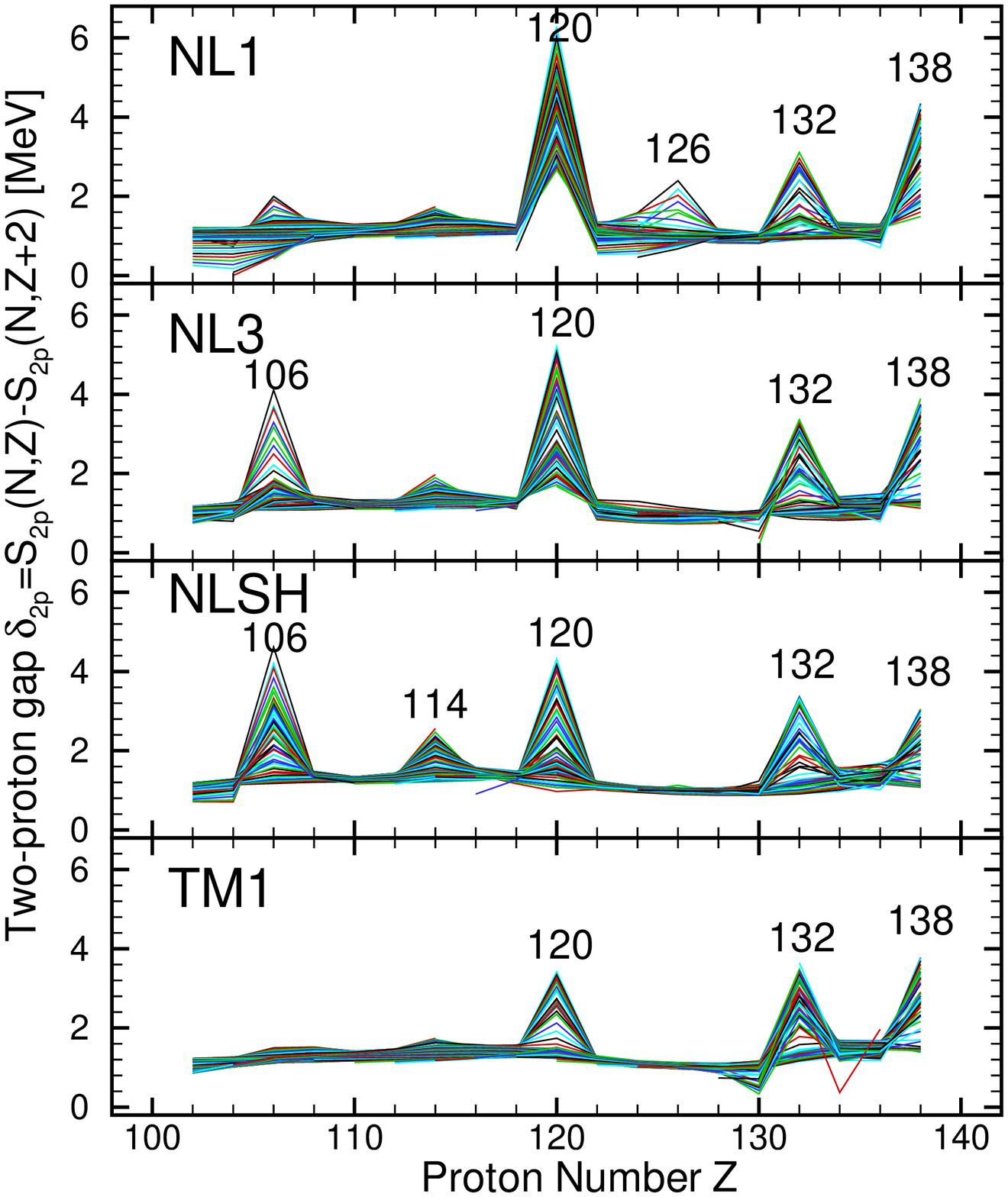}
\includegraphics[scale=0.5]{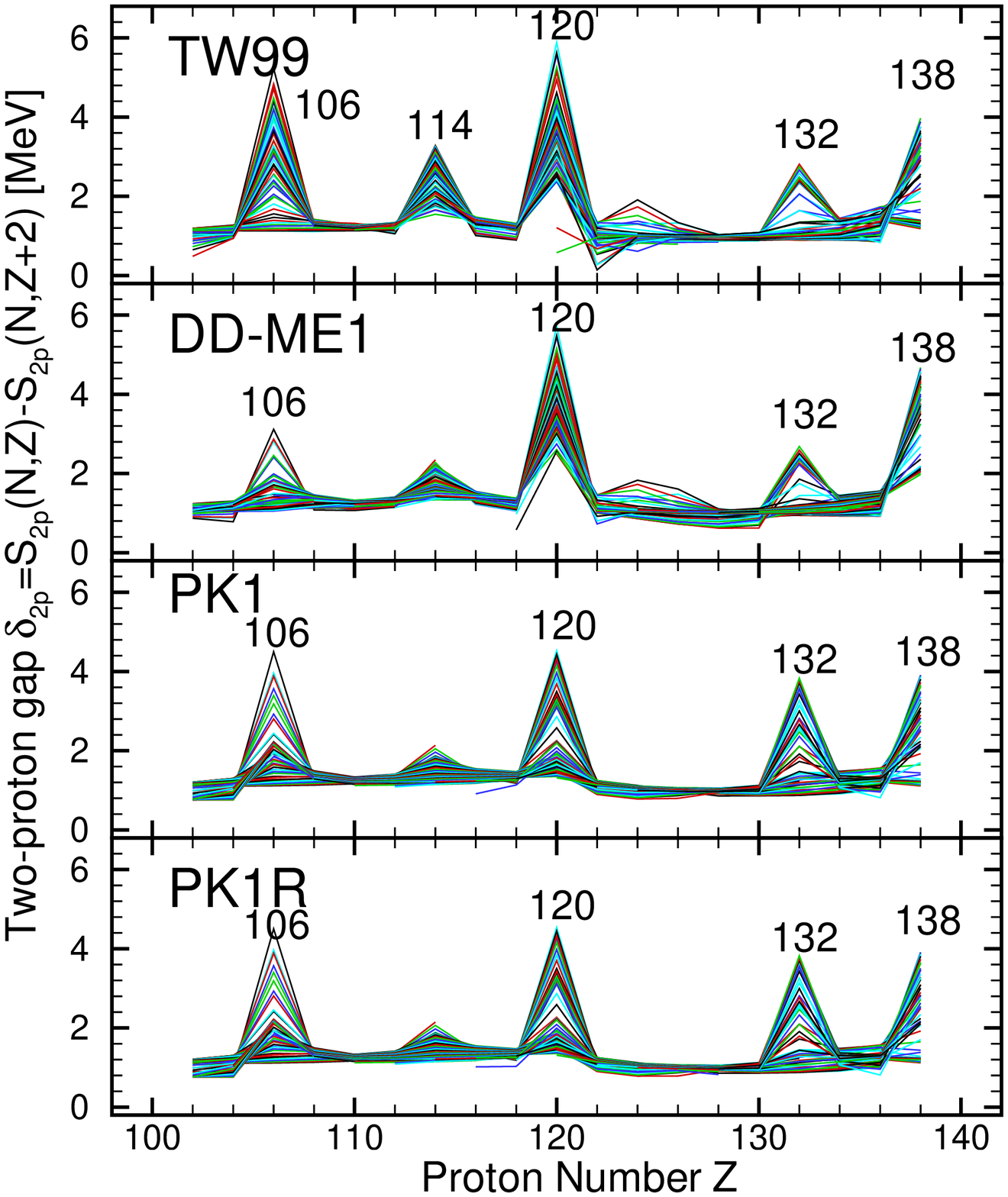}
\caption{The two-proton gaps
$\delta_{2p}(N,Z)=S_{2p}(N,Z)-S_{2p}(N,Z+2)$ as a function of
proton number obtained by RCHB calculation with effective
interactions NL1, NL3, NL-SH, TM1, TW-99, DD-ME1, PK1, and PK1R,
respectively.} \label{ds2p}
\end{sidewaysfigure}

The two-proton gaps $\delta_{2p}$ from the RCHB calculation for
even-even nuclei with $Z$=102 - 138 as a function of $Z$ with the
above eight effective interactions are shown in Fig. \ref{ds2p}.
The two-proton gaps $\delta_{2p}$ with the same $N$ are connected
as a curve. A peak at certain $Z$ in the curve suggests the
existence of magic proton number. The sharpness of the peaks
represent the goodness of the magic numbers while the quenching
effects are associated with the bundle of the curves at the
certain $Z$. The size of the gaps of two-proton separation
energies $S_{2p}$ in Fig. \ref{s2p} correspond to the magnitude of
the peaks of two-proton gaps $\delta_{2p}$ in Fig. \ref{ds2p} .
Generally the magic numbers indicated by two-proton separation
energies $S_{2p}$ are supported by two-proton gaps $\delta_{2p}$.
It is observed that common magic proton numbers $Z$=120, 132, and
138 exist for all effective interactions while $Z$=106 is observed
only for NL3, NLSH, TW99, DD-ME1, PK1, and PK1R. Furthermore, the
peak at $Z$=114 for NLSH and TW99 and the peak at $Z$=126 for NL1
are also observed, though they are not so obvious as that at
$Z$=120.

\begin{sidewaysfigure}[htbp]
\centering
\includegraphics[scale=0.6]{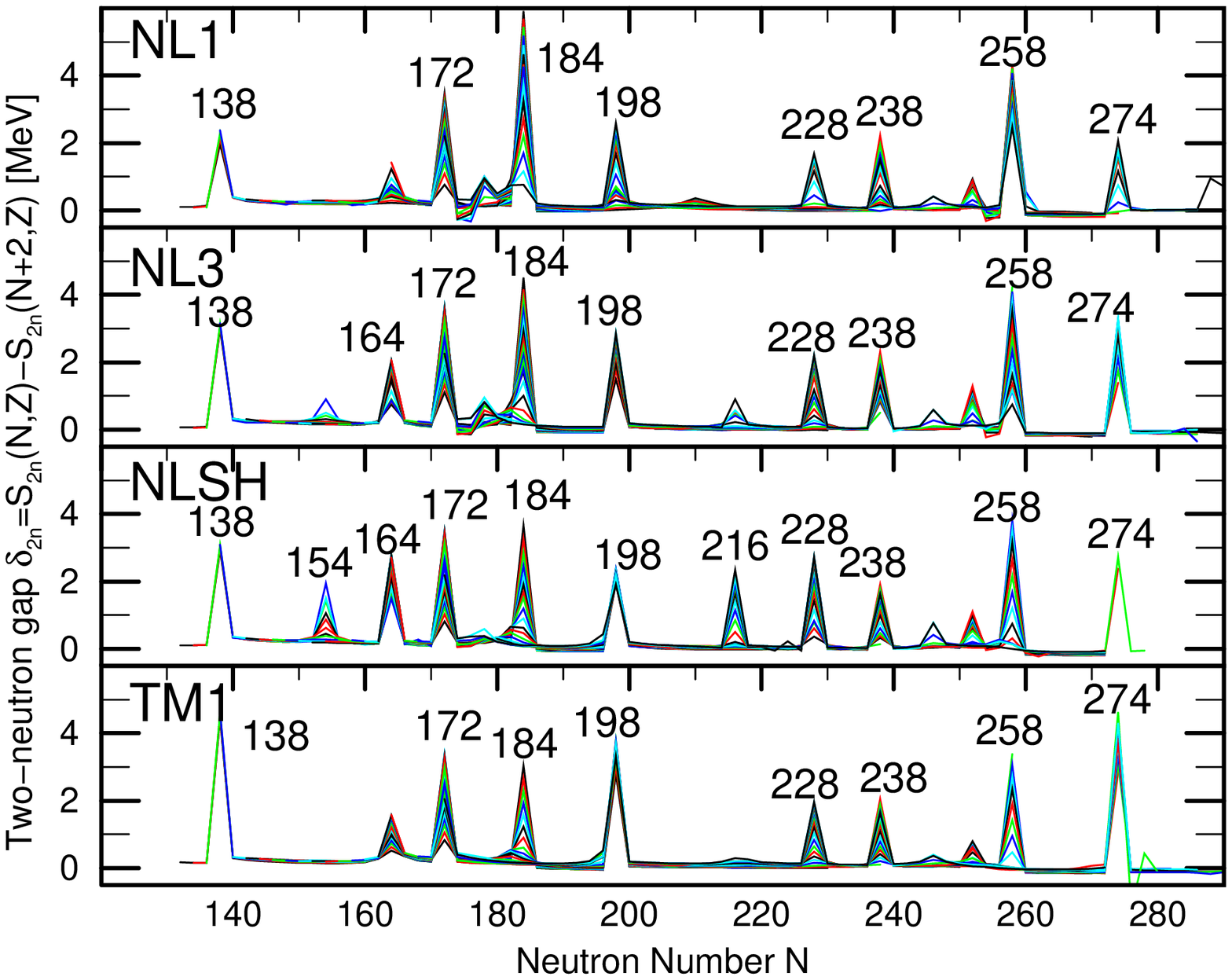}\hspace{0.5cm}
\includegraphics[scale=0.6]{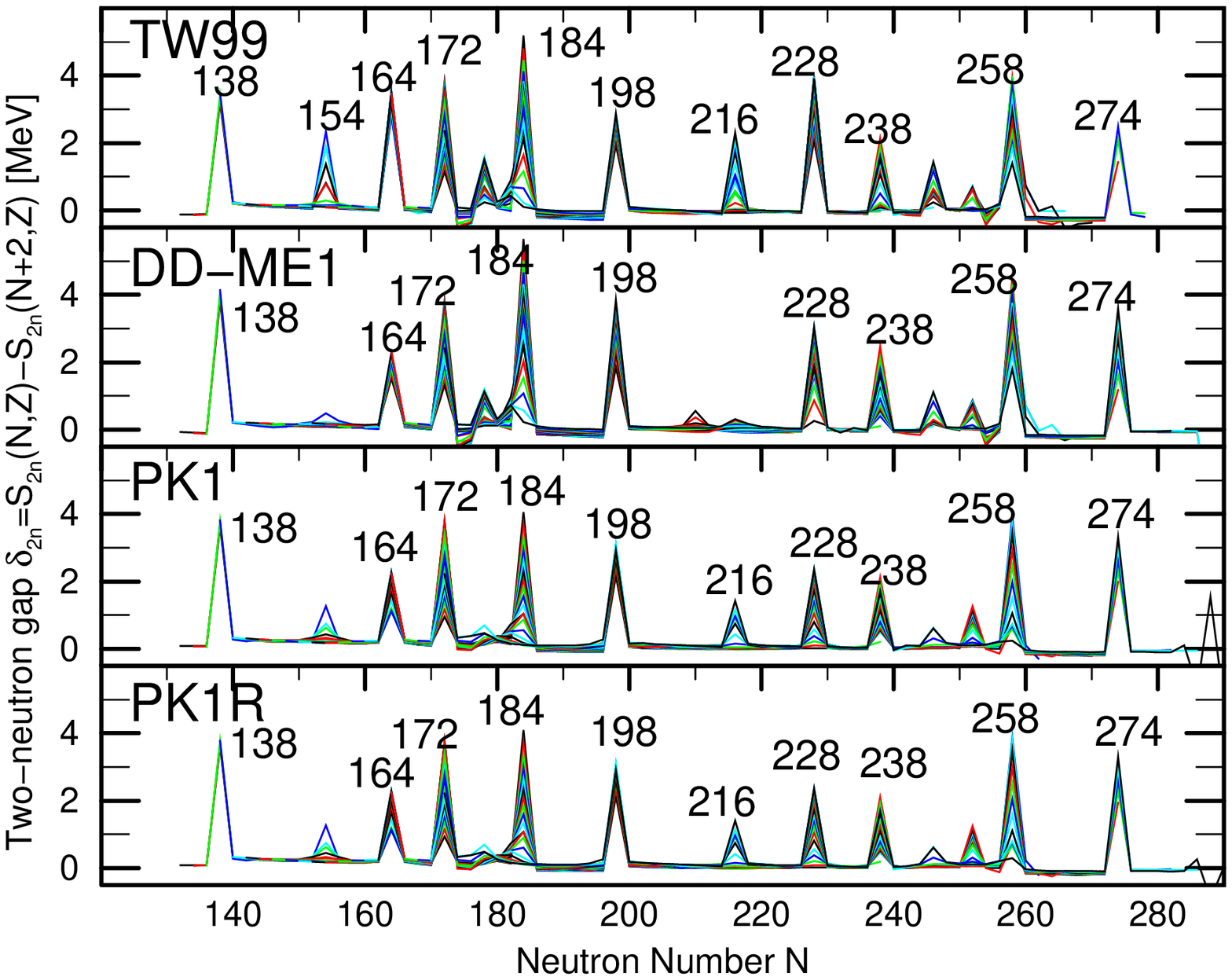}
\caption{The two-neutron gaps
$\delta_{2n}(N,Z)=S_{2n}(N,Z)-S_{2n}(N+2,Z)$ as a function of
neutron number obtained by RCHB calculation with effective
interactions NL1, NL3, NL-SH, TM1, TW-99, DD-ME1, PK1, and PK1R,
respectively.} \label{ds2n}
\end{sidewaysfigure}

As counterpart of two-proton gaps $\delta_{2p}$, two-neutron gaps
$\delta_{2n}$ from the RCHB calculation for even-even nuclei
$N$=132 - 308 with effective interactions NL1, NL3, NLSH, TM1,
TW99, DD-ME1, PK1, and PK1R are shown in Fig. \ref{ds2n}. There
are interaction-independent peaks of two-neutron gaps
$\delta_{2n}$ at $N$=138, 172, 184, 198, 228, 238, 258, and 274
which are consistent with the common gaps of two-neutron
separation energies $S_{2n}$ in Fig. \ref{s2n}. Moreover, small
peaks are also observed for $N$=154 (NLSH and TW99), $N$=164 (NL3,
NLSH, TW99, DD-ME1, PK1, and PK1R), and $N$=216 (NLSH, TW99, PK1,
and PK1R).

In general, based on two-nucleon gaps $\delta_{2p}$ and
$\delta_{2n}$, the magic numbers $Z$=120, 132, and 138, and
$N$=138, 172, 184, 198, 228, 238, 258, and 274 can also be
observed for RCHB calculation with all effective interactions.
Apart from $Z$=106 and $N$=164 and 216, additional magic numbers
$Z$=114 and 126, $N$=154 appear in the two-nucleon gaps
$\delta_{2p}$ and $\delta_{2n}$ in Figs. \ref{ds2p} and \ref{ds2n}
for some effective interactions.

\subsection{Shell correction energies}
The shell correction energy, representing the overall behavior of
the single-particle spectra, may be a good candidate to identify
the magicity. This quantity, which is derived from the
single-particle spectra, is defined as the difference between the
total single-particle energy $E$ and the smooth single-particle
energy $\bar {E}$: \beq
  \label{microeq1}
  E_{shell} = E - \bar {E}
            = \sum\limits_{i = 1}^{N(Z)} {e_i }
            - 2\int\limits_{ -\infty }^{\bar{\lambda}} {e\bar {g}(e)de}.
\eeq
where $N(Z)$ is the particle number, $e_i$ is the
single-particle energy, $\bar{\lambda}$ is the smoothed Fermi
level and $\bar{g}(e)$ is the smoothed level density. The shell
correction energy provides an indicator about the deviation in the
level structure of nuclei away from uniformly distributed ones. A
large negative shell correction energy corresponds to shell
closure at particular nucleon number.

The shell correction energy can be obtained by the Strutinsky
procedure, in which the smoothed Fermi level $\bar{\lambda}$ is
determined by the particle number equation
 $N(Z) = 2\int\limits_{ - \infty}^{\bar{\lambda}} {\bar {g}(e)de}$.
The smoothed level density $\bar {g}(e)$ takes the form:
 \beq
  \bar{g}(e)= \dfrac{1}{\gamma}  \int\limits_{ -\infty }^\infty
             {\left(\sum\limits_{i = 1}^\infty
             {\delta(e'- e_i)}\right) f(\frac{e' - e}{\gamma })de'}
            = \dfrac{1}{\gamma}  \sum\limits_{i = 1}^\infty
             {f(\dfrac{e_i - e}{\gamma }) }.
 \eeq
where $\gamma$ is the smoothing range, the folding function $f(x)
= \dfrac{1}{\sqrt \pi}e^{ - x^2}P(x)$, and $P(x)$ is an associated
Laguerre polynomial $L_s^{1 / 2}(x^2)$.

In present calculation, the smoothing range is $\gamma = 1.2$ and
the order of the associated Laguerre polynomial is $s=3$. The unit
of $\gamma$ is ${\raisebox{0em}[2em][2em]{}}$
$41~A^{-1/3}\left(1\pm\dfrac{1}{3}\dfrac{N-Z}{A}\right)$ MeV where
the plus (minus) sign holds for neutrons (protons). Then the
particle number equation can be evaluated as:
 \beq
  \label{microeq4}
  N(Z) = 2\sum\limits_{i = 1}^\infty    {[\dfrac{1 - \mbox{erf}(t_i)}{2}
         - \dfrac{e^{ - t_i^2}}{48\sqrt \pi }(57t_i - 32t_i^3 + 4t_i^5 )]},
 \eeq
where $t_i = \dfrac{e_i - \bar{\lambda }}{\gamma}$ and erf($t_i$)
is the error function. The corresponding smooth single-particle
energy $\bar {E}$ takes the form:
 \beq
  \label{microeq5}
  \bar {E}=2 \sum\limits_{i = 1}^\infty {\{e_i [\dfrac{1 - \mbox{erf}(t_i )}{2} -
                    \dfrac{e^{ - t_i^2 }}{48\sqrt \pi }(57t_i - 32t_i^3 + 4t_i^5 )]
           - \gamma \dfrac{e^{ - t_i^2 }}{96\sqrt \pi }(15 - 90t_i^2 + 60t_i^4 - 8t_i^6 )\}} .
 \eeq

\begin{sidewaysfigure}[htbp]
\centering
\includegraphics[scale=0.5]{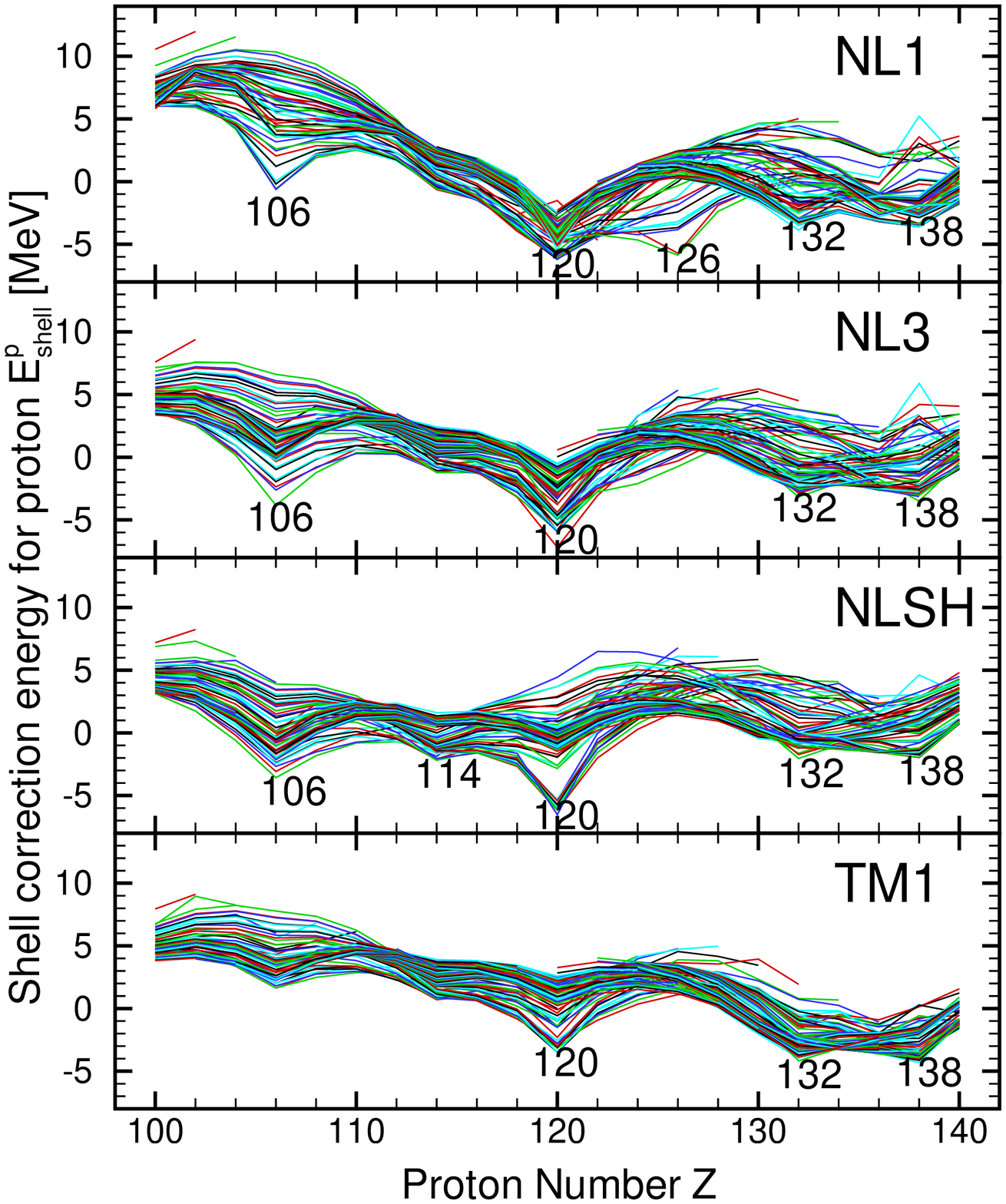}
\includegraphics[scale=0.5]{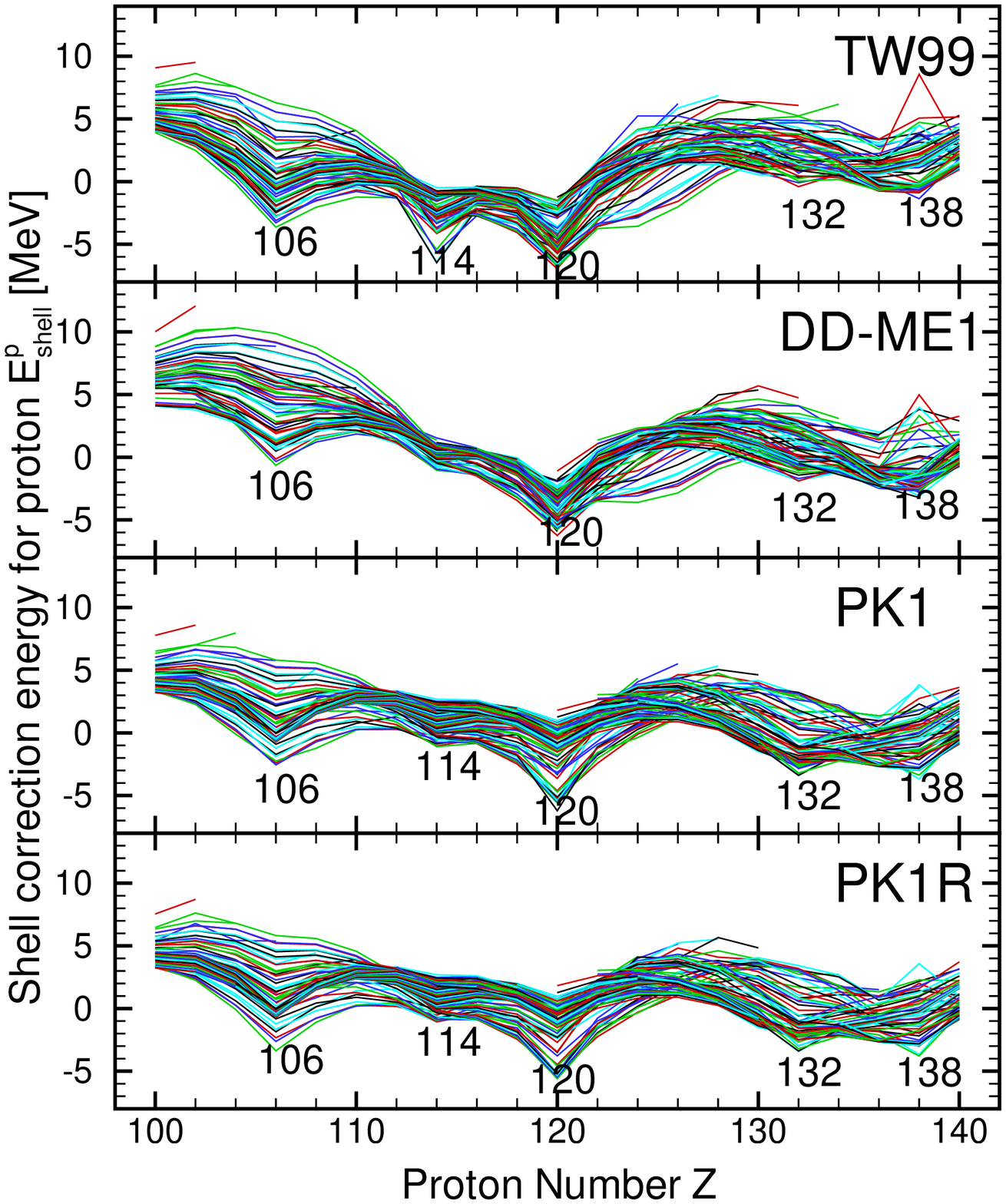}
\caption{The shell correction energies for proton $E_{shell}^{p}$
as a function of proton number obtained by RCHB calculation with
effective interactions NL1, NL3, NL-SH, TM1, TW-99, DD-ME1, PK1
and PK1R, respectively.} \label{scp}
\end{sidewaysfigure}

With the proton single-particle spectra in canonical basis from
the RCHB calculation for even-even nuclei with $Z$=100 - 140 as a
function of $Z$ with effective interactions NL1, NL3, NLSH, TM1,
TW99, DD-ME1, PK1, and PK1R, the shell correction energies for
proton $E_{shell}^{p}$ are shown in Fig. \ref{scp}. The nuclei
with the same $N$ are connected as a curve. A deep valley at
certain $Z$ on a curve hints the magicity. The valleys at $Z$=120,
132, and 138 are conspicuous for all effective interactions.
Possible shell closure at $Z$=106 for NL1, NL3, NLSH, TW99,
DD-ME1, PK1, and PK1R, at $Z$=114 for NLSH, TW99, PK1, and PK1R,
and at $Z$=126 for NL1 are also observed. The shell closure from
the shell correction energies for proton $E_{shell}^{p}$ in Fig.
\ref{scp} are consistent with the peaks of the two-proton gaps
$\delta_{2p}$ in Fig. \ref{ds2p}, though the magnitude of magicity
slightly differs. Similar to two-proton gaps $\delta_{2p}$, the
spread of the valleys of the shell correction energies for proton
$E_{shell}^{p}$ indicates the quenching of the magic number. In
view of the shell correction energies for proton $E_{shell}^{p}$,
the quenching phenomena of magic proton numbers in RCHB
calculation are universal. These facts suggest that the magic
proton numbers depend also on neutron number $N$.

\begin{sidewaysfigure}[htbp]
\centering
\includegraphics[scale=0.6]{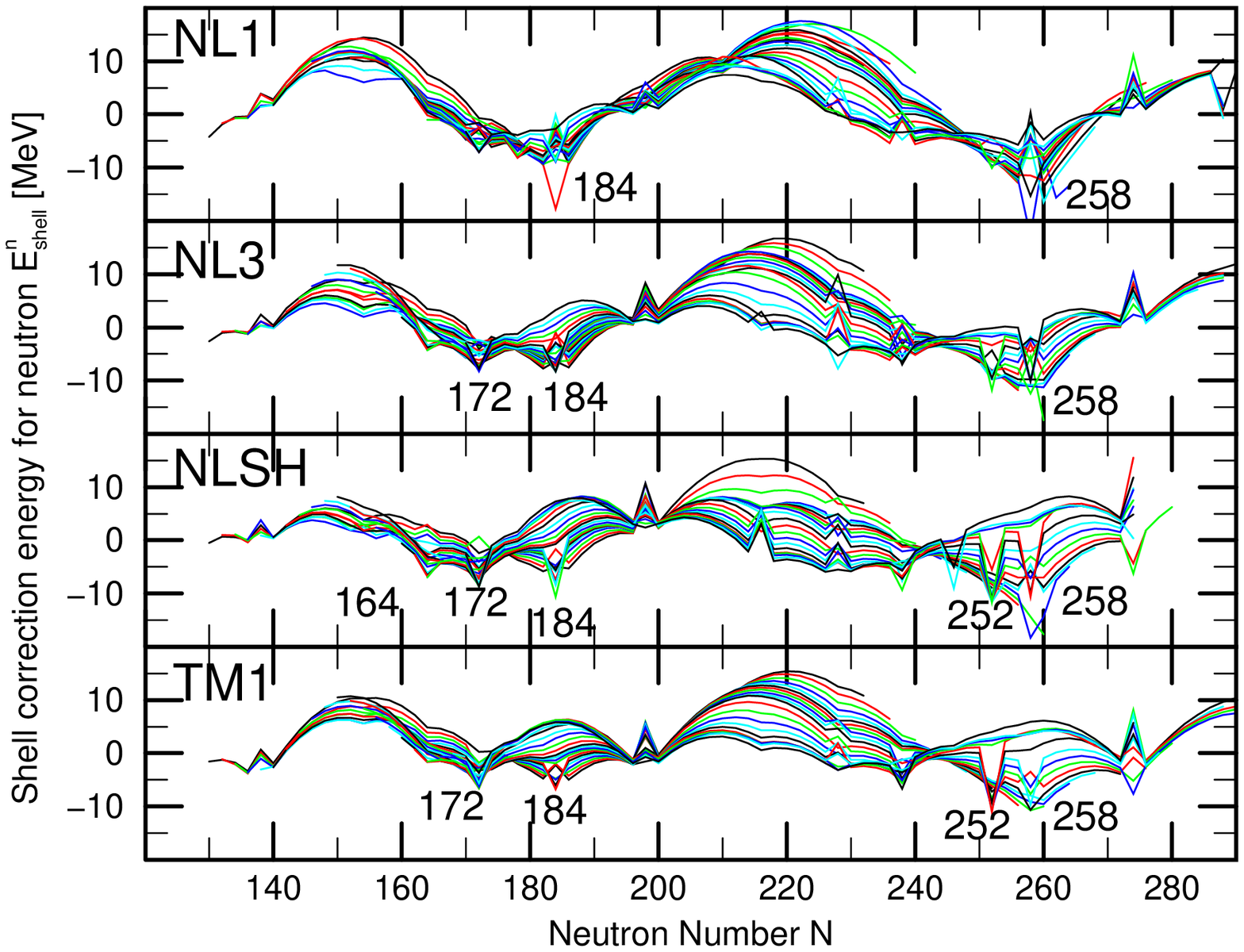}\hspace{0.5cm}
\includegraphics[scale=0.6]{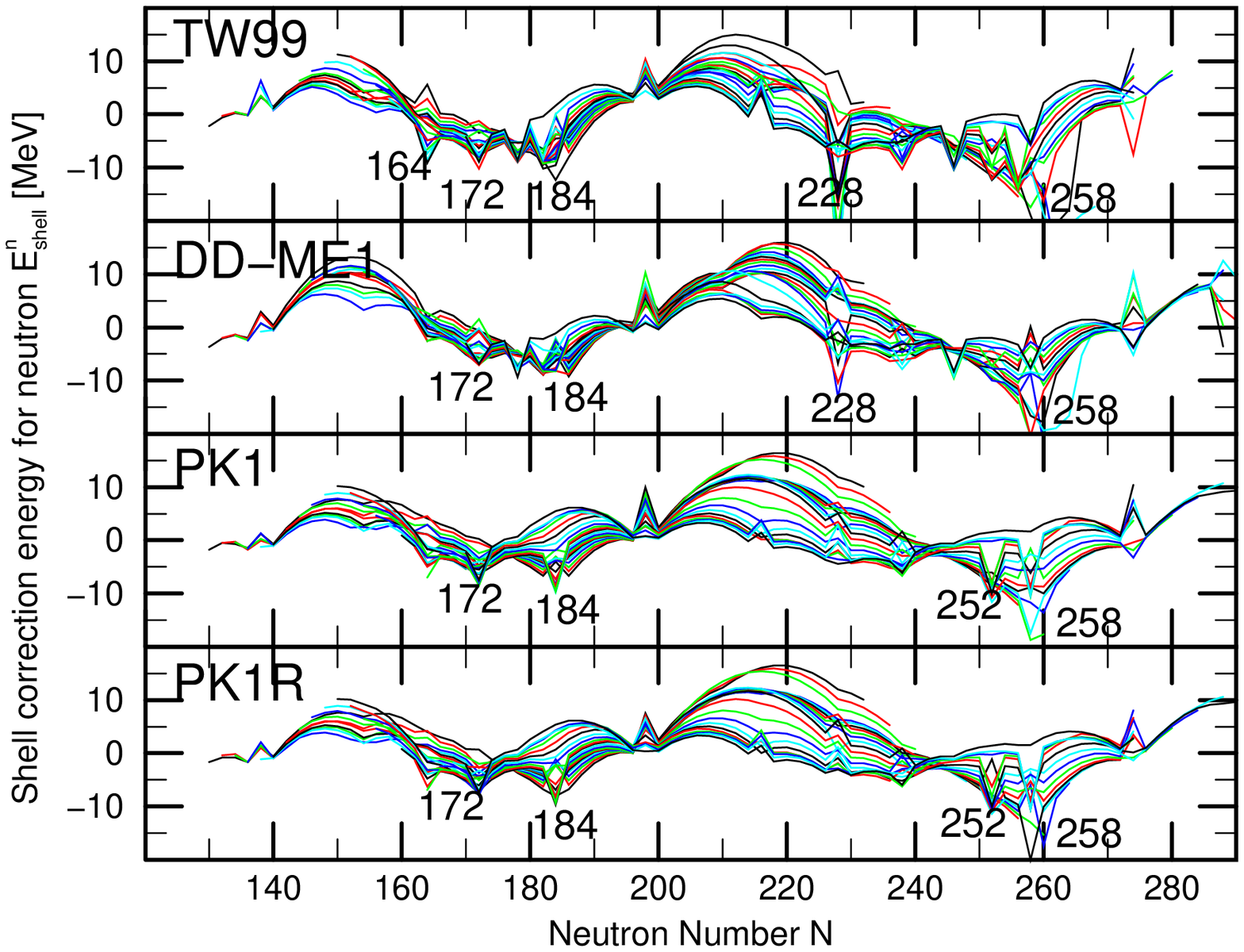}
\caption{The shell correction energies for neutron $E_{shell}^{n}$
as a function of neutron number obtained by RCHB calculation with
effective interactions NL1, NL3, NL-SH, TM1, TW-99, DD-ME1, PK1
and PK1R, respectively.} \label{scn}
\end{sidewaysfigure}

The corresponding shell correction energies for neutron
$E_{shell}^{n}$ from the RCHB calculation for even-even nuclei
with $N$=130 - 312 as a function of $N$ are demonstrated in Fig.
\ref{scn}. As the neutron numbers extends from $N$=130 to $N$=290,
the valleys for shell correction energies for neutron
$E_{shell}^{n}$ in Fig. \ref{scn} are not so obvious as those for
shell correction energies for proton $E_{shell}^{p}$ in Fig.
\ref{scp}. However, similar conclusions as two-neutron separation
energies $S_{2n}$ in Fig. \ref{s2n} and two-neutron gaps
$\delta_{2n}$ in Fig. \ref{ds2n} for magic neutron numbers can
also be drawn. For all effective interactions, $N$=184 and 258 are
found to be the common magic numbers. Similar calculation has also
be done in Ref. \cite{Bender2001} with SHF and RMF. From the shell
correction energies for neutron, although the magic neutron
numbers are not so sharp as those got from two-neutron separation
energies $S_{2n}$ and two-neutron gaps $\delta_{2n}$, however, the
small irregularity in the shell correction energies still
correspond to the magic number discussed above.

The first valley locates in $N$=184 for all effective interactions
and involves with $N$=172 (NL3, NLSH, TM1, TW99, DD-ME1, PK1, and
PK1R), even with $N$=164 (NLSH and TW99). And the other valley is
the mixture of the valleys at $N$=228 (TW99 and DD-ME1), $N$=252
(NLSH, TM1, PK1, and PK1R) and $N$=258. These fine structures
suggest the smearing of magic neutron numbers.

Summarizing the above results, we may say that based on shell
correction energies, the shell closures are smeared than those for
two-nucleon separation energies $S_{2p}$ and $S_{2n}$ and
two-nucleon gaps $\delta_{2p}$ and $\delta_{2n}$, however we can
observe magic numbers at $Z$=120, 132, and 138 are common and
strongly quenched, the shell closures near $N$=172 and $N$=184,
and $N$=228, $N$=252, and $N$=258 are blurred. The other magic
numbers from the shell correction energies, such as $Z$=106, 114,
and 126, appear only for some effective interactions.

\subsection{Pairing energies and pairing gaps}
In this subsection we will investigate the pairing effects, which
can also provide reliable information of nuclear magic number,
including the pairing energy and the effective pairing gap. The
pairing energy in RHB theory is given by
 \beq
E_{pair}=-\dfrac{1}{2} {\rm Tr} \Delta \kappa,
 \eeq
where $\Delta$  and $\kappa$ are the pairing potential and the
pairing tensor respectively. Generally speaking the pairing
energies vanish at the closed shell and have a maximum values in
the middle of two closed shells. Similarly, the effective pairing
gap defined as
 \beq
\Delta_{n(p)}
 =\dfrac{\sum\limits_i \Delta_i  v^{2}_{i} (2j+1)} {N(Z)},
 \eeq
can also be used to denote the shell closure, where the sum $i$ is
over the single particle levels of neutron or proton in the
canonical basis\cite{Meng1998a}, $2j+1 $ is the degeneracy number
of the corresponding energy level $i$ and $N(Z)$ is the particle
number, $\Delta_i$ is the pairing gap for level $i$ in canonical
basis determined by $(E_{i}-\lambda)\cdot 2 u_{i} v_{i}=\Delta_i
(v^{2}_{i}-u^{2}_{i})$ where $\lambda$ is the Fermi energy and the
particle and hole amplitude $u_{i},\,v_{i}$ satisfies
$u^{2}_{i}+v^{2}_{i}=1$.

\begin{sidewaysfigure}[htbp]
\centering
\includegraphics[scale=0.5]{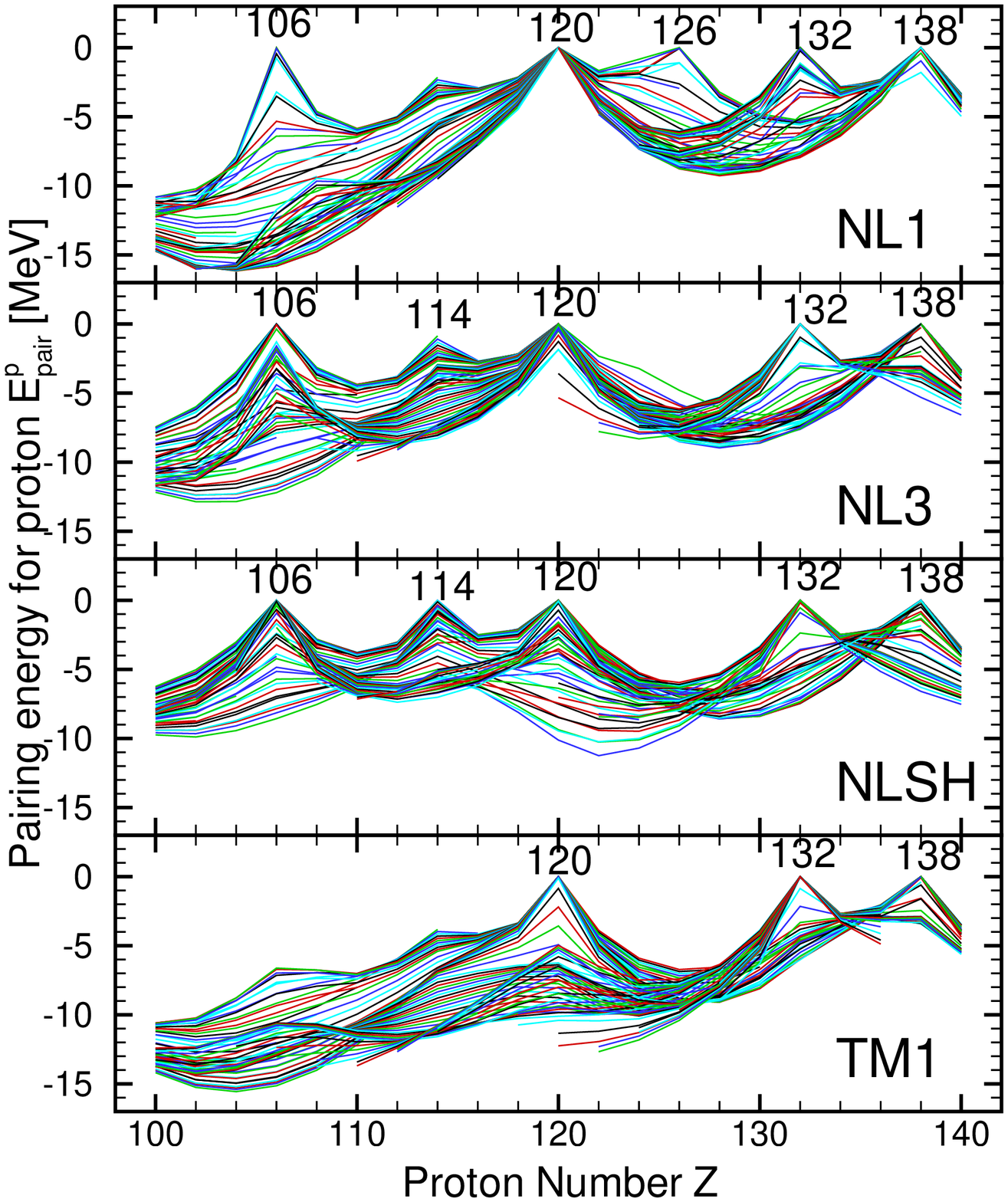}
\includegraphics[scale=0.5]{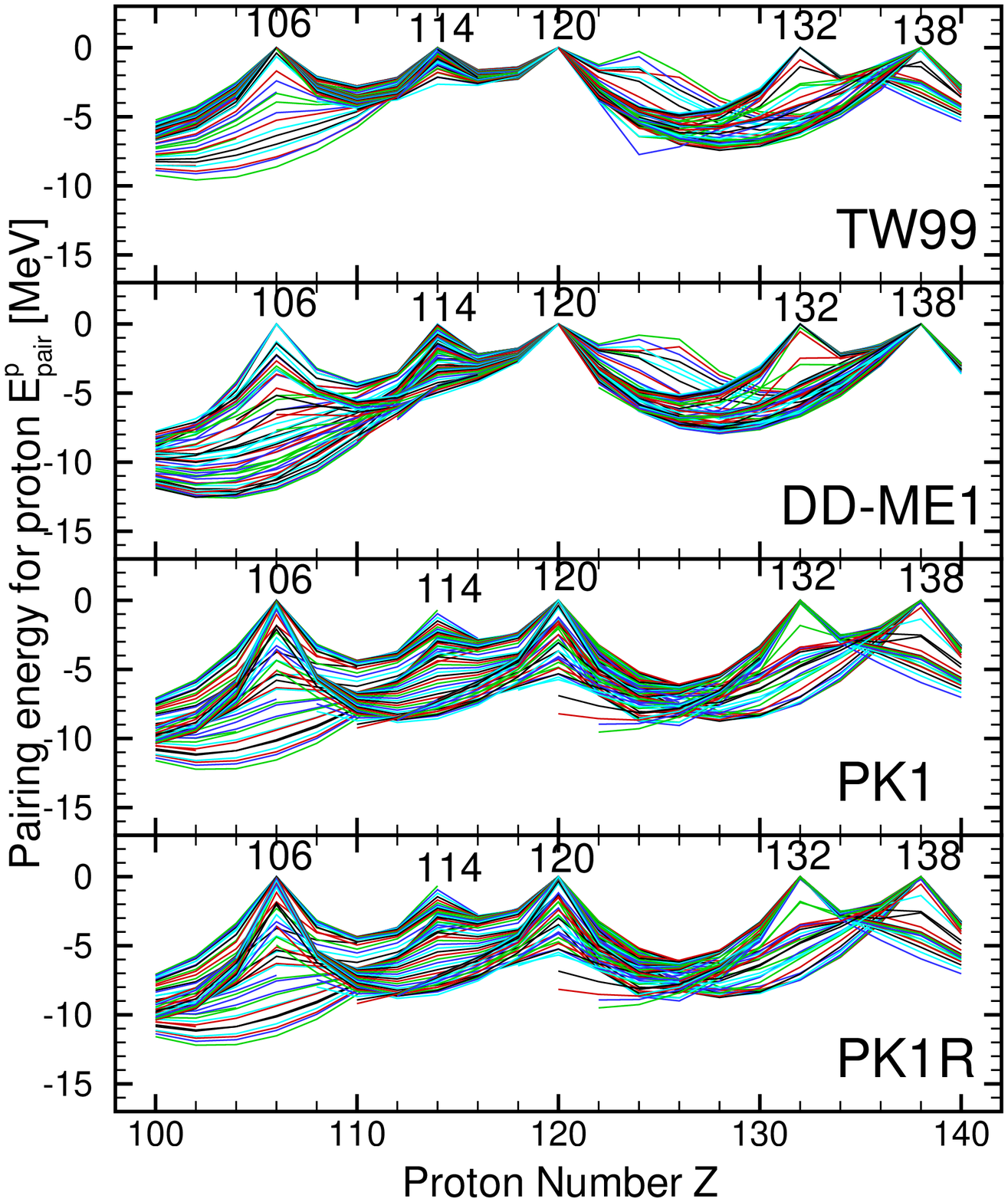}
\caption{The pairing energies for proton $E_{pair}^{p}$ as a
function of proton number obtained by RCHB calculation with
effective interactions NL1, NL3, NL-SH, TM1, TW-99, DD-ME1, PK1
and PK1R, respectively.} \label{pp}
\end{sidewaysfigure}

\begin{sidewaysfigure}[htbp]
\centering
\includegraphics[scale=0.5]{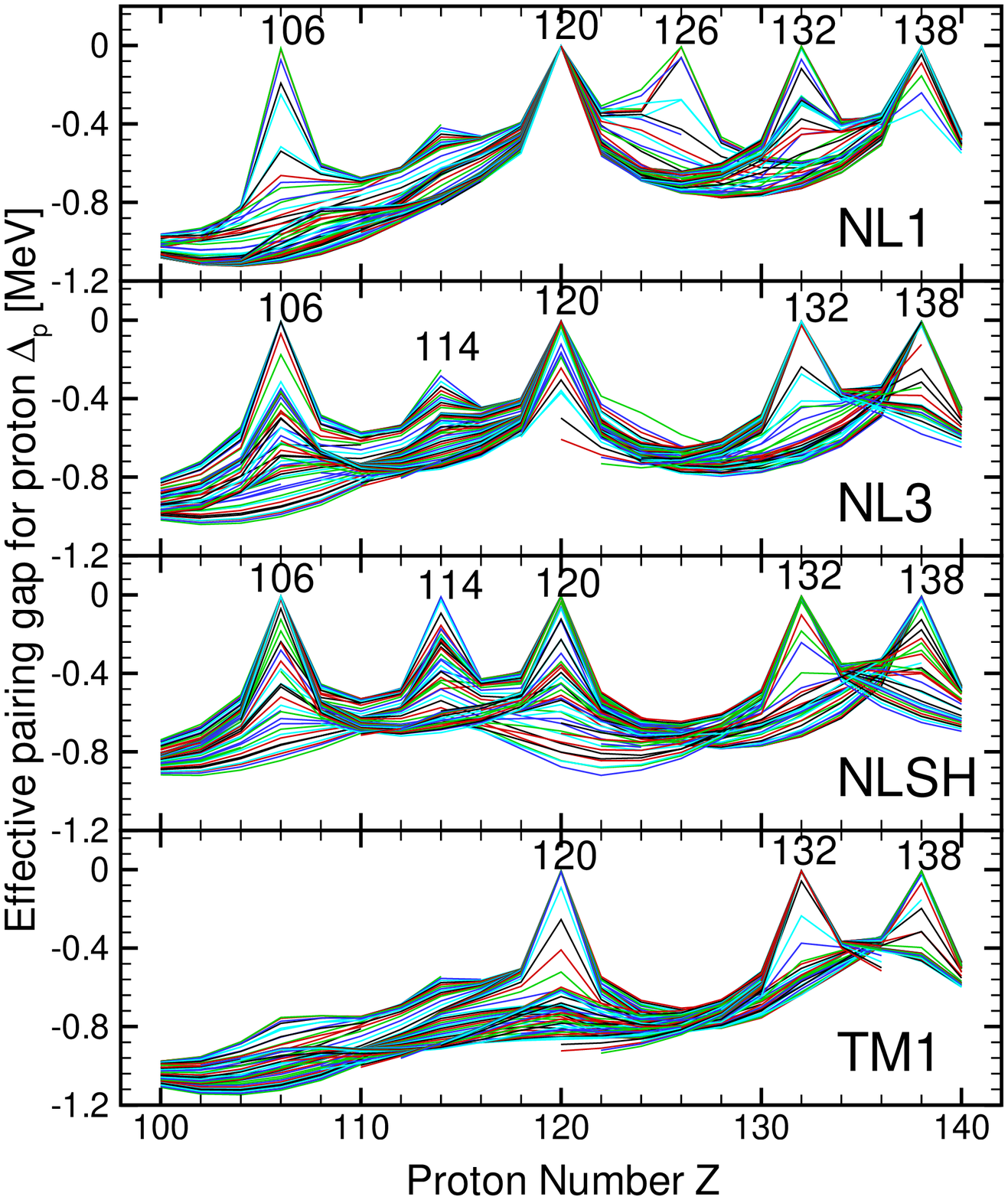}
\includegraphics[scale=0.5]{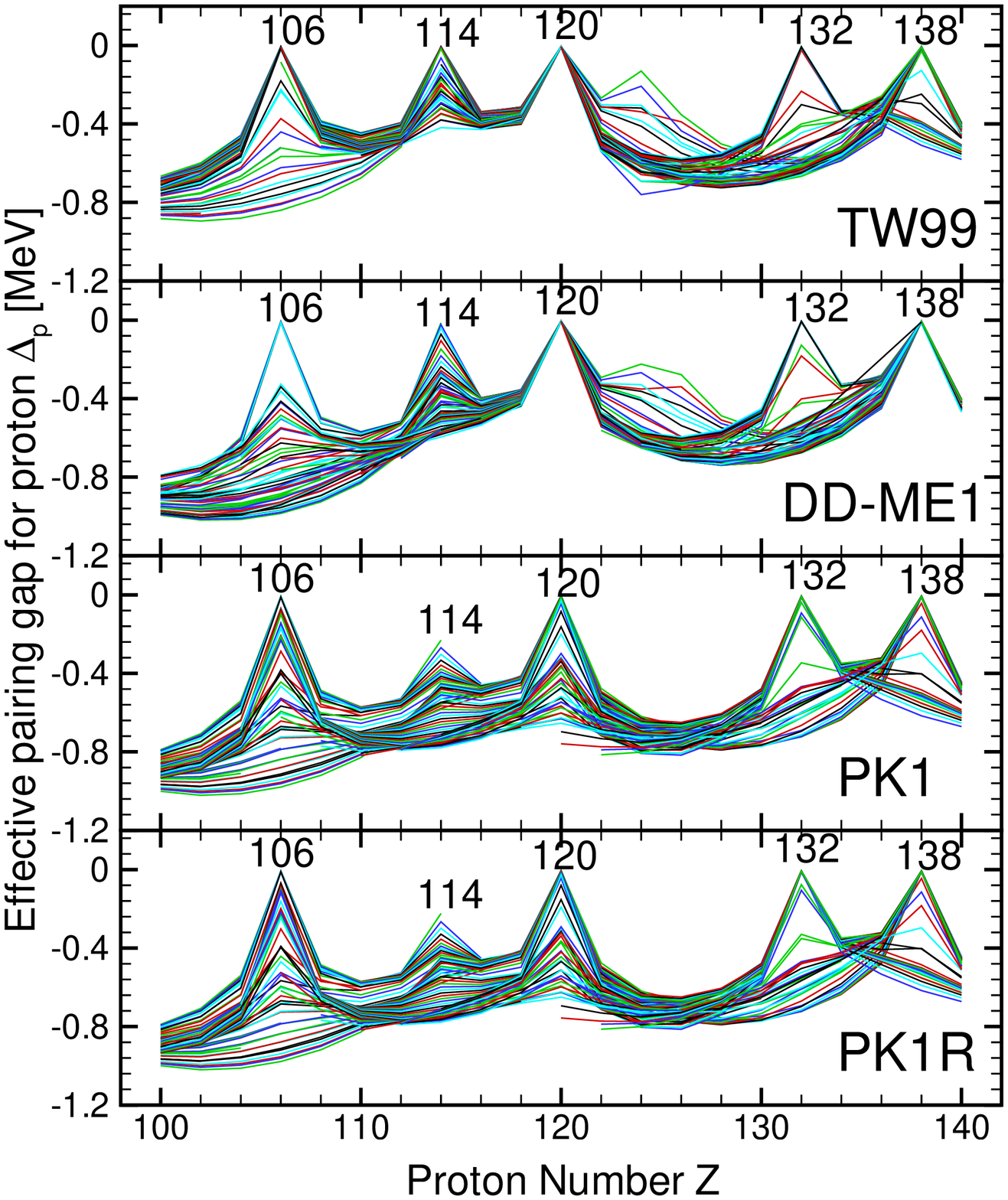}
\caption{The effective pairing gaps for proton $\Delta_{p}$ as a
function of proton number obtained by RCHB calculation with
effective interactions NL1, NL3, NL-SH, TM1, TW-99, DD-ME1, PK1
and PK1R, respectively.} \label{dp}
\end{sidewaysfigure}

The pairing energies for proton $E_{pair}^{p}$ and the effective
pairing gaps for proton $\Delta_{p}$ from the RCHB calculation for
even-even nuclei with $Z$=100 - 140 as a function of $Z$ with
effective interactions NL1, NL3, NLSH, TM1, TW99, DD-ME1, PK1, and
PK1R are shown in Figs. \ref{pp} and \ref{dp} respectively. Both
figures has the same pattern while the shell closures in Fig.
\ref{dp} are more clear. It is observed that common magic proton
numbers $Z$=120, 132, and 138 exist for all effective interactions
while $Z$=106 exists only for NL1, NL3, NLSH, TW99, DD-ME1, PK1,
and PK1R, $Z$=114 exists only for NL3, NLSH, TW99, DD-ME1, PK1,
and PK1R, and $Z$=126 exists only for NL1, TW99, DD-ME1. Generally
the shell closures are quenched for all magic proton numbers. It
should be noted that the peak at $Z$=120 for NL1, TW99, and DD-ME1
is stable for all isotopes while for NLSH, TM1, PK1, and PK1R, it
is quenched.

\begin{sidewaysfigure}[htbp]
\centering
\includegraphics[scale=0.6]{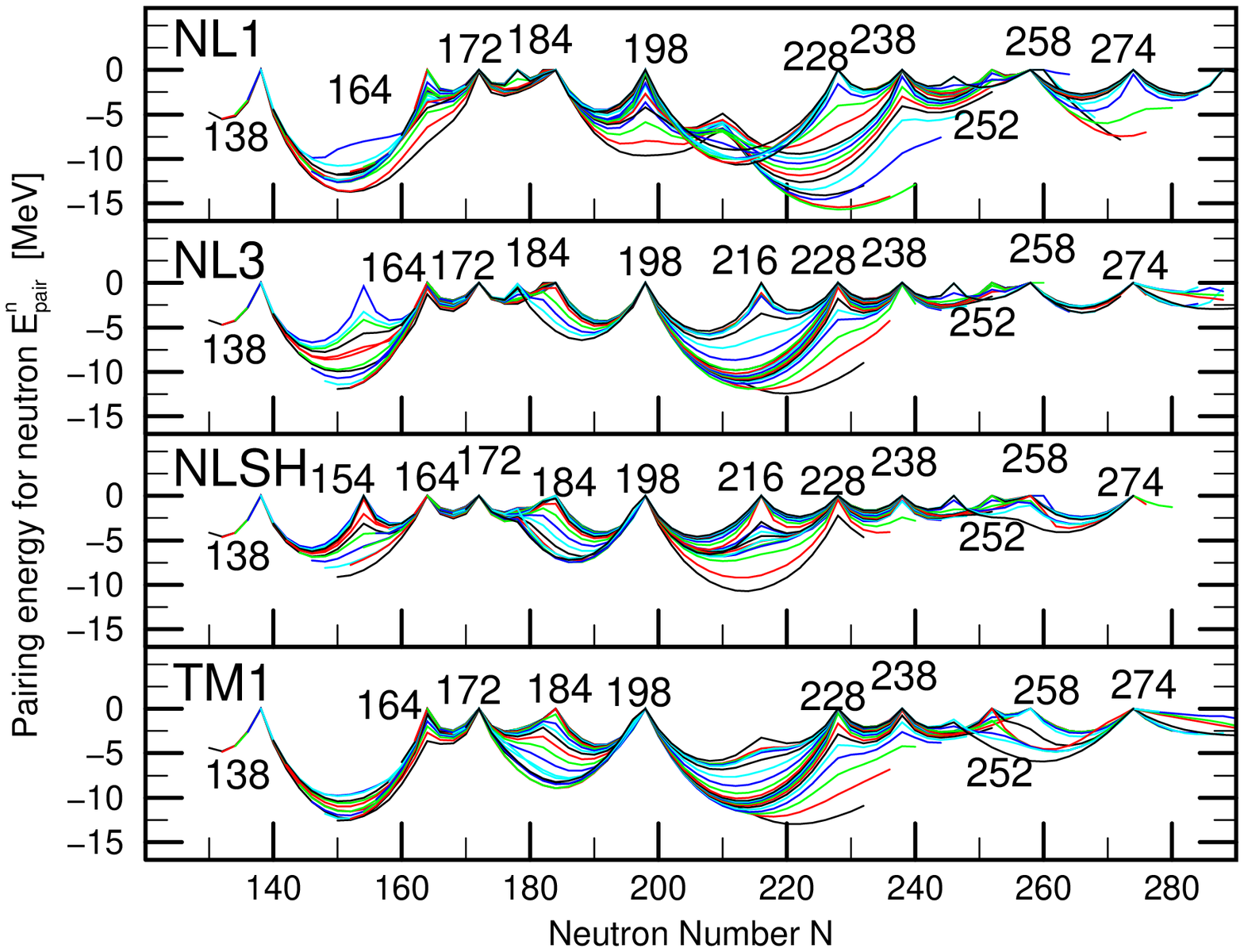}\hspace{0.5cm}
\includegraphics[scale=0.6]{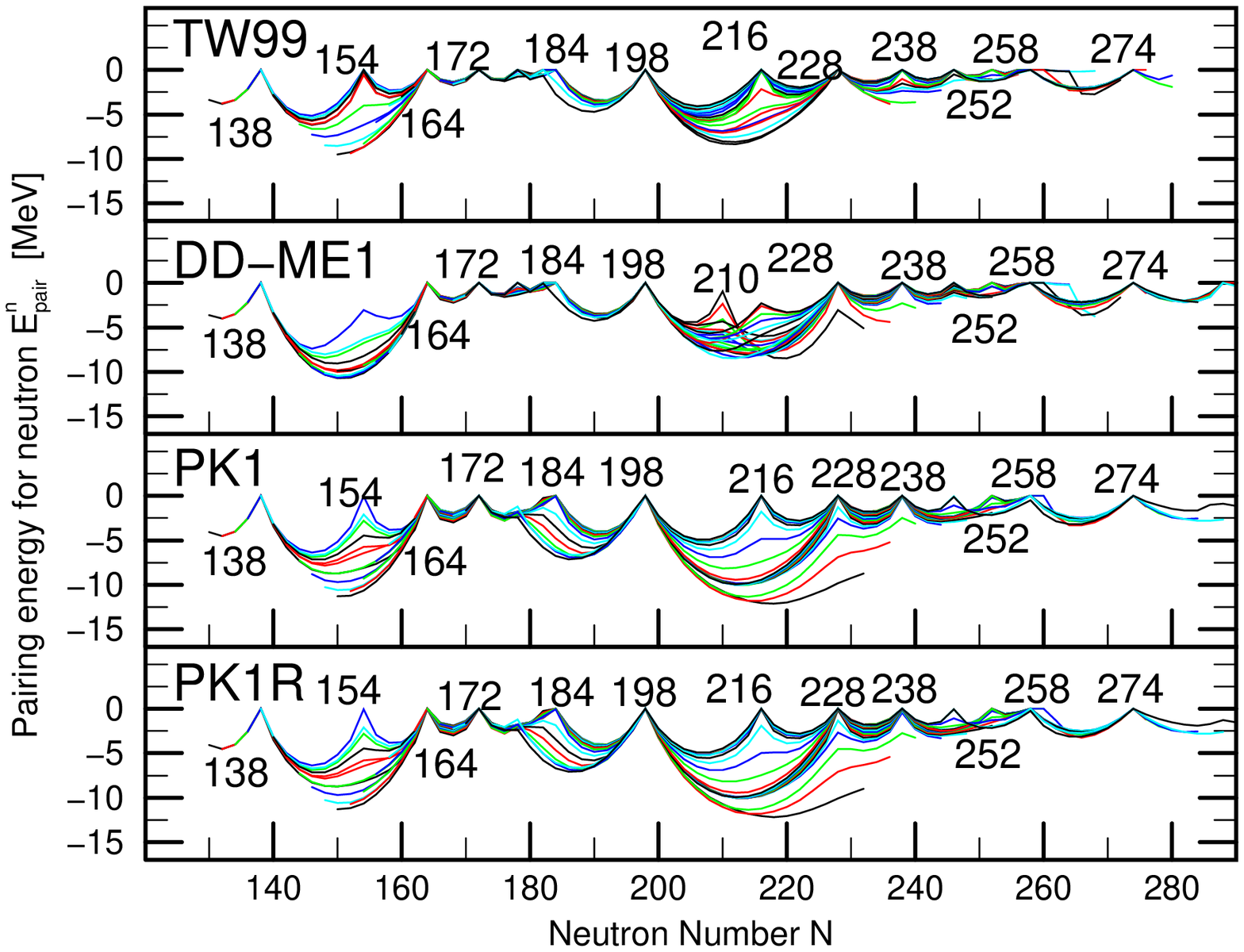}
\caption{The pairing energies for neutron $E_{pair}^{n}$ as a
function of neutron number obtained by RCHB calculation with
effective interactions NL1, NL3, NL-SH, TM1, TW-99, DD-ME1, PK1
and PK1R, respectively.} \label{pn}
\end{sidewaysfigure}

\begin{sidewaysfigure}[htbp]
\centering
\includegraphics[scale=0.6]{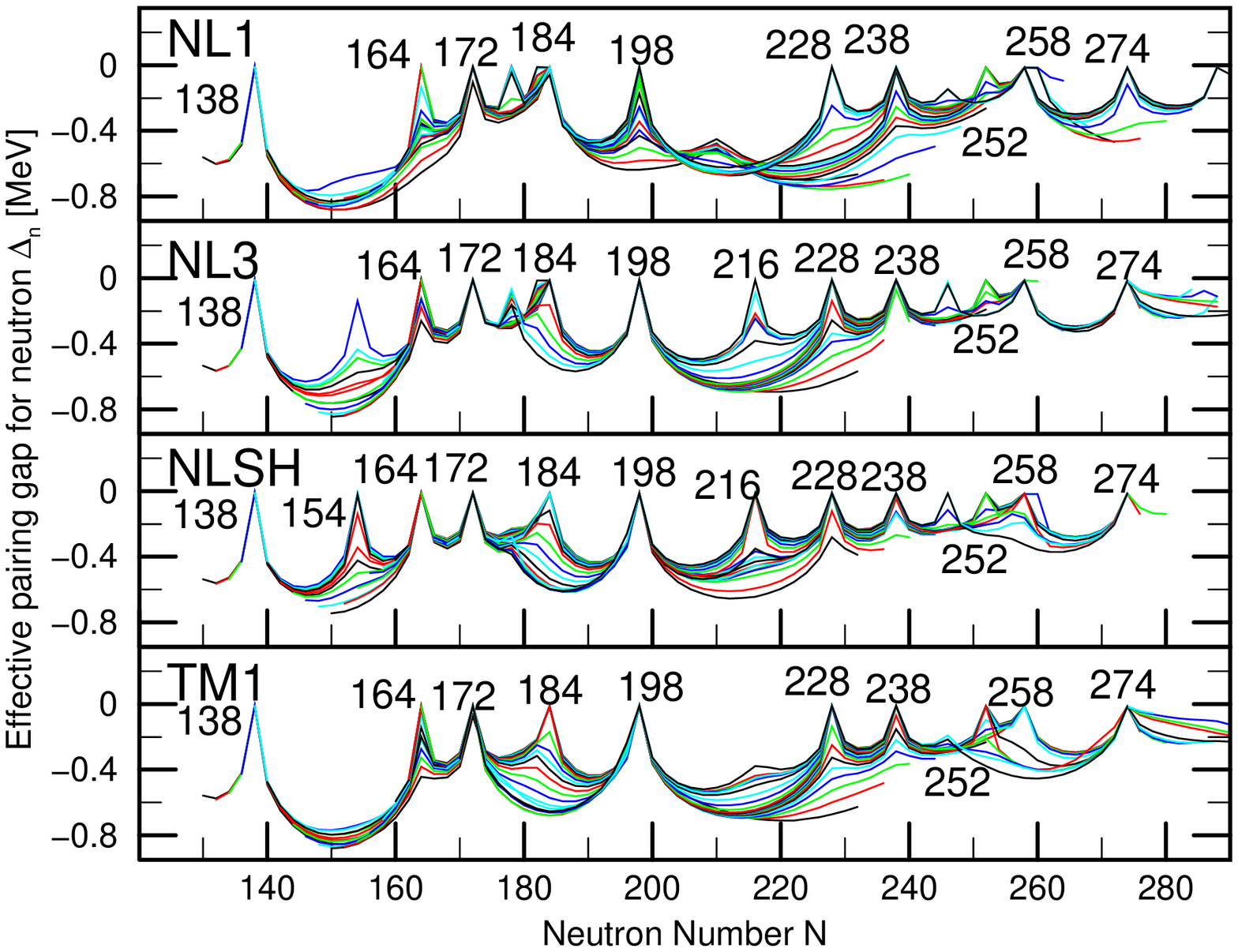}\hspace{0.5cm}
\includegraphics[scale=0.6]{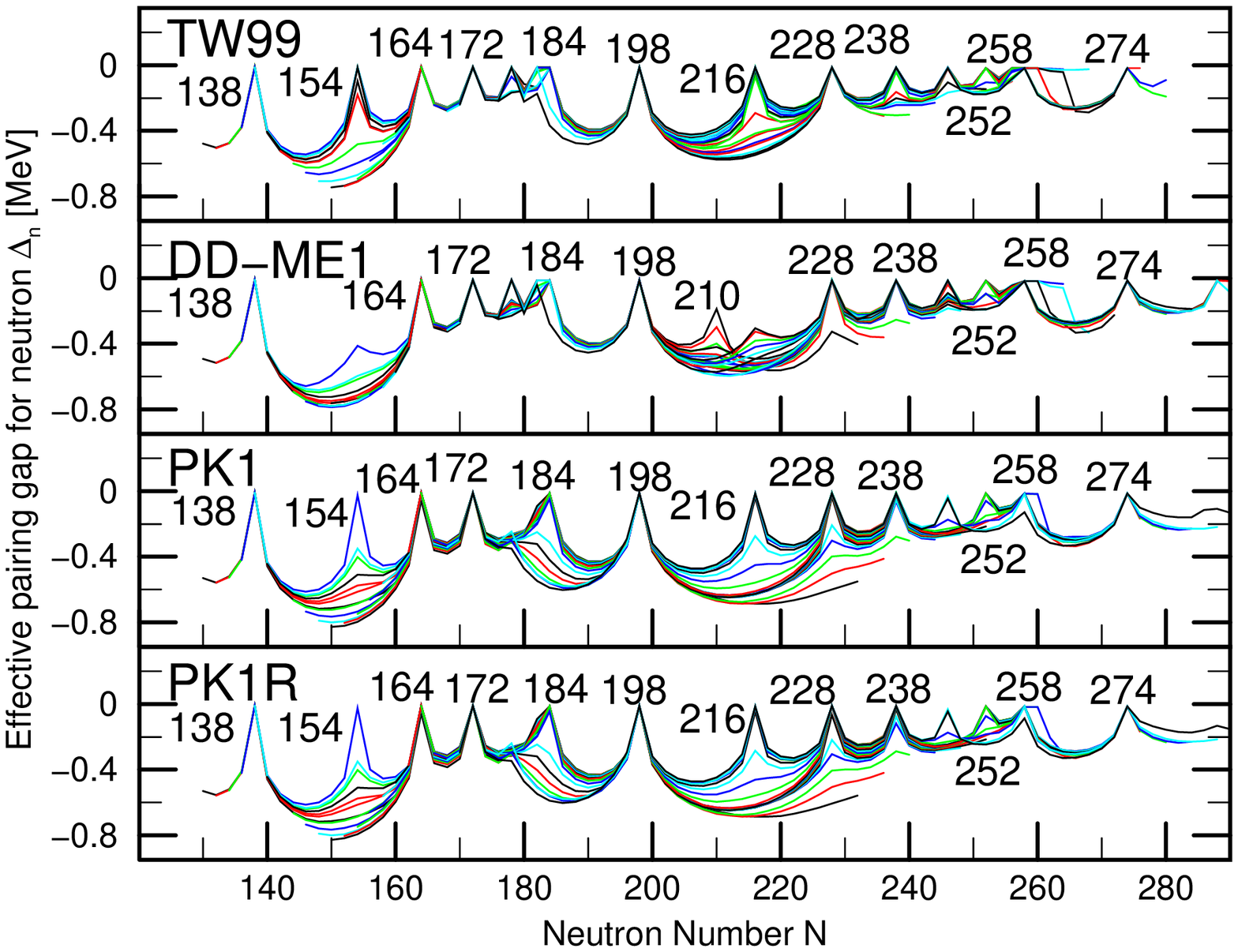}
\caption{The effective pairing gaps for neutron  $\Delta_{n}$ as a
function of neutron number obtained by RCHB calculation with
effective interactions NL1, NL3, NL-SH, TM1, TW-99, DD-ME1, PK1
and PK1R, respectively.} \label{dn}
\end{sidewaysfigure}

Fig. \ref{pn} shows the pairing energies for neutron
$E_{pair}^{n}$ from the RCHB calculation for even-even nuclei with
$N$=130 - 312 as a function of $N$ with eight effective
interactions. The vanishing of pairing at $N$=138, 164, 172, 184,
198, 228, 238, 252, 258, and 274 are common for all interactions
while the vanishing of pairing at $N$=154 and $N$=216 exist only
for NL3, NLSH, TW99, PK1, and PK1R, and while the vanishing of
pairing  at $N$=210 exists only for DD-ME1. The same pattern is
observed by the effective pairing gaps for neutron $\Delta_{n}$ in
Fig. \ref{dn}.

From the above discussion, the magic numbers $Z$=120, 132, and 138
and $N$=138, 164, 172, 184, 198, 228, 238, 252, 258, and 274 are
supported by the pairing energies and the effective pairing gaps.

\subsection{Alpha decay half-lives}
Alpha decay is one of the most predominant decay modes for
superheavy nuclei. For an area of enhanced stability, the
$\alpha$-decay half-lives are expected to be longer than its
neighbors. The phenomenological formula of Viola and Seaborg
\cite{Viola1966} for calculation of $\alpha$-decay half-lives with
the parameter in Ref. \cite{Sobiczewski1989}:
 \beq
    log_{10}T_{\alpha}
        =(1.66175Z_{P}-8.5166) Q_{\alpha}^{-1/2}-(0.20228Z_{P}+33.9069)
 \label{vs}
 \eeq
are used, where $Z_{P}$ is the proton number of a parent nucleus,
$Q_{\alpha}$ is the $\alpha$-decay energy
$Q_{\alpha}=E_B(Z-2,N-2)-E_B(Z,N)+28.295673$ in MeV, and
$T_{\alpha}$ is in seconds. It should be noted that Eq.(\ref{vs})
is based on the WKB approximation, and provides only a rather
crude estimate of $T_{\alpha}$ since it disregards many structure
effects such as deformation, and configuration changes, etc.

\begin{figure}[htbp]
\centering
\includegraphics[scale=0.3]{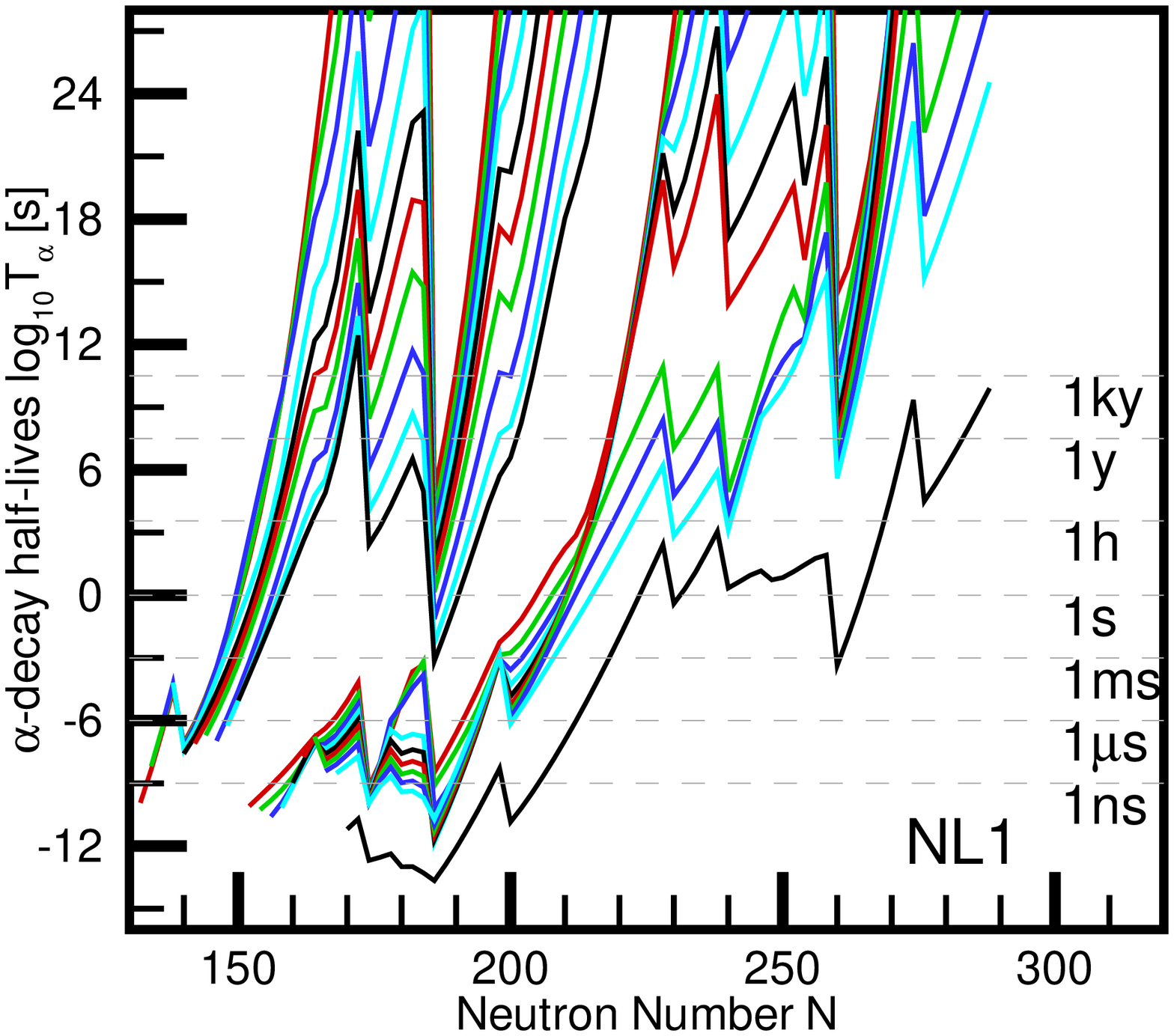}
\includegraphics[scale=0.3]{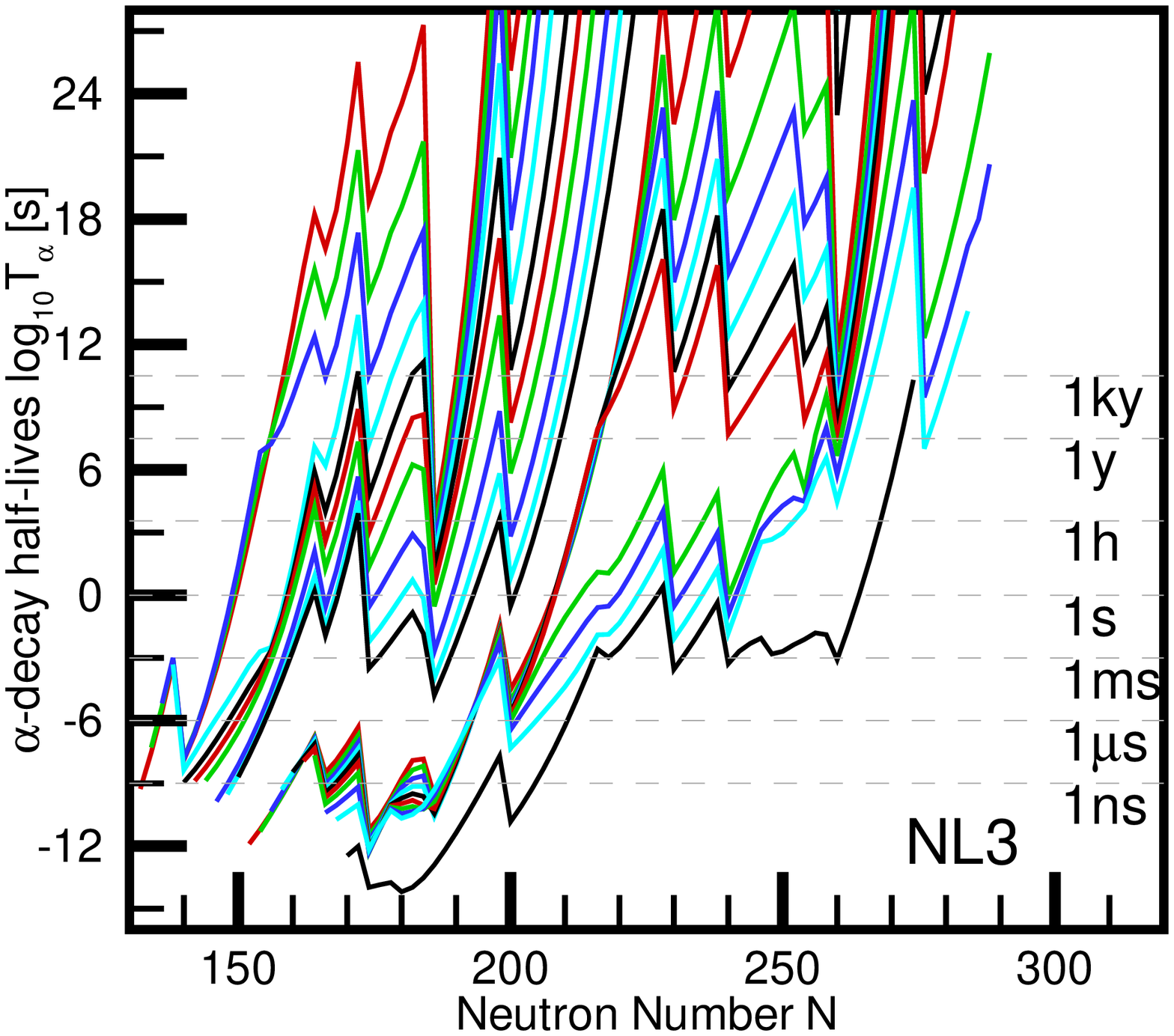}
\includegraphics[scale=0.3]{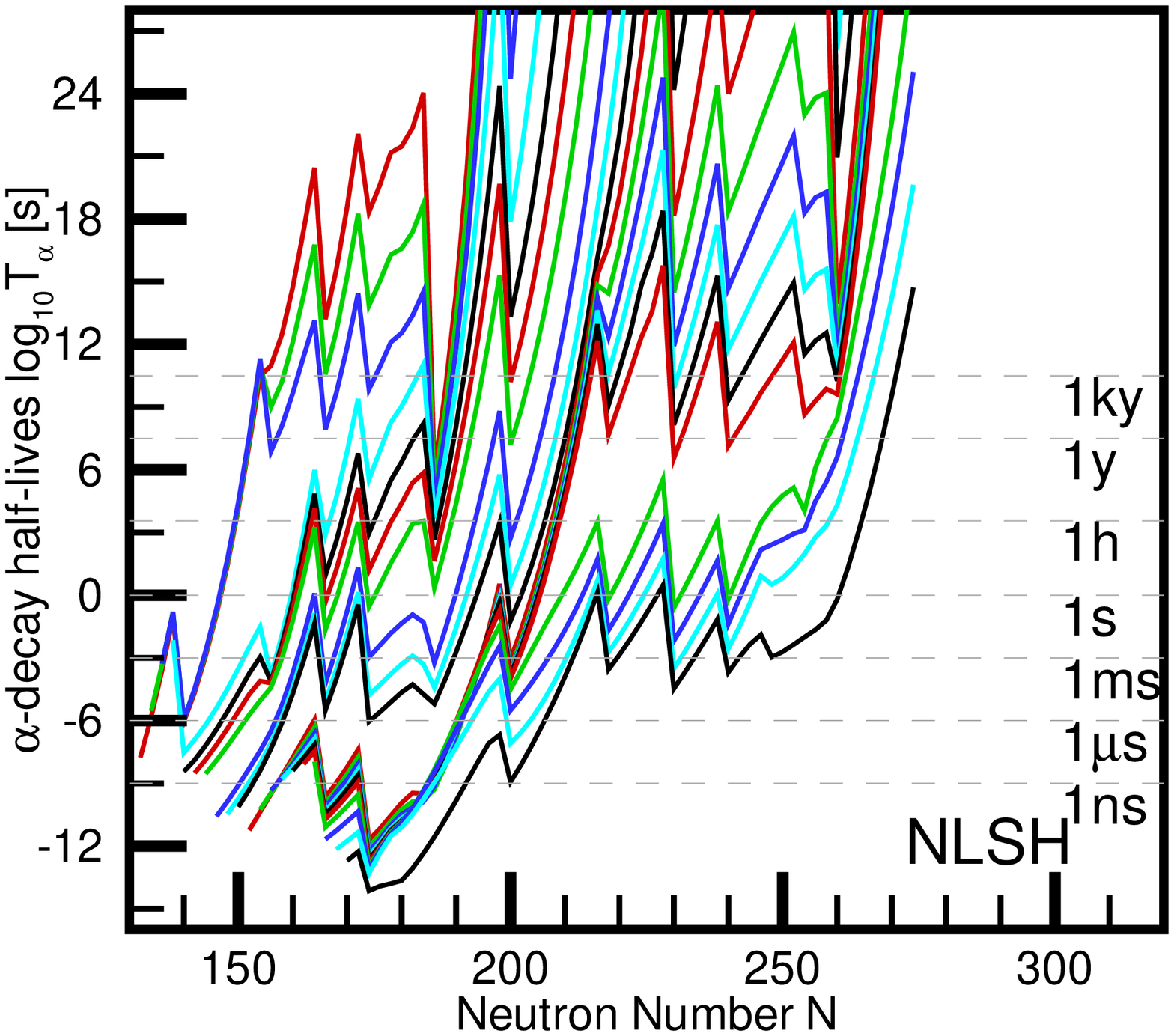}
\includegraphics[scale=0.3]{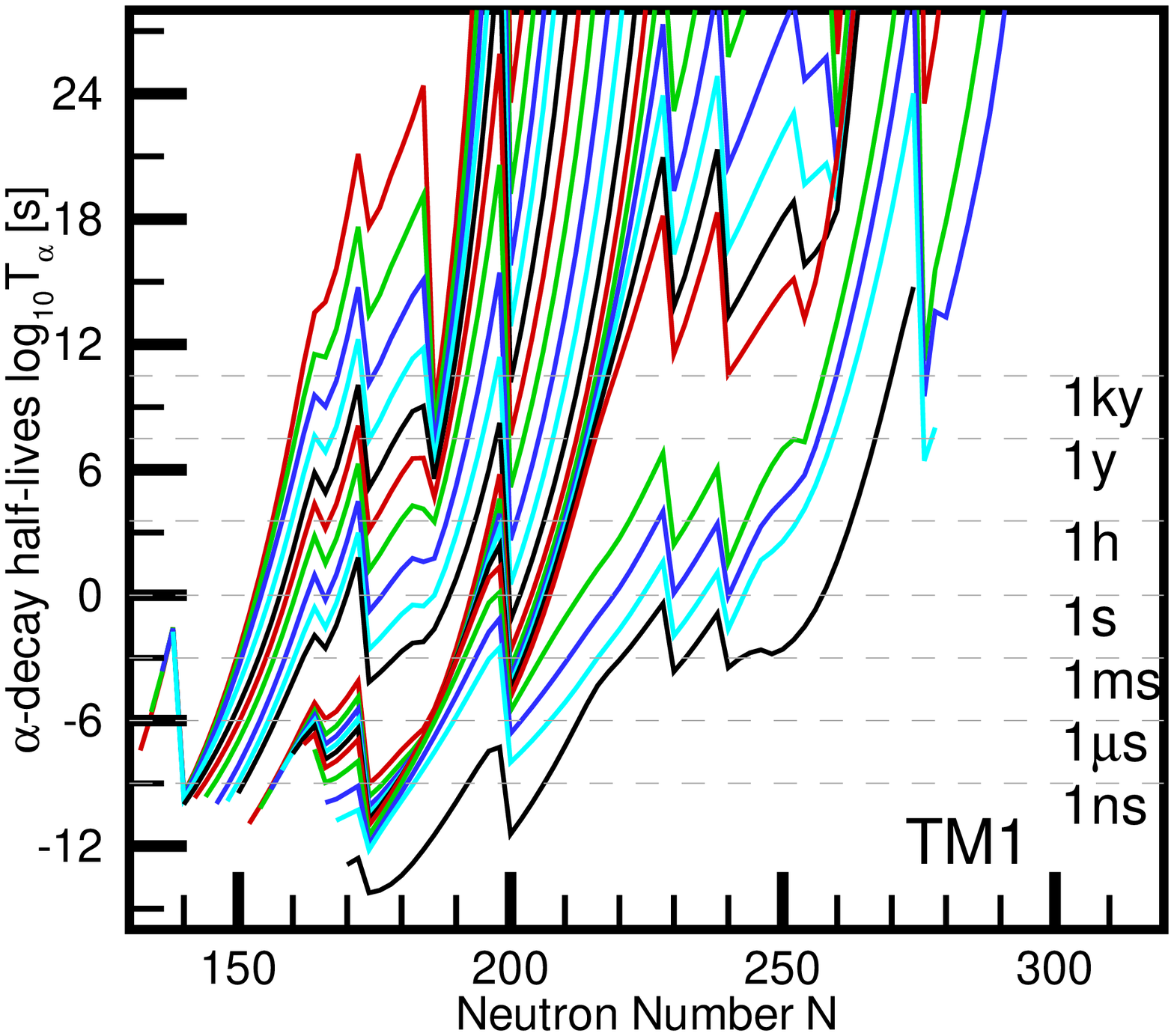}
\includegraphics[scale=0.3]{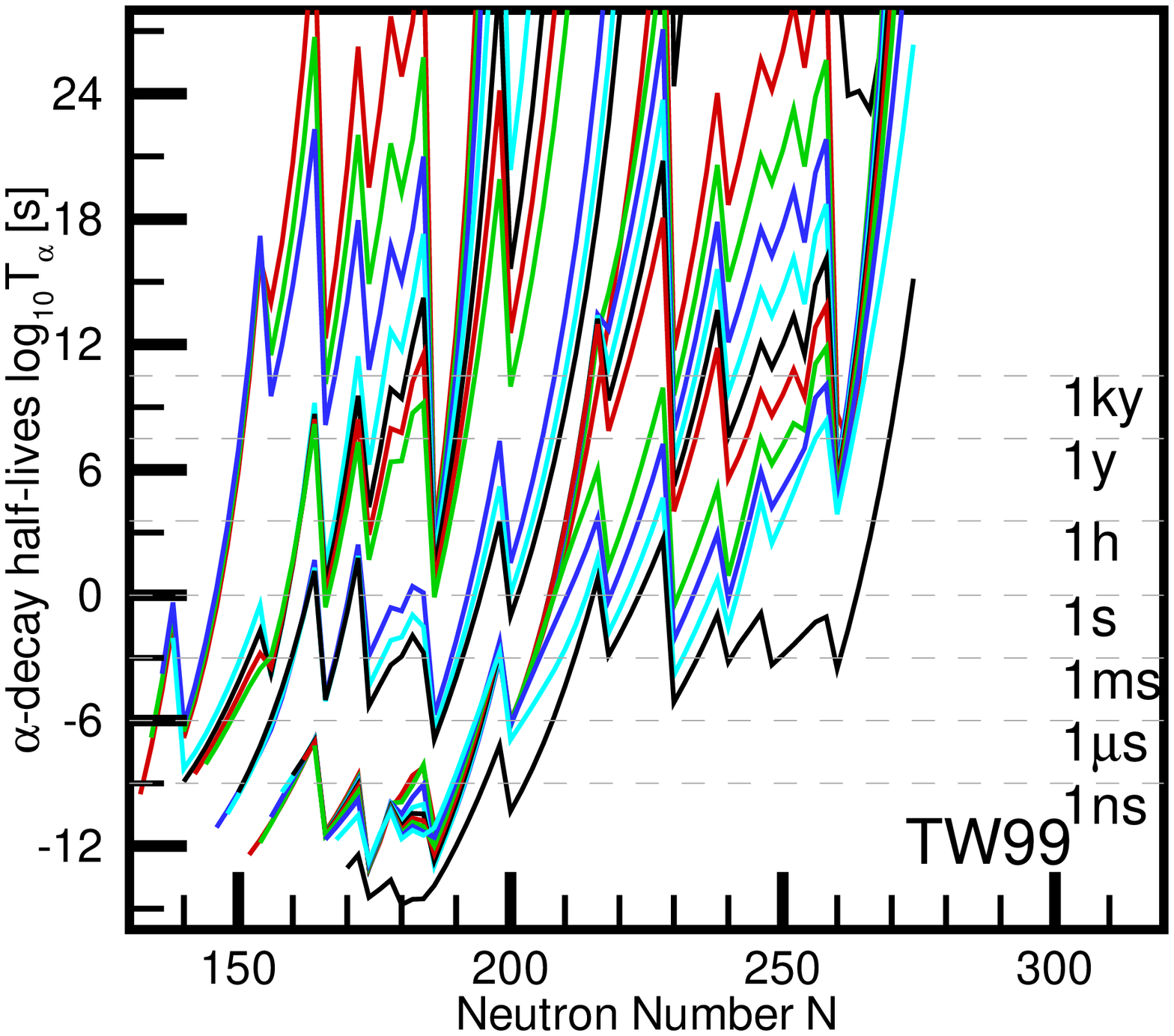}
\includegraphics[scale=0.3]{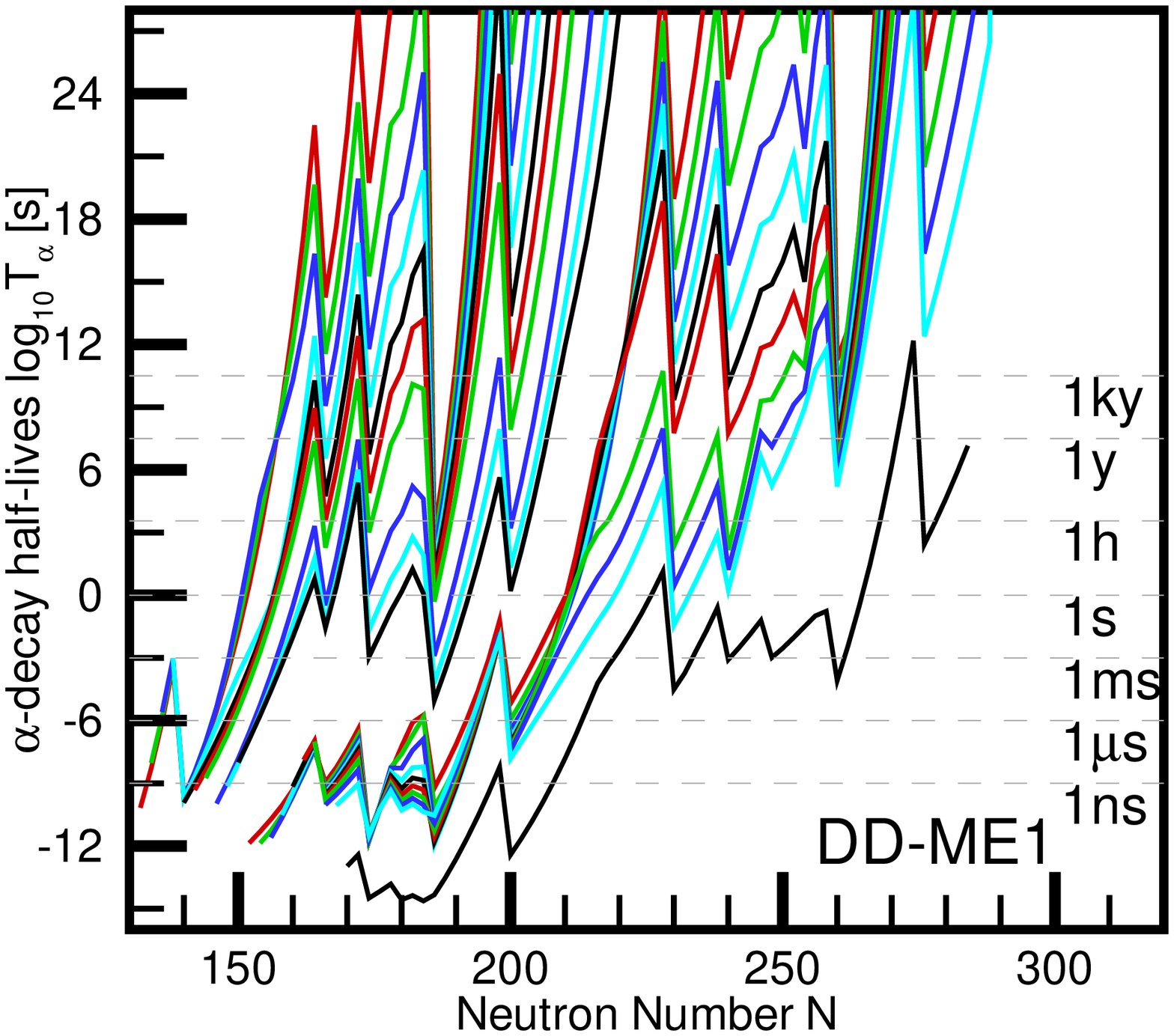}
\includegraphics[scale=0.3]{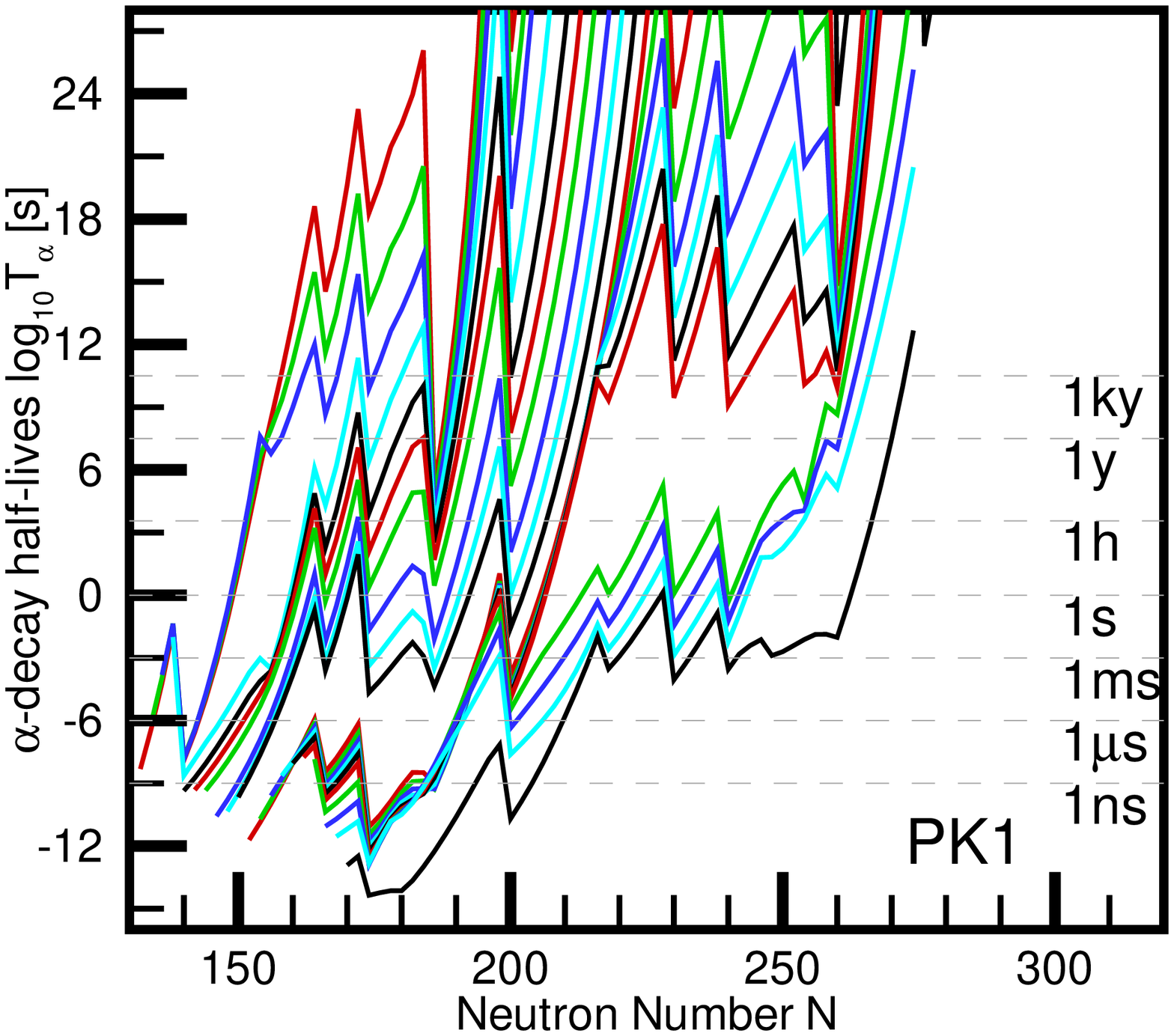}
\includegraphics[scale=0.3]{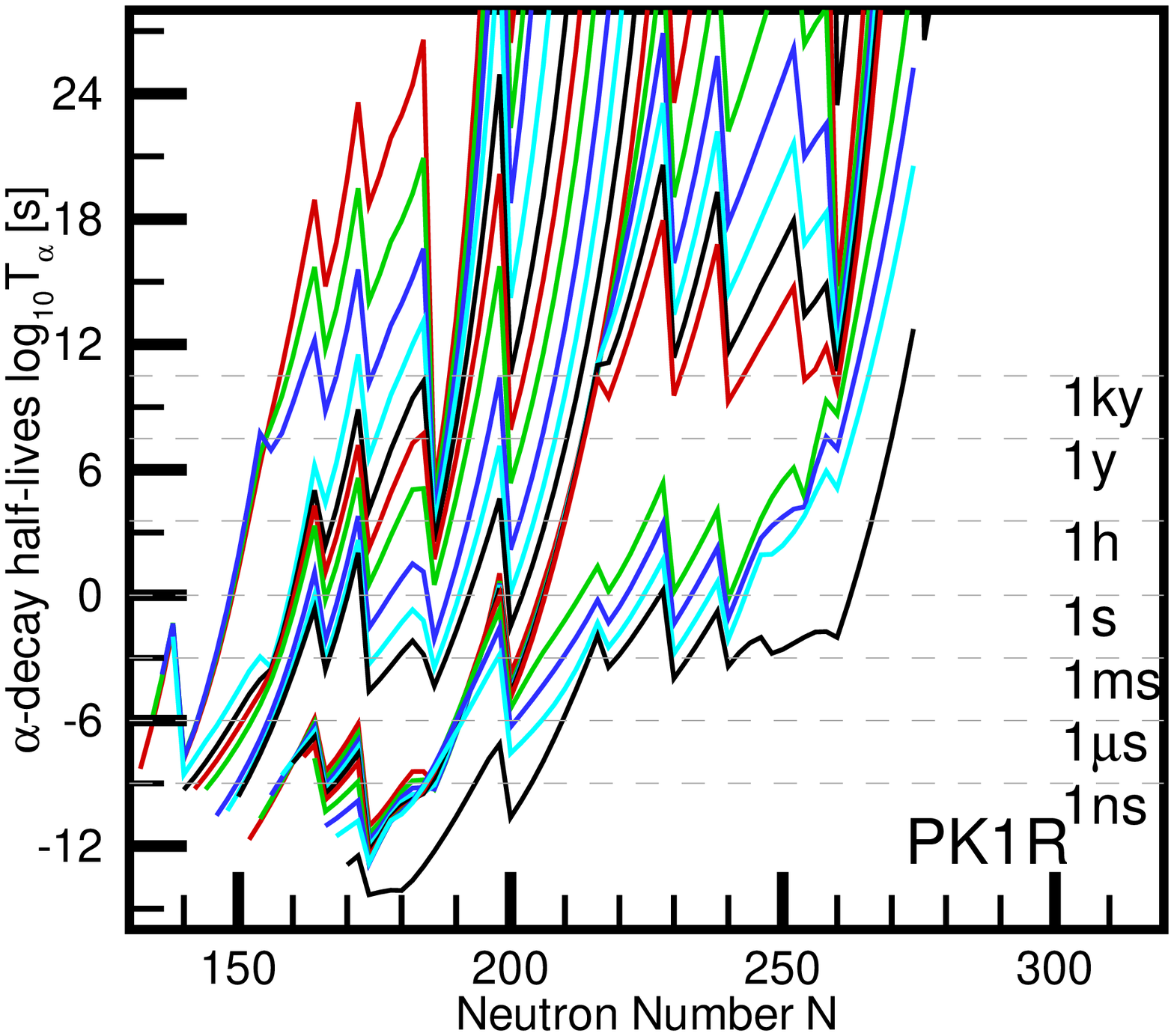}
\caption{The $\alpha$-decay half-lives $T_{\alpha}$ as a function
of neutron number obtained by RCHB calculation with effective
interactions NL1, NL3, NL-SH, TM1, TW-99, DD-ME1, PK1, and PK1R,
respectively.} \label{t}
\end{figure}

In Fig. \ref{t}, we plot the half-lives $T_{\alpha}$ from the RCHB
calculation in logarithm scale as a function of neutron number $N$
with effective interactions NL1, NL3, NLSH, TM1, TW99, DD-ME1, PK1
and PK1R. The half-lives correspond to 1ns, 1$\mu$s, 1ms, 1s, 1h,
1y, and 1ky are marked by the dashed lines. Each curve in this
figure corresponds to an isotopic chain in the region $Z$=102-140.
In Fig. \ref{t}, the half-lives $log_{10}T_{\alpha}$ increase with
$N$. The jumps of the curves correspond to magic proton numbers
$Z$=106, 120, 132, and 138, which depend on the effective
interactions. The peaks of each curve correspond to magic neutron
numbers, i.e., $N$=172, 184, 228, 238 and 258. It can be seen that
the magic numbers suggested by two-neutron separation energies
$S_{2n}$(two-neutron gaps $\delta_{2n}$) or two-proton separation
energies $S_{2p}$(two-proton gaps $\delta_{2p}$) can also be found
here.

\vspace{1cm}

So far, the RCHB theory with eight different interactions, namely
NL1, NL3, NLSH, TM1, TW99, DD-ME1, PK1£¬ and PK1R has been applied
to the superheavy region with proton number $Z$=100 - 140 and
neutron number $N$= ($Z$+30) - (2$Z$+32). After detailed
analysises on two-nucleon separation energies $S_{2p}$ and
$S_{2n}$, two-nucleon gaps $\delta_{2p}$ and $\delta_{2n}$, the
shell correction energies $E_{shell}^{p}$ and $E_{shell}^{n}$, as
well as the pairing energies $E_{pair}^{p}$ and $E_{pair}^{n}$,
the effective pairing gaps $\Delta_{p}$ and $\Delta_{n}$, the
magic numbers are predicted. Table \ref{magicp} and Table
\ref{magicn} list the possible magic proton and neutron numbers
suggested by the above quantities and interactions. In these
tables, the physical quantity (row-wise) corresponding to the
magicity of certain proton number (column-wise) is marked as a
black square for each interaction. According to Table \ref{magicp}
and Table \ref{magicn}, the doubly magic superheavy nuclei can be
the combination of magic proton numbers $Z$=120, 132, and 138 and
magic neutron numbers $N$=172, 184, 198, 228, 238, and 258. Such
magic numbers are presented as gray lines in Fig. \ref{nuch}. It
should be noted that the crosses of lines in Fig. \ref{nuch} do
not necessary mean doubly magic nuclei as there are also shell
quenching effects. However, these crosses include all the possible
doubly magic nuclei in the state-of-the-art relativistic approach.

\section{Doubly magic nuclei in superheavy nuclei}

The magic numbers obtained before with the RCHB theory are based
on the assumption of spherical geometrical configuration. However,
the assumption is not always true for superheavy nuclei. In fact,
most superheavy nuclei found experimentally are known to be
deformed. It is worthy to investigate the potential energy
surfaces in order to see the validity of spherical configuration.
For this purpose, the deformation-constrained RMF calculations
have been done. The deformation parameter $\beta_2$ of the
harmonic oscillator basis (with 20 shells) is set to the expected
deformation to obtain high accuracy and reduce the computing
times. Although such calculation is very lengthy for superheavy
nuclei, we have systematically calculated the potential energy
surfaces for all the possible doubly magic nuclei with the
effective interaction NL3. The calculation with other interactions
for some typical doubly magic nuclei are also done for comparison.
To save space, only the results for $Z$=120 isotopes with NL3 are
presented here.

\begin{figure}[htbp]
\centering
\includegraphics[scale=0.45]{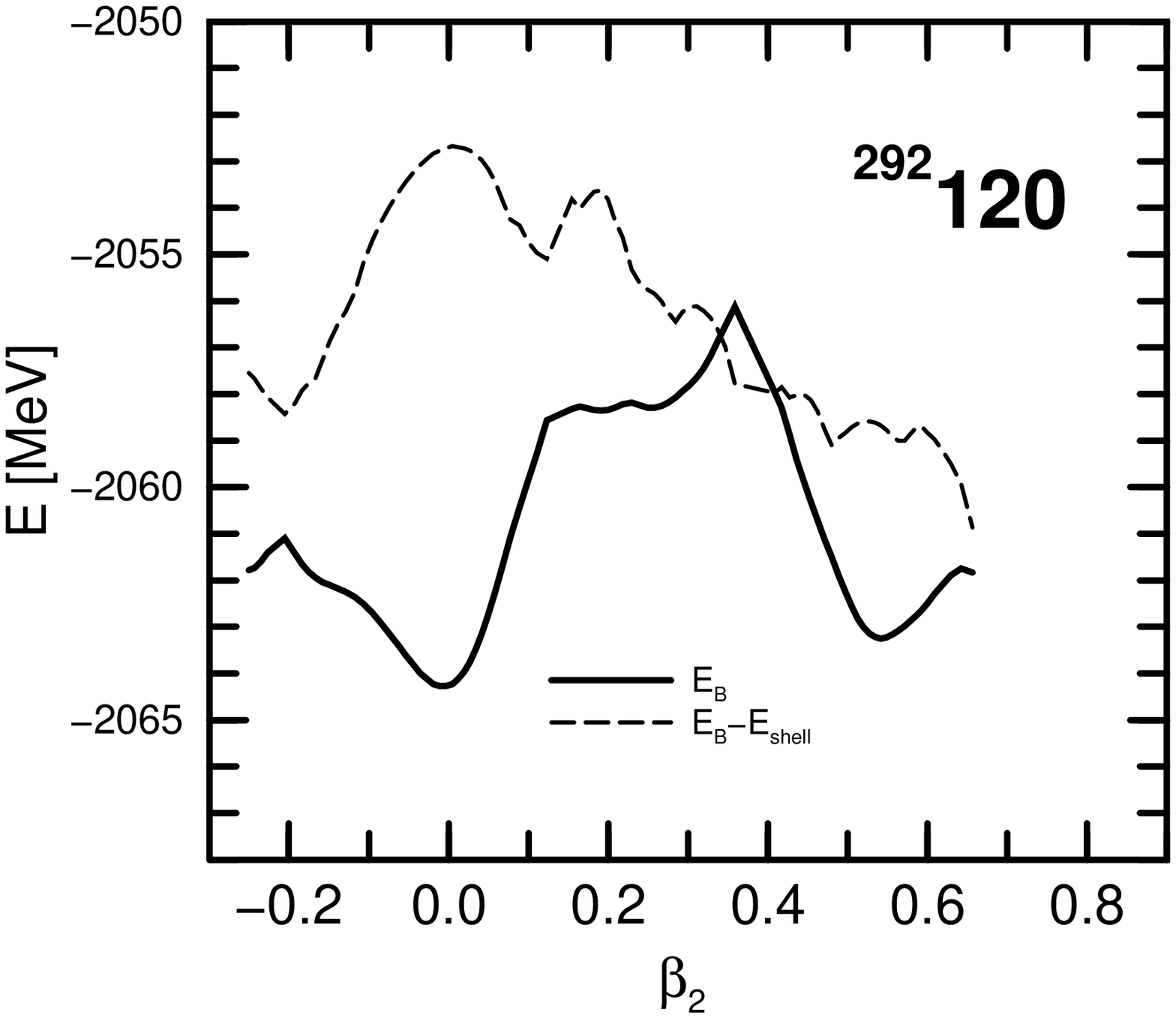}
\includegraphics[scale=0.45]{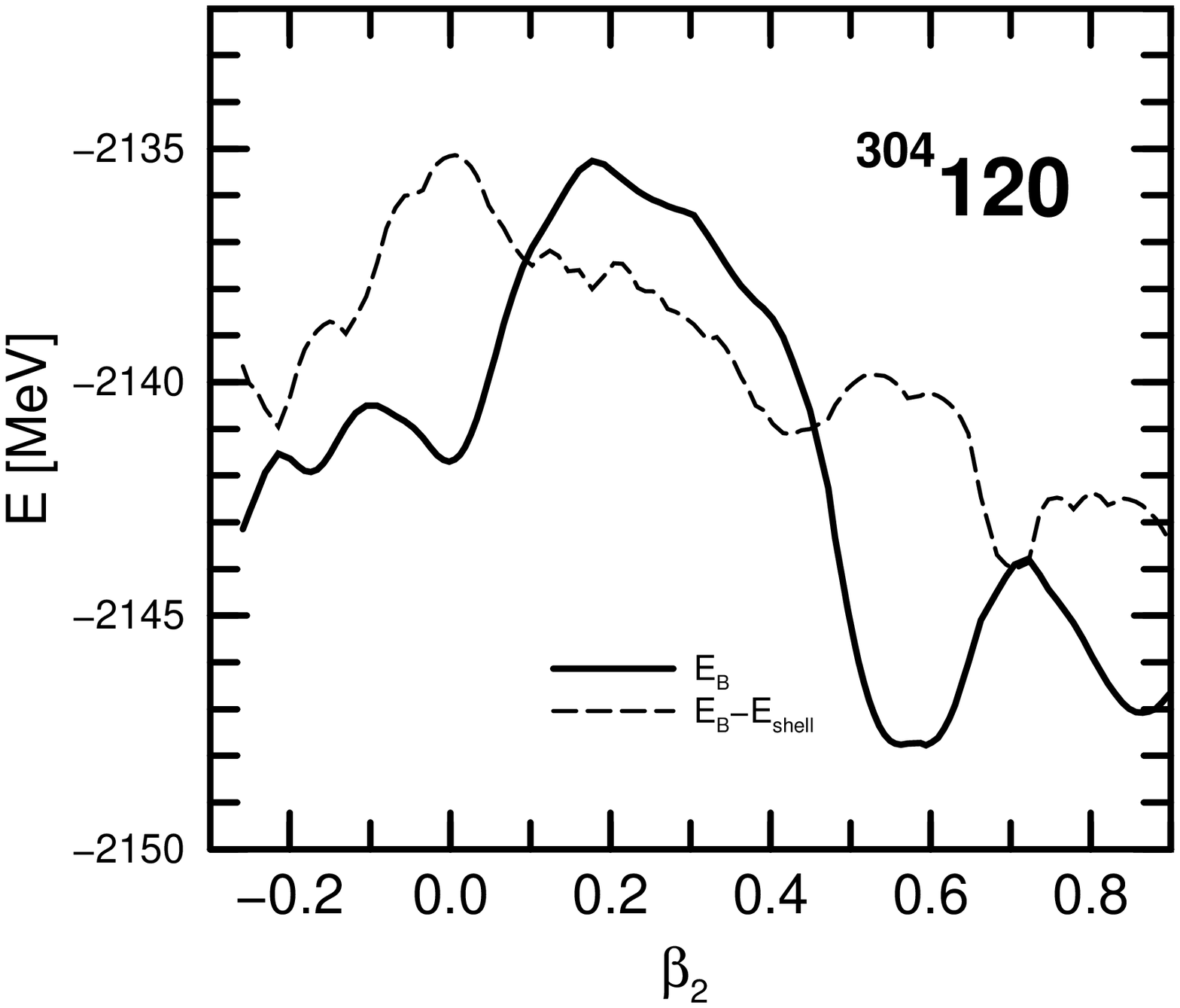}
\includegraphics[scale=0.45]{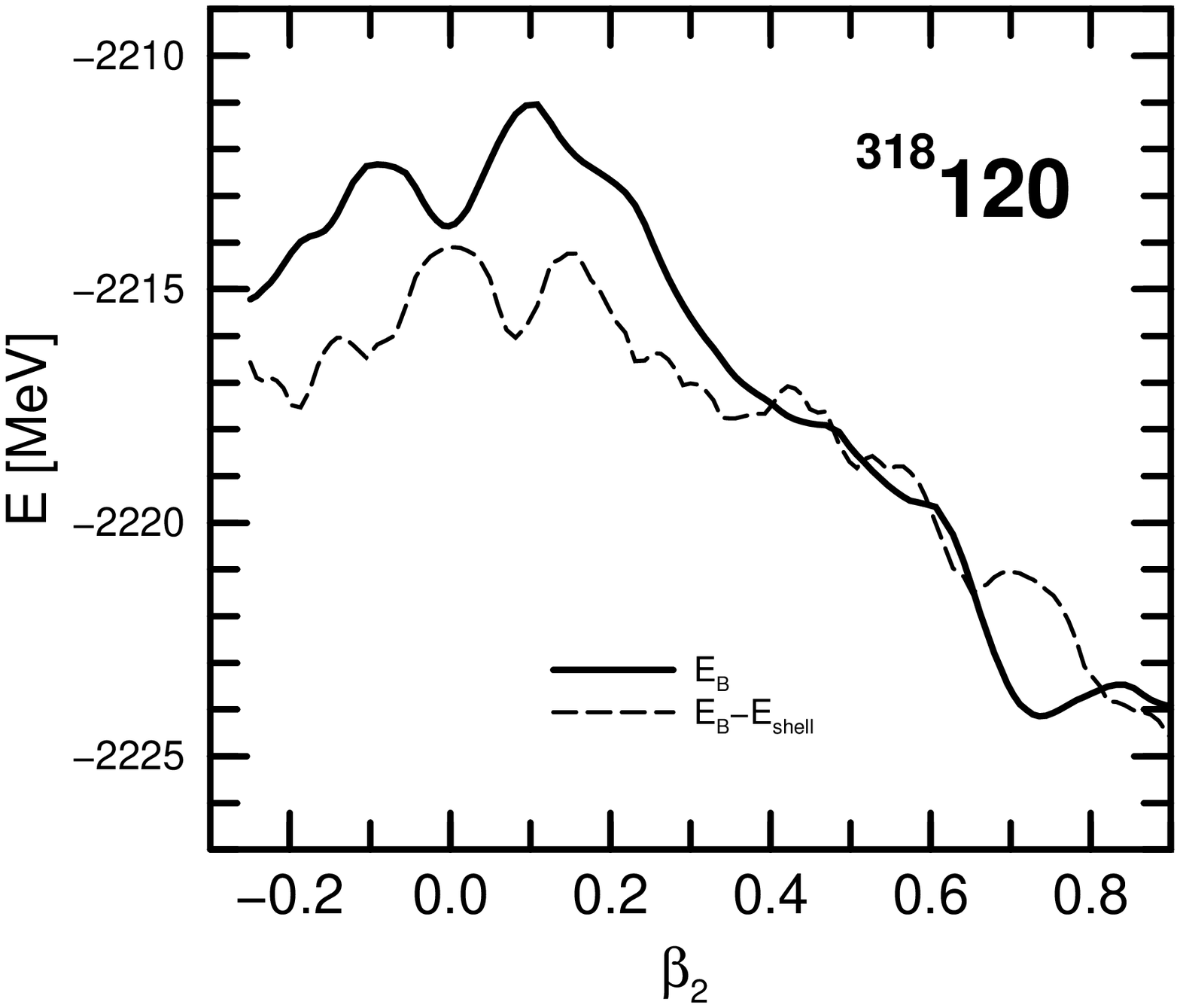}
\includegraphics[scale=0.45]{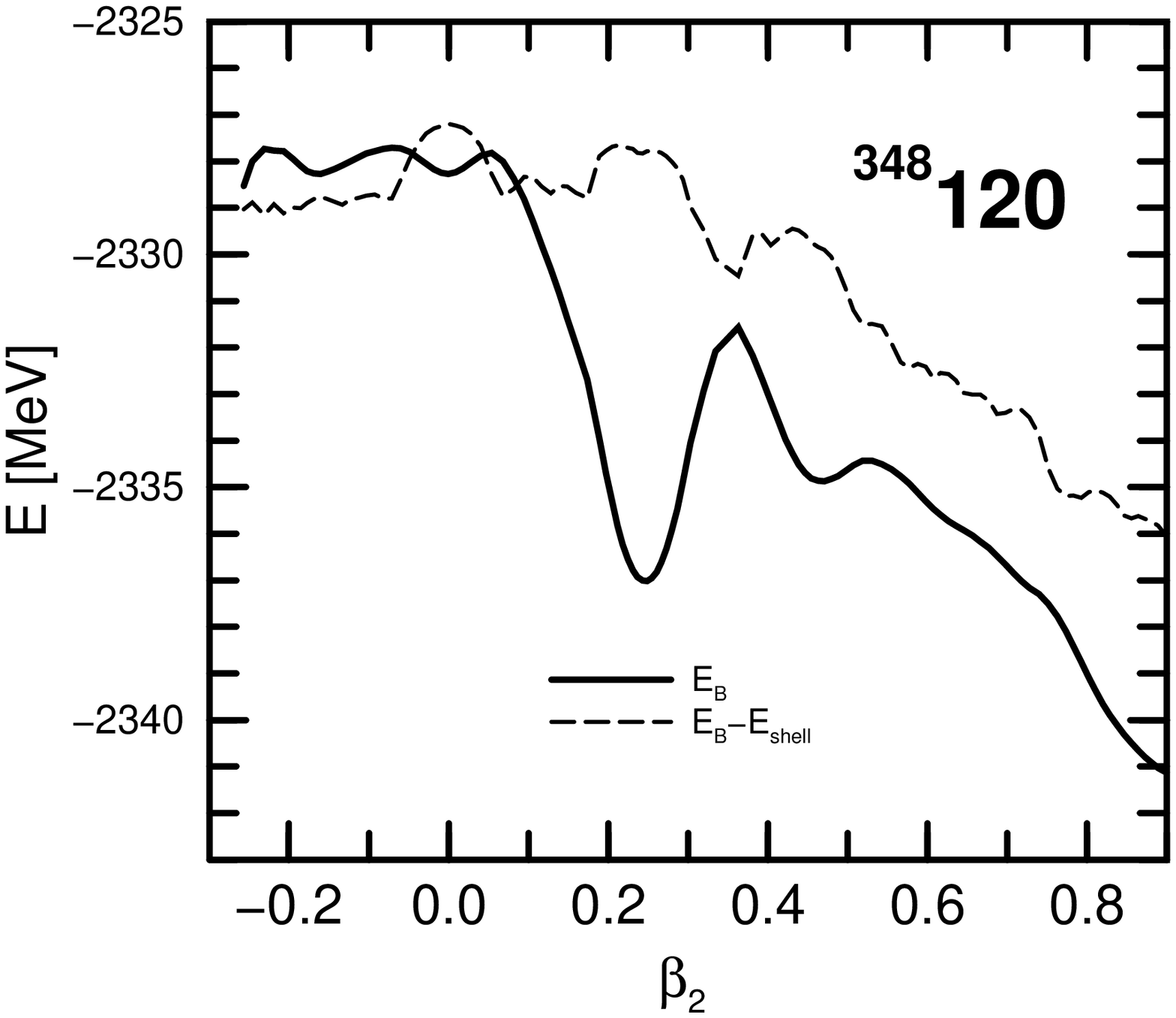}
\includegraphics[scale=0.45]{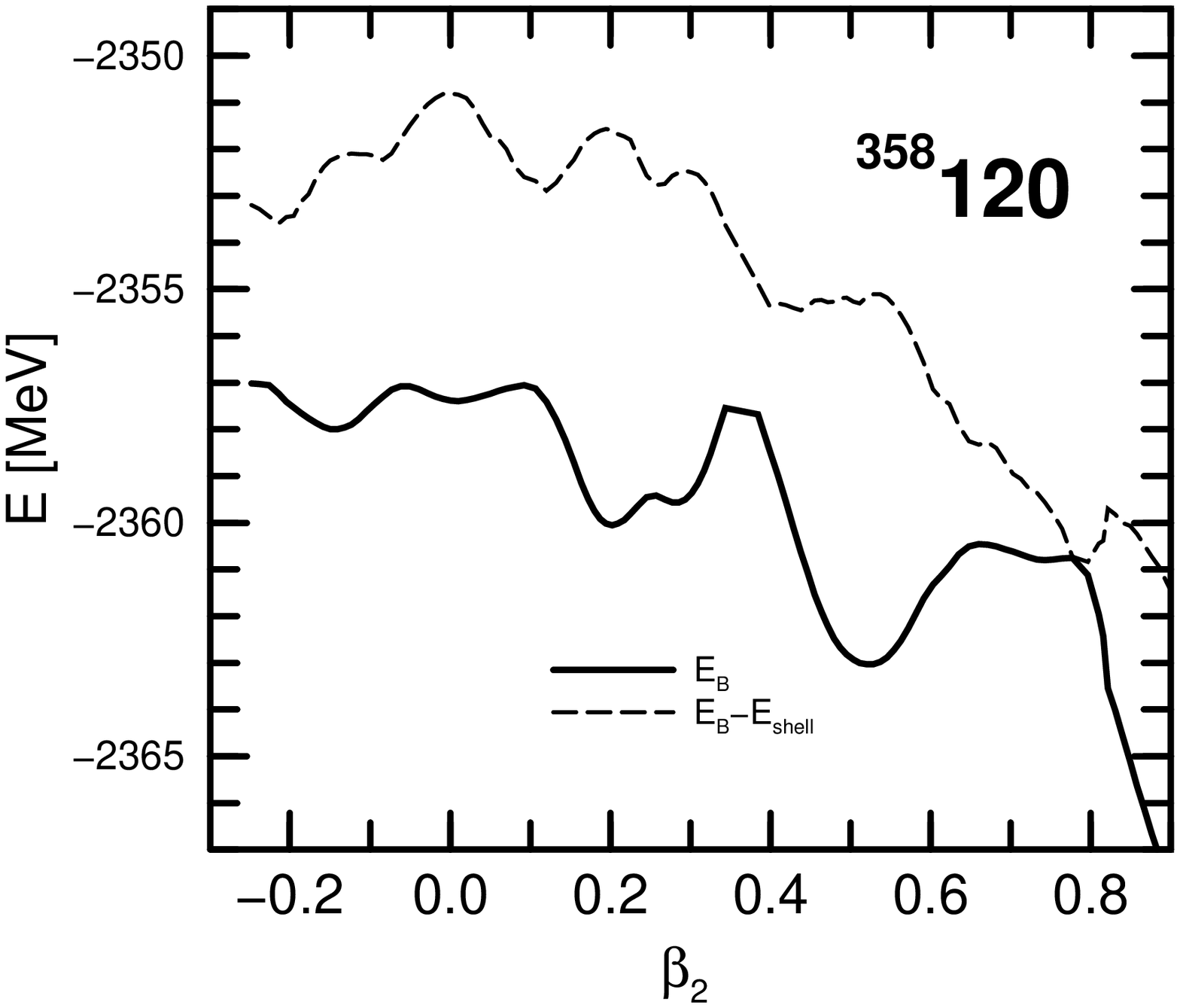}
\includegraphics[scale=0.45]{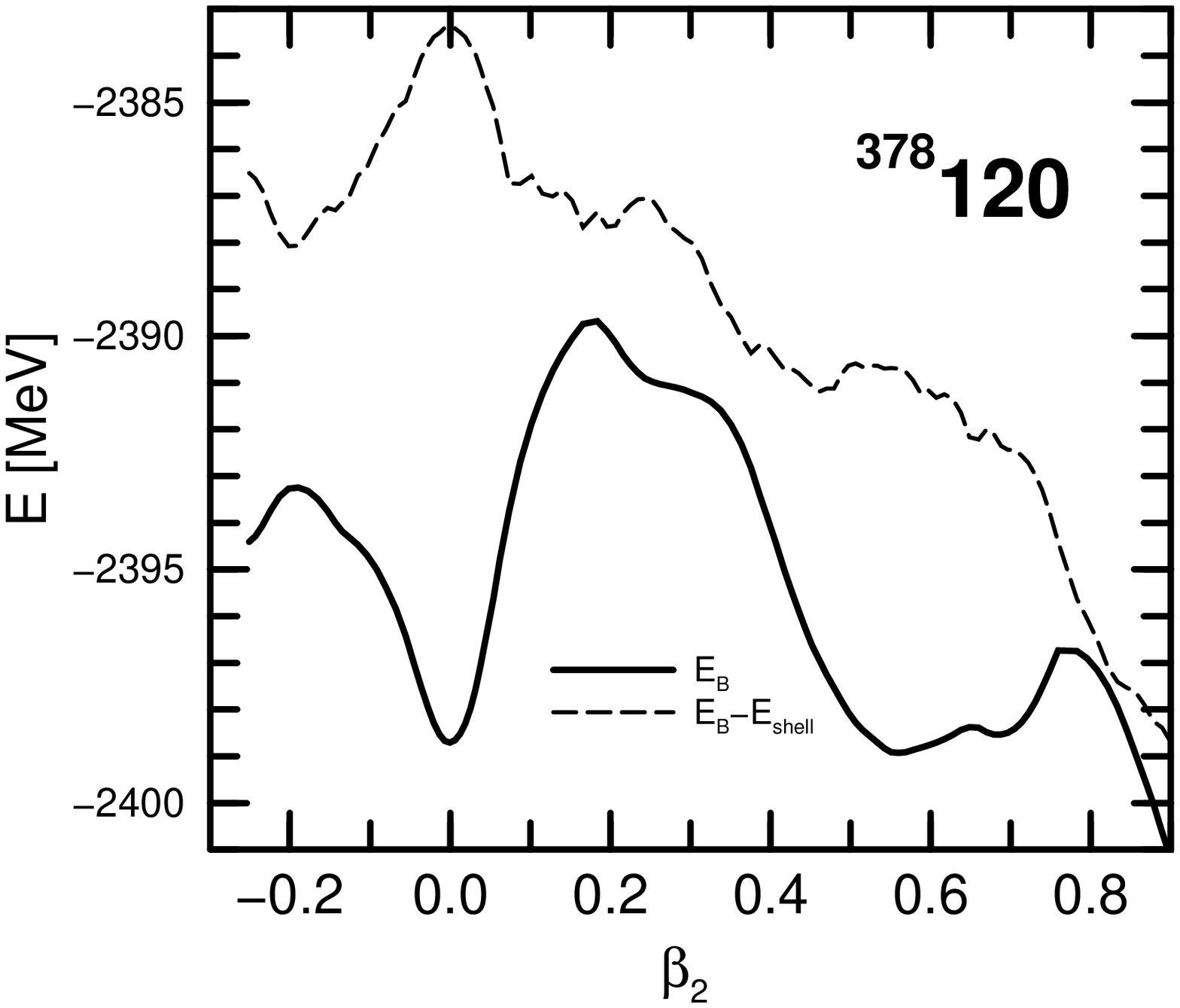}
\caption{The binding energy and the equivalent macroscopic energy
of $^{292,304,318,348,358,378}$120 calculated in the constrained
RMF theory with effective interaction NL3.}
\label{120NL3es}
\end{figure}

In Fig. \ref{120NL3es}, the potential energy surfaces for nuclei
$^{292}$120, $^{304}$120, $^{318}$120, $^{348}$120, $^{358}$120,
and $^{378}$120 are presented with the solid lines in
corresponding subfigures. These nuclei are clearly shown as the
crosses with $Z=120$ in Fig. \ref{nuch}. Although these nuclei
have the same proton number, their potential energy surfaces are
quite different from each other. For nuclei $^{292}$120,
$^{304}$120 and $^{378}$120, there is an obvious local minimum
with the spherical configuration $\beta_2 \sim 0$, while another
local minimum with large deformation $\beta_2 \sim 0.6$ can be
also clearly seen. For $^{318}$120, the spherical local minimum is
very shallow, and for $^{348}$120 and $^{358}$120, the spherical
minimum is hardly seen and a local minimum with $\beta_2 \sim
0.25$ appears instead. In addition to the local minima discussed,
we should make remark on the absolute minima for these nuclei. For
$^{292}$120 and $^{378}$120, the two minima with the different
deformation are almost with the same energies, i.e. the so-called
``shape coexistence", may exist. In particular, the spherical
minimum is indeed the absolute minimum for $^{292}$120 ($E_B=$
-2064.3 MeV with $\beta_2 \sim 0$ vs. $E_B=$ -2063.2 MeV with
$\beta_2 \sim 0.6$), while the ground state of $^{378}$120 is
quite delicate with $E_B=$ -2398.7 MeV for $\beta_2 \sim 0$ and
$E_B=$ -2398.9 MeV for $\beta_2 \sim 0.6$. For $^{304}$120, the
absolute minimum at $\beta_2\sim 0.6$ is much deeper ($\sim$ 6.1
MeV) than the spherical configuration. In addition for
$^{348}$120, the absolute minimum lies at $\beta_2 \sim 0.2$, and
it is a well-deformed nucleus according to the present
calculation.

\begin{figure}[htbp]
\centering
\includegraphics[scale=0.45]{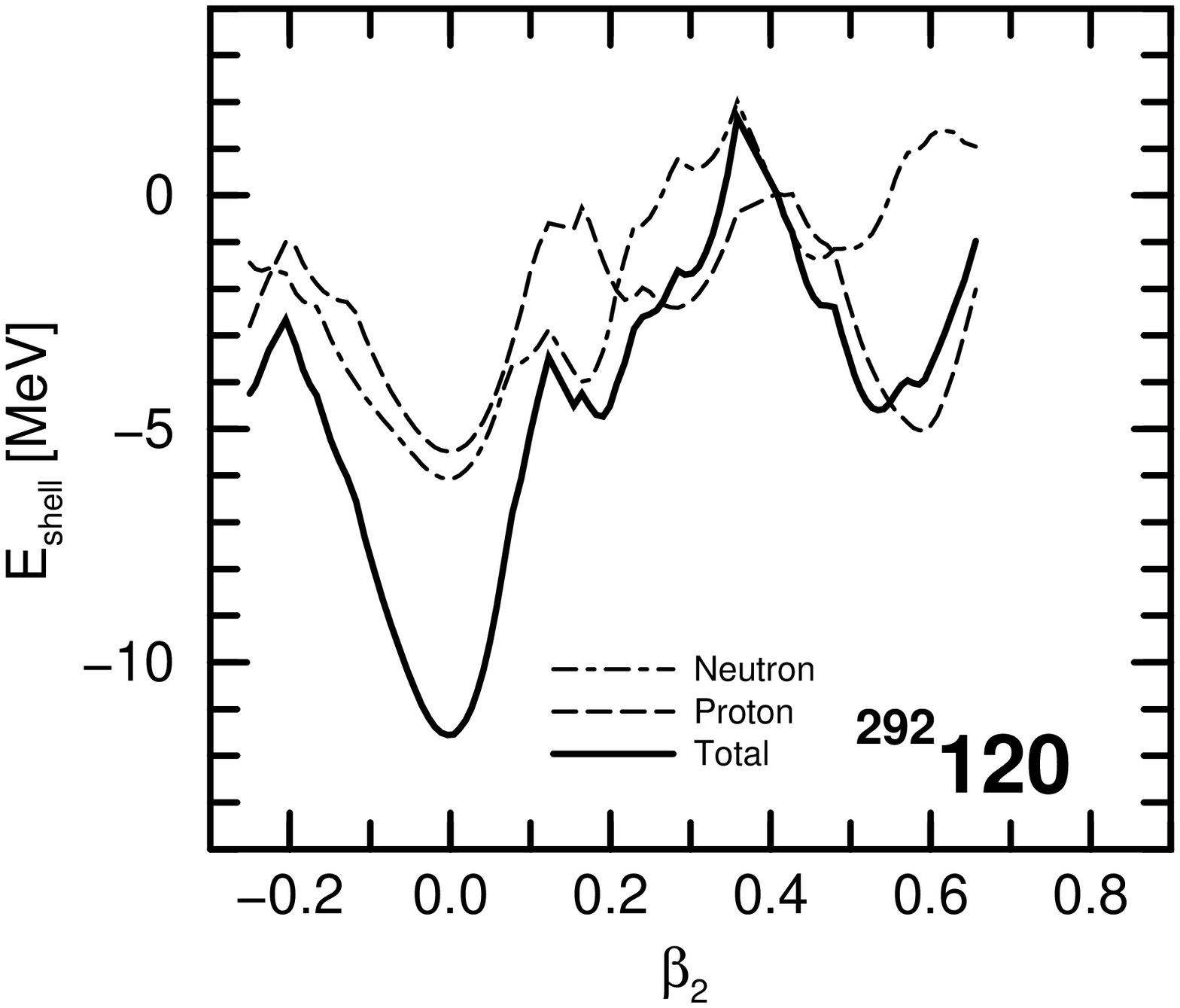}
\includegraphics[scale=0.45]{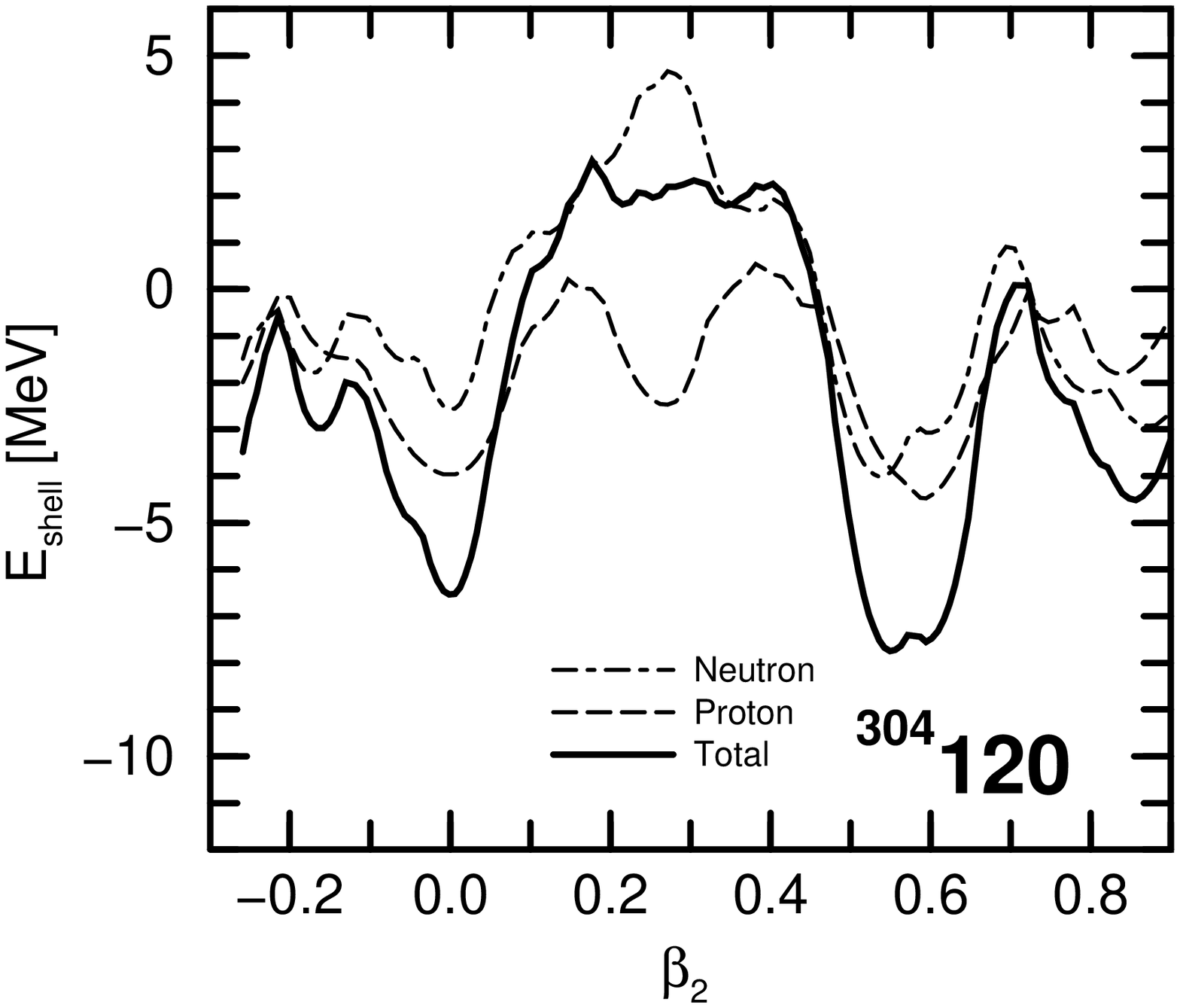}
\includegraphics[scale=0.45]{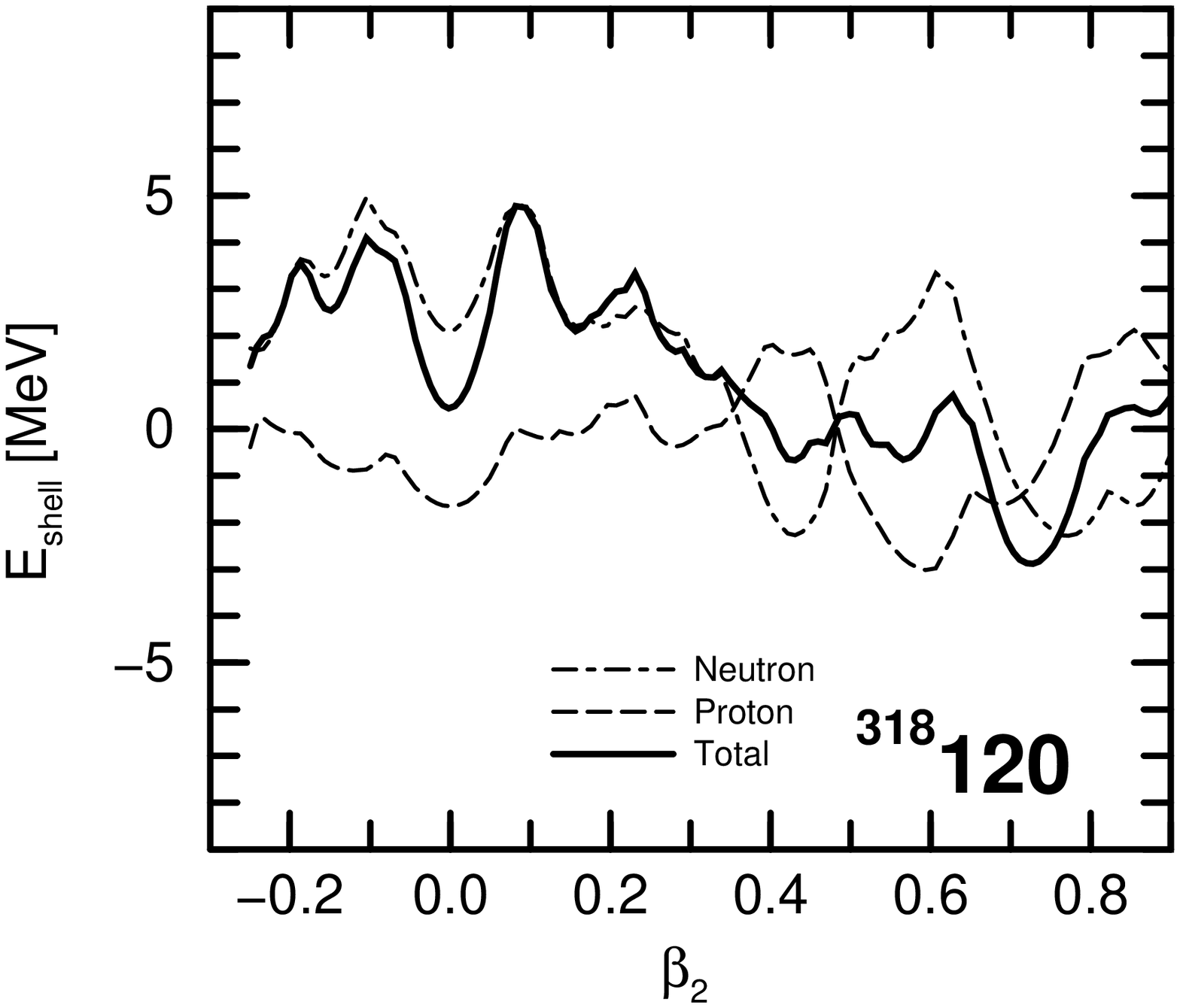}
\includegraphics[scale=0.45]{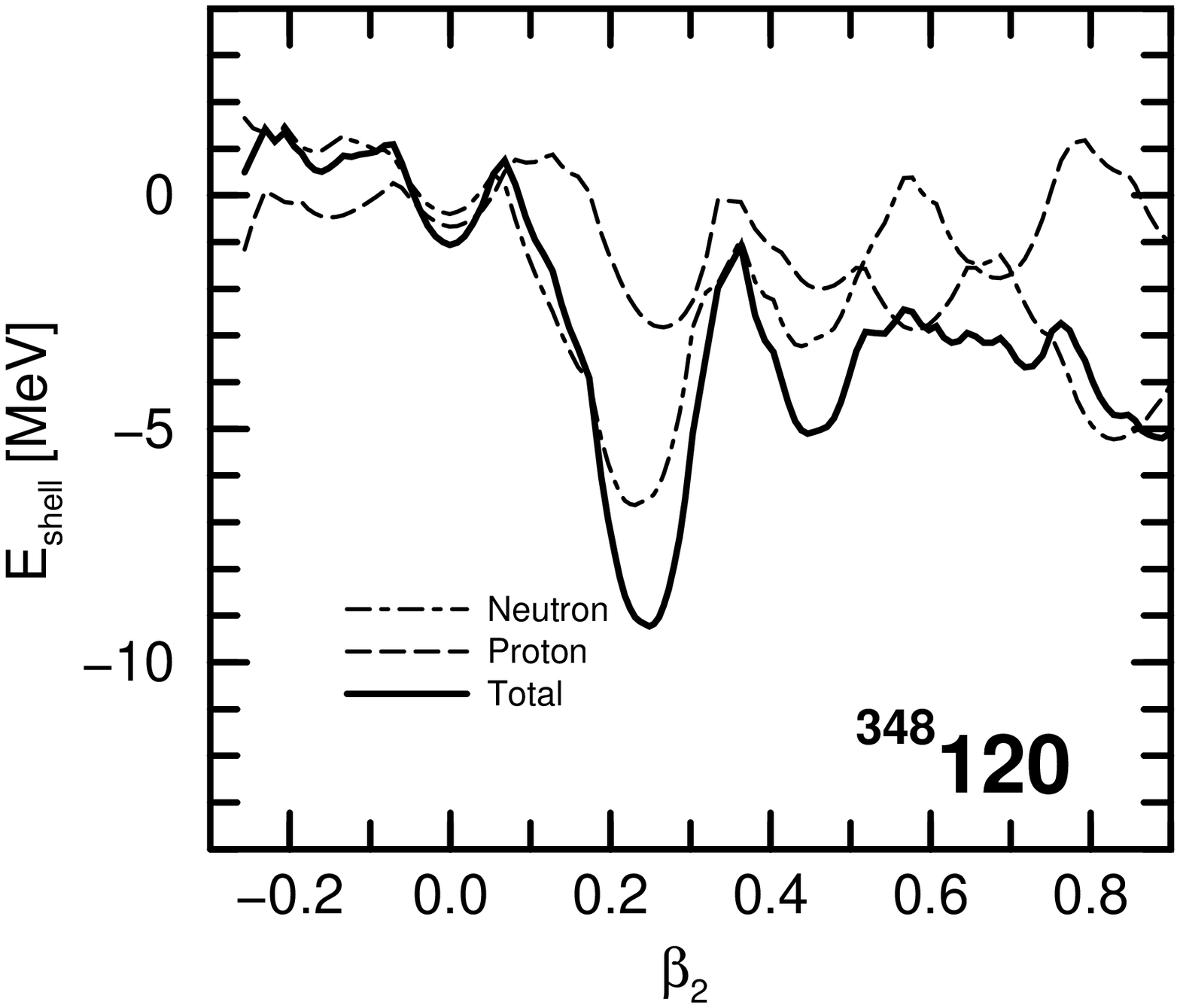}
\includegraphics[scale=0.45]{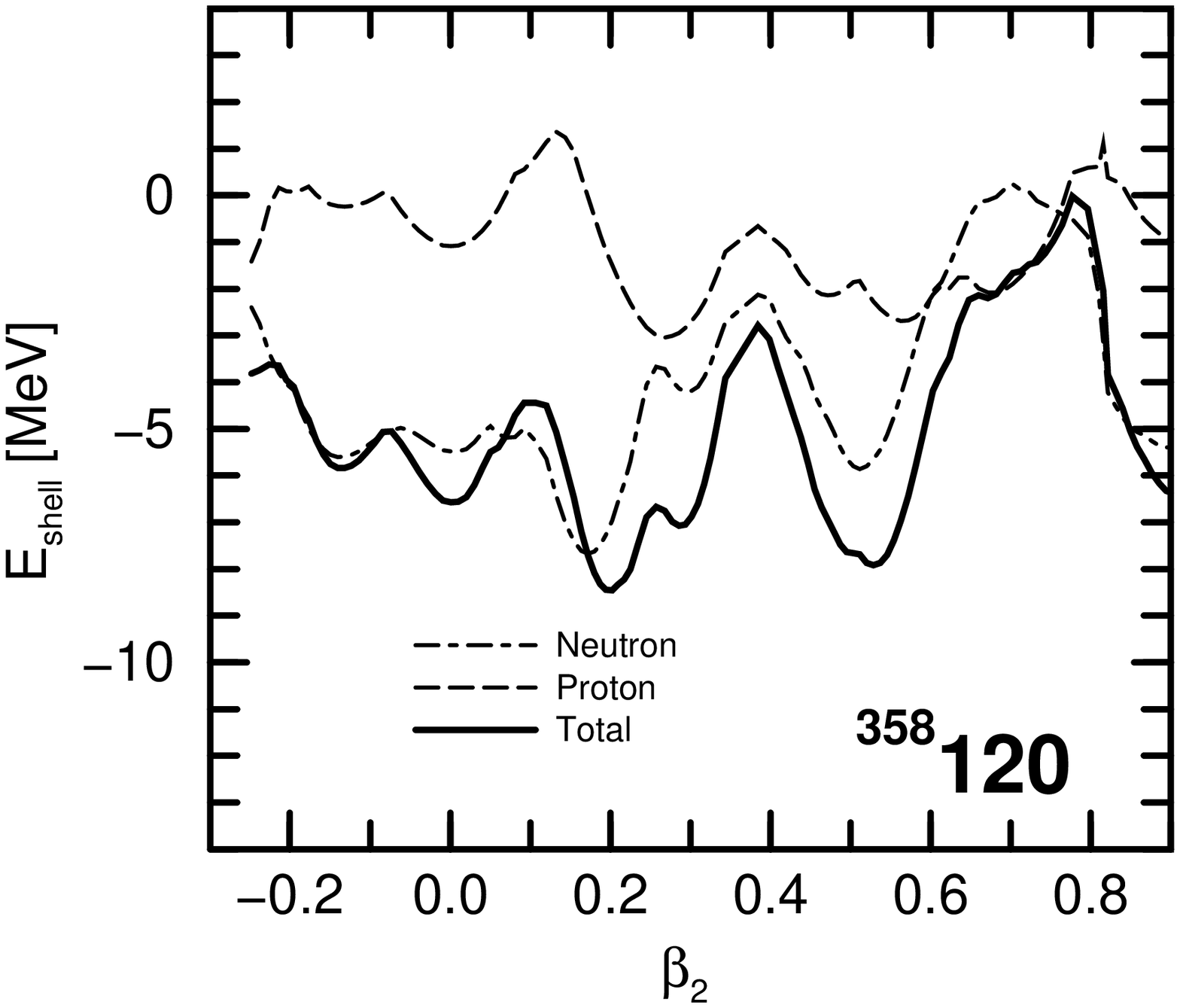}
\includegraphics[scale=0.45]{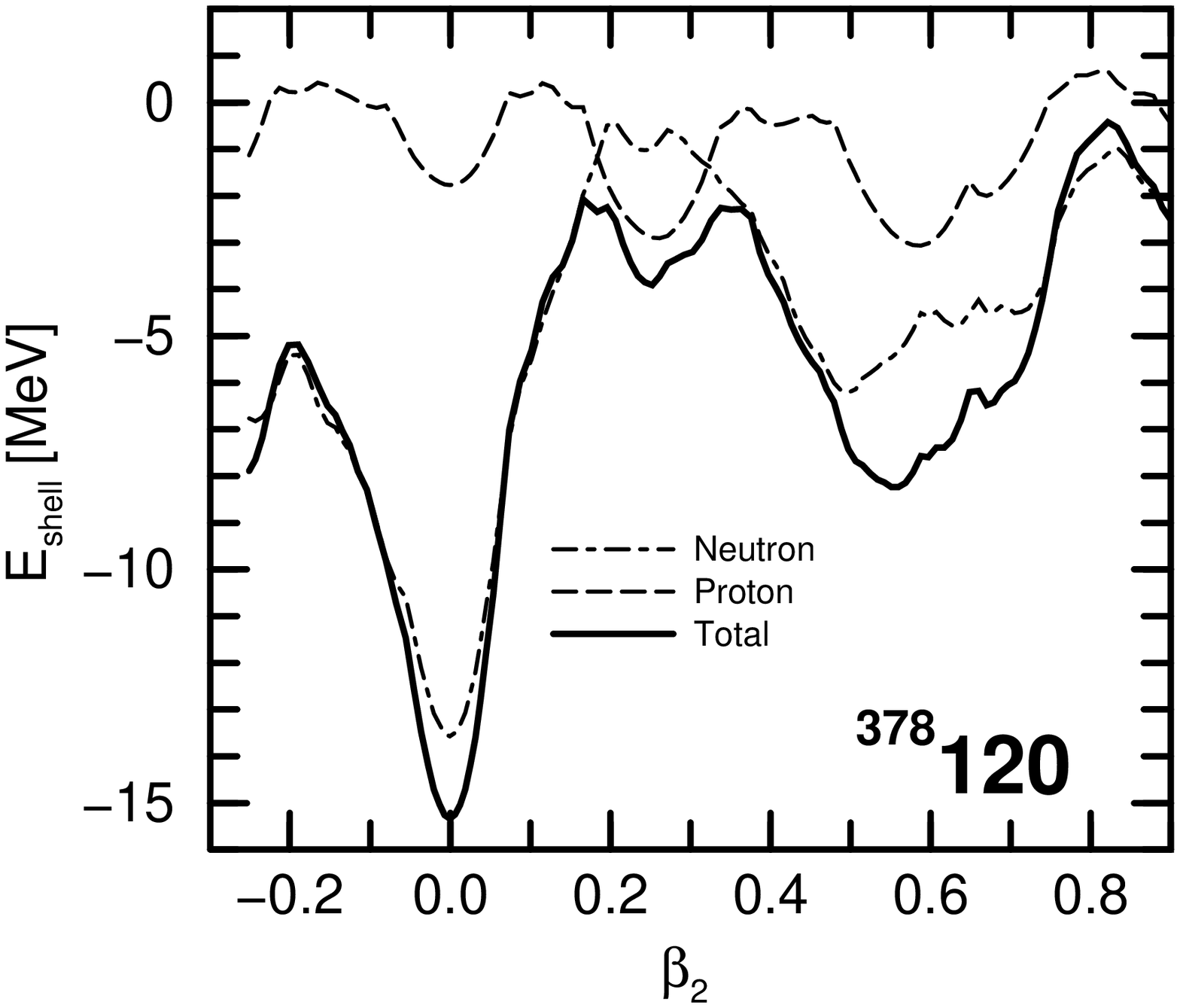}
\caption{The shell correction energy of
$^{292,304,318,348,358,378}$120 calculated in the constrained RMF
theory with effective interaction NL3.}
\label{120NL3s}
\end{figure}

To investigate the role of shell effects for superheavy nuclei,
the corresponding macroscopic energies as a function of
deformation $\beta_2$ with the effective interaction NL3 are
presented as the dashed lines in Fig. \ref{120NL3es}. The
macroscopic energy is defined as the difference between the
binding energy and the total shell correction energy which is
obtained similarly as in Section III.C. It can be seen that
without the shell effects, the superheavy nuclei hardly exist. It
is the shell effects that play an essential role to stabilize the
superheavy nuclei against the fission. In Fig. \ref{120NL3s}, the
shell correction energies for neutrons, protons, and nucleons are
respectively presented as the dot-dashed, dashed lines, and solid
lines. With the same proton number $Z$=120 , the shell correction
energies $E_{shell}$ in these nuclei sensitively depends on the
deformation $\beta_2$. The shell correction energies for protons
$E_{shell}^p$ at $\beta_2=0$ are respective $-5.5$, $-4.0$,
$-1.6$, $-0.7$, $-1.1$, and $-1.8$ MeV for $^{292}$120,
$^{304}$120, $^{318}$120, $^{348}$120, $^{358}$120, and
$^{378}$120, which partly explain the spherical configuration in
Fig. \ref{120NL3es}. In contrast to $E_{shell}^p$, the shell
correction energies for neutrons $E_{shell}^n$ at $\beta_2=0$ are
respective $-6.1$, $-2.6$, $2.1$, $-0.4$, $-5.5$, and $-13.5$ MeV.
Due to the large $E_{shell}^p$ and / or $E_{shell}^n$ energies,
the outstanding spherical local minima for $^{292}$120,
$^{304}$120 and $^{378}$120 are seen in Fig. \ref{120NL3es}. From
the constrained RMF calculation, we can conclude that for the
$Z=120$ isotope chain, $^{292}$120 and $^{378}$120 would be doubly
magic nuclei with spherical configuration.

Inspiring by the spherical absolute minimum of $^{292}$120
calculated by constrained RMF theory with NL3, we use the simple
cold fusion model with the doubly magic nuclei $^{208}$Pb as a
target to estimate the formation cross section $\sigma$ for
$^{292}$120 and $^{304}$120, although it is noticed that the
deformed actinide nuclei are recommended to synthesize superheavy
nuclei with A $>$ 112 \cite{Armbruster2003}. The properties of the
nuclei including the microscopic energy extracted from the
constrained RMF calculation are applied in the phenomenological
model \cite{Smolanczuk1999} with the reference reaction chosen as
$^{208}$Pb($^{48}$Ca,1n)$^{255}$No. The free parameter $C$ in the
model is determined by assuming the formation cross section
$\sigma$ of the reference reaction, i.e., $\sigma(C_{1})$=500nb
and $\sigma(C_{2})$=260nb. In Table \ref{120sigma}, the $Q$ value,
the height of the static fission barrier $B_f^{stat}$(ER), the
optimal excitation energy $E_{opt}^\ast$, and the formation cross
section $\sigma$ at $E_{opt}^\ast$ with the input from the
constrained RMF for effective interactions NL1, NL3, NLSH, and TM1
are listed. The strong dependence of the cross section $\sigma$ on
the RMF effective interactions can be seen. The largest cross
section $\sigma$ is given by NL1. Considered that the experimental
measure limit of cross section $\sigma$ is above 1pb, the two
possible doubly magic nuclei $^{292}$120 and $^{304}$120 can be
synthesized in the laboratory with the view for NL1. However the
results from NL1 are not supported by the other RMF interactions.
It should be mentioned that the results of NL3 are a little
different from those of NLSH and TM1, it gives nearly accessible
cross section ($\sigma\sim 1$ pb) for $^{292}$120.

\section{Summary}

We have investigated the possible doubly magic nuclei in the
superheavy region with proton number $Z$=100 - 140 and neutron
number $N$= ($Z$+30) - (2$Z$+32) within the RCHB theory with
interactions NL1, NL3, NLSH, TM1, TW99, DD-ME1, PK1, and PK1R.
Shell closure are quantified in terms of the two-nucleon
separation energies $S_{2p}$ and $S_{2n}$, the two-nucleon gaps
$\delta_{2p}$ and $\delta_{2n}$, the shell correction energies
$E_{shell}^{p}$ and $E_{shell}^{n}$, the pairing energies
$E_{pair}^{p}$ and $E_{pair}^{n}$ and the pairing gaps
$\Delta_{p}$ and $\Delta_{n}$ obtained in the RCHB theory. The
$\alpha$-decay half-lives are also discussed. $Z$=120, 132, and
138 and $N$=172, 184, 198, 228, 238, and 258 are inferred to be
magic numbers. In addition, the spherical configuration of the
doubly magic nuclei $^{292}$120 is supported by examining the
potential energy surfaces in the deformation-constrained RMF
theory and is consistent with other studies applying relativistic
forces\cite{Rutz1997}. The shell effects stabilizing the
superheavy nuclei are emphasized by extracting the shell
correction energies form the deformation-constrained RMF
calculation. Finally the formation cross sections of $^{292}$120
and $^{304}$120 with $^{208}$Pb as a target are estimated. With
the view of effective interaction NL1, $^{292}_{172}$120 may be
synthesized in the current experimental setup.

\begin{acknowledgments}

This work is partly supported by the Major State Basic Research
Development Program Under Contract Number G2000077407, the
National Natural Science Foundation of China under Grant Nos.
10025522, and 10221003, and the Doctoral Program Foundation from
the Ministry of Education in China.

\end{acknowledgments}

\clearpage
\renewcommand{\baselinestretch}{1}\small \normalsize
\bcen
\begin{table}[htbp]
\caption{The effective interactions PK1 and PK1R\cite{Long2003}.
The mass of the isovector vector meson is taken as $m_\rho$=763
MeV for PK1 and PK1R.}
\label{pk} \btab{cccccccccc}
\multicolumn{10}{c}{\vspace{0.2cm}}\\
\hline\hline
Effective   &$m_\sigma$ &$m_\omega$ &$g_\sigma$ &$g_\omega$ &$g_\rho$ &$g_2$     &$g_3$     &$c_3$     &$d_3$\\
interactions&[MeV]      &[MeV]      &           &           &$[fm^{-1}]$&        &          &          &     \\
\hline
PK1  &514.089 &784.254 &10.322  &13.013  &4.530 &-8.169 &-9.998 &55.636 &  0\\
PK1R &514.087 &784.222 &10.322  &13.013  &4.550 &-8.156 &-10.198&54.446 &350\\
\hline\hline \etab
\end{table}
\ecen

\bcen
\begin{table}
\caption{The possible magic proton number suggested by two-proton
separation energies, two-proton gaps, shell correction energies,
pairing energies and effective pairing gaps for proton with
interactions NL1, NL3, NL-SH, TM1, TW-99, DD-ME1, PK1 and PK1R,
respectively.}
\label{magicp} \btab{c|c|*{6}{c}}
\multicolumn{8}{c}{\vspace{0.2cm}}\\
\hline\hline \multirow{2} {3cm}   {Effective interactions} &
\multirow{2} {1.5cm} {Quantity}      &
\multicolumn{6}{c}{Magic proton number candidates}\\
\cline{3-8}
                         &               &106&114&120&126&132&138\\
\hline
\multirow{5} {1.5cm}{NL1}&$S_{2p}$       &   &   &\ks&   &\ks&\ks\\
                         &$\delta_{2p}$  &   &   &\ks&\ks&\ks&\ks\\
                         &$E_{shell}^{p}$&\ks&   &\ks&\ks&\ks&\ks\\
                         &$E_{pair}^{p}$ &\ks&   &\ks&\ks&\ks&\ks\\
                         &$\Delta_{p}$   &\ks&   &\ks&\ks&\ks&\ks\\
\hline
\multirow{5} {1.5cm}{NL3}&$S_{2p}$       &\ks&   &\ks&   &\ks&\ks\\
                         &$\delta_{2p}$  &\ks&   &\ks&   &\ks&\ks\\
                         &$E_{shell}^{p}$&\ks&   &\ks&   &\ks&\ks\\
                         &$E_{pair}^{p}$ &\ks&\ks&\ks&   &\ks&\ks\\
                         &$\Delta_{p}$   &\ks&\ks&\ks&   &\ks&\ks\\
\hline
\multirow{5} {2cm} {NLSH}&$S_{2p}$       &\ks&   &\ks&   &\ks&\ks\\
                         &$\delta_{2p}$  &\ks&\ks&\ks&   &\ks&\ks\\
                         &$E_{shell}^{p}$&\ks&\ks&\ks&   &\ks&\ks\\
                         &$E_{pair}^{p}$ &\ks&\ks&\ks&   &\ks&\ks\\
                         &$\Delta_{p}$   &\ks&\ks&\ks&   &\ks&\ks\\
\hline
\multirow{5} {1.5cm}{TM1}&$S_{2p}$       &   &   &\ks&   &\ks&\ks\\
                         &$\delta_{2p}$  &   &   &\ks&   &\ks&\ks\\
                         &$E_{shell}^{p}$&   &   &\ks&   &\ks&\ks\\
                         &$E_{pair}^{p}$ &   &   &\ks&   &\ks&\ks\\
                         &$\Delta_{p}$   &   &   &\ks&   &\ks&\ks\\
\hline
\multirow{5} {2cm} {TW99}&$S_{2p}$       &\ks&   &\ks&   &\ks&\ks\\
                         &$\delta_{2p}$  &\ks&\ks&\ks&   &\ks&\ks\\
                         &$E_{shell}^{p}$&\ks&\ks&\ks&   &\ks&\ks\\
                         &$E_{pair}^{p}$ &\ks&\ks&\ks&\ks&\ks&\ks\\
                         &$\Delta_{p}$   &\ks&\ks&\ks&\ks&\ks&\ks\\
\hline
\multirow{5}{2cm}{DD-ME1}&$S_{2p}$       &   &   &\ks&   &\ks&\ks\\
                         &$\delta_{2p}$  &\ks&   &\ks&   &\ks&\ks\\
                         &$E_{shell}^{p}$&\ks&   &\ks&   &\ks&\ks\\
                         &$E_{pair}^{p}$ &\ks&\ks&\ks&\ks&\ks&\ks\\
                         &$\Delta_{p}$   &\ks&\ks&\ks&\ks&\ks&\ks\\
\hline
\multirow{5} {1.5cm}{PK1}&$S_{2p}$       &\ks&   &\ks&   &\ks&\ks\\
                         &$\delta_{2p}$  &\ks&   &\ks&   &\ks&\ks\\
                         &$E_{shell}^{p}$&\ks&\ks&\ks&   &\ks&\ks\\
                         &$E_{pair}^{p}$ &\ks&\ks&\ks&   &\ks&\ks\\
                         &$\Delta_{p}$   &\ks&\ks&\ks&   &\ks&\ks\\
\hline
\multirow{5} {2cm} {PK1R}&$S_{2p}$       &\ks&   &\ks&   &\ks&\ks\\
                         &$\delta_{2p}$  &\ks&   &\ks&   &\ks&\ks\\
                         &$E_{shell}^{p}$&\ks&\ks&\ks&   &\ks&\ks\\
                         &$E_{pair}^{p}$ &\ks&\ks&\ks&   &\ks&\ks\\
                         &$\Delta_{p}$   &\ks&\ks&\ks&   &\ks&\ks\\
\hline\hline \etab
\end{table}
\ecen

\bcen
\begin{table}
\caption{The possible magic neutron number suggested by
two-neutron separation energies, two-neutron gaps, shell
correction energies, pairing energies and effective pairing gaps
for neutron with interactions NL1, NL3, NL-SH, TM1, TW-99, DD-ME1,
PK1 and PK1R, respectively.}
\label{magicn} \btab{c|c|*{11}{c}}
\multicolumn{13}{c}{\vspace{0.2cm}}\\
\hline\hline \multirow{2} {3cm}   {Effective interactions} &
\multirow{2}{1.5cm} {Quantity}      &
\multicolumn{11}{c}{Magic neutron number candidates}\\
\cline{3-13}
                         &               &138&164&172&184&198&216&228&238&252&258&274\\
\hline
\multirow{5} {1.5cm}{NL1}&$S_{2n}$       &\ks&   &\ks&\ks&\ks&   &\ks&\ks&\ks&\ks&\ks\\
                         &$\delta_{2n}$  &\ks&   &\ks&\ks&\ks&   &\ks&\ks&   &\ks&\ks\\
                         &$E_{shell}^{n}$&   &   &   &\ks&   &   &   &   &   &\ks&   \\
                         &$E_{pair}^{n}$ &\ks&\ks&\ks&\ks&\ks&   &\ks&\ks&\ks&\ks&\ks\\
                         &$\Delta_{n}$   &\ks&\ks&\ks&\ks&\ks&   &\ks&\ks&\ks&\ks&\ks\\
\hline
\multirow{5} {1.5cm}{NL3}&$S_{2n}$       &\ks&   &\ks&\ks&\ks&   &\ks&\ks&\ks&\ks&\ks\\
                         &$\delta_{2n}$  &\ks&\ks&\ks&\ks&\ks&   &\ks&\ks&   &\ks&\ks\\
                         &$E_{shell}^{n}$&   &   &\ks&\ks&   &   &   &   &   &\ks&   \\
                         &$E_{pair}^{n}$ &\ks&\ks&\ks&\ks&\ks&\ks&\ks&\ks&\ks&\ks&\ks\\
                         &$\Delta_{n}$   &\ks&\ks&\ks&\ks&\ks&\ks&\ks&\ks&\ks&\ks&\ks\\
\hline
\multirow{5} {2cm} {NLSH}&$S_{2n}$       &\ks&\ks&\ks&\ks&\ks&\ks&\ks&\ks&\ks&\ks&\ks\\
                         &$\delta_{2n}$  &\ks&\ks&\ks&\ks&\ks&\ks&\ks&\ks&   &\ks&\ks\\
                         &$E_{shell}^{n}$&   &\ks&\ks&\ks&   &   &   &   &\ks&\ks&   \\
                         &$E_{pair}^{n}$ &\ks&\ks&\ks&\ks&\ks&\ks&\ks&\ks&\ks&\ks&\ks\\
                         &$\Delta_{n}$   &\ks&\ks&\ks&\ks&\ks&\ks&\ks&\ks&\ks&\ks&\ks\\
\hline
\multirow{5} {1.5cm}{TM1}&$S_{2n}$       &\ks&   &\ks&\ks&\ks&   &\ks&\ks&\ks&\ks&\ks\\
                         &$\delta_{2n}$  &\ks&   &\ks&\ks&\ks&   &\ks&\ks&   &\ks&\ks\\
                         &$E_{shell}^{n}$&   &   &\ks&\ks&   &   &   &   &   &\ks&   \\
                         &$E_{pair}^{n}$ &\ks&\ks&\ks&\ks&\ks&   &\ks&\ks&\ks&\ks&\ks\\
                         &$\Delta_{n}$   &\ks&\ks&\ks&\ks&\ks&   &\ks&\ks&\ks&\ks&\ks\\
\hline
\multirow{5} {2cm} {TW99}&$S_{2n}$       &\ks&\ks&\ks&\ks&\ks&\ks&\ks&\ks&\ks&\ks&\ks\\
                         &$\delta_{2n}$  &\ks&\ks&\ks&\ks&\ks&\ks&\ks&\ks&   &\ks&\ks\\
                         &$E_{shell}^{n}$&   &\ks&\ks&\ks&   &\ks&   &   &   &\ks&   \\
                         &$E_{pair}^{n}$ &\ks&\ks&\ks&\ks&\ks&\ks&\ks&\ks&\ks&\ks&\ks\\
                         &$\Delta_{n}$   &\ks&\ks&\ks&\ks&\ks&\ks&\ks&\ks&\ks&\ks&\ks\\
\hline
\multirow{5}{2cm}{DD-ME1}&$S_{2n}$       &\ks&\ks&\ks&\ks&\ks&   &\ks&\ks&\ks&\ks&\ks\\
                         &$\delta_{2n}$  &\ks&\ks&\ks&\ks&\ks&   &\ks&\ks&   &\ks&\ks\\
                         &$E_{shell}^{n}$&   &   &\ks&\ks&   &\ks&   &   &   &\ks&   \\
                         &$E_{pair}^{n}$ &\ks&\ks&\ks&\ks&\ks&   &\ks&\ks&\ks&\ks&\ks\\
                         &$\Delta_{n}$   &\ks&\ks&\ks&\ks&\ks&   &\ks&\ks&\ks&\ks&\ks\\
\hline
\multirow{5} {1.5cm}{PK1}&$S_{2n}$       &\ks&\ks&\ks&\ks&\ks&\ks&\ks&\ks&\ks&\ks&\ks\\
                         &$\delta_{2n}$  &\ks&\ks&\ks&\ks&\ks&\ks&\ks&\ks&   &\ks&\ks\\
                         &$E_{shell}^{n}$&   &   &\ks&\ks&   &   &   &   &   &\ks&   \\
                         &$E_{pair}^{n}$ &\ks&\ks&\ks&\ks&\ks&\ks&\ks&\ks&\ks&\ks&\ks\\
                         &$\Delta_{n}$   &\ks&\ks&\ks&\ks&\ks&\ks&\ks&\ks&\ks&\ks&\ks\\
\hline
\multirow{5} {2cm} {PK1R}&$S_{2n}$       &\ks&\ks&\ks&\ks&\ks&\ks&\ks&\ks&\ks&\ks&\ks\\
                         &$\delta_{2n}$  &\ks&\ks&\ks&\ks&\ks&\ks&\ks&\ks&   &\ks&\ks\\
                         &$E_{shell}^{n}$&   &   &\ks&\ks&   &   &   &   &   &\ks&   \\
                         &$E_{pair}^{n}$ &\ks&\ks&\ks&\ks&\ks&\ks&\ks&\ks&\ks&\ks&\ks\\
                         &$\Delta_{n}$   &\ks&\ks&\ks&\ks&\ks&\ks&\ks&\ks&\ks&\ks&\ks\\
\hline\hline \etab
\end{table}
\ecen

\begin{table}[htbp]
\caption{The $Q$ value, the height of the static fission barrier
$B_f^{stat}$(ER), the optimal excitation energy $E_{opt}^\ast$ and
$C$-dependent formation cross section $\sigma$ at $E_{opt}^\ast$
for the cold fusion reaction $^{208}$Pb($^{85}$Sr,1n)$^{292}$120
and $^{208}$Pb($^{97}$Sr,1n)$^{304}$120 with the input from the
constrained RMF for effective interactions NL1, NL3, NLSH and TM1}
\label{120sigma} \btab{ccccccc}
\multicolumn{7}{c}{\vspace{0.2cm}}\\
\hline\hline \multirow{2} {4cm}{Cold fusion reactions}&
RMF&$Q$&$B_f^{stat}$(ER)&$E_{opt}^\ast$&\multicolumn{2}{c}{$\sigma$}\\
\cline{6-7}
&Interactions&[MeV]&[MeV]&[MeV]&~~$\sigma(C_{1})$~~&~~$\sigma(C_{2})$~~\\
\hline \multirow{4} {4cm}{$^{208}$Pb($^{85}$Sr,1n)$^{292}$120}
&  NL1&    296.51&     10.62&     14.78&    15.8pb&     9.5pb\\
&  NL3&    302.84&      8.15&     14.92&     1.1pb&     720fb\\
& NLSH&    297.88&      7.21&     13.94&     0.26fb&   0.16fb\\
&  TM1&    296.83&      5.74&     12.83&     0.47fb&   0.29fb\\
\hline \multirow{4} {4cm}{$^{208}$Pb($^{97}$Sr,1n)$^{304}$120}
&  NL1&    308.42&     12.62&      9.96&    86.8nb&    55.1nb\\
&  NL3&    305.09&      3.98&      9.55&    0.64fb&    0.40fb\\
& NLSH&    300.37&      5.15&     10.49&    13.8fb&     8.6fb\\
&  TM1&    299.73&      5.51&     11.21&    20.5fb&    12.6fb\\
\hline\hline \etab \end{table}

\newpage






\end{document}